\documentclass[reprint,amsmath,amssymb,aps,showkeys]{revtex4-1}

\usepackage{graphicx}
\usepackage{bm}
\usepackage{dcolumn}
\usepackage{soul}

\begin{document}
\title{Introduction to Supersymmetric Theory of Stochastics}

\author{Igor V. Ovchinnikov}

\affiliation{Department of Electrical Engineering, University of California at Los Angeles, Los Angeles, CA  90095, USA}
\email{igor.vlad.ovchinnikov@gmail.com} 

\begin{abstract}
Many natural and engineered dynamical systems, including all living objects, exhibit signatures of what can be called spontaneous dynamical long-range order (DLRO). This order's omnipresence has long been recognized by the scientific community, as evidenced by a myriad of related concepts, theoretical and phenomenological frameworks, and experimental phenomena such as turbulence, $1/f$ noise, dynamical complexity, chaos and the butterfly effect, the Richter scale for earthquakes and the scale-free statistics of other sudden processes, self-organization and pattern formation, self-organized criticality, \emph{etc}. Although several successful approaches to various realizations of DLRO have been established, the universal theoretical understanding of this phenomenon remained elusive. The possibility of constructing a unified theory of DLRO has emerged recently within the approximation-free supersymmetric theory of stochastics (STS). There, DLRO is the spontaneous breakdown of the topological or de Rham supersymmetry that all stochastic differential equations (SDEs) possess. This theory may be interesting to researchers with very different backgrounds because the ubiquitous DLRO is a truly interdisciplinary entity. The STS is also an~interdisciplinary construction. This theory is based on dynamical systems theory, cohomological field theories, the theory of pseudo-Hermitian operators, and the conventional theory of SDEs. Reviewing the literature on all these mathematical disciplines can be time consuming. As~such, a~concise and self-contained introduction to the STS, the goal of this paper, may be useful.\\
\end{abstract}


\maketitle

\section{Introduction}
\vspace{-6pt}
\label{Chap:Intro}

\subsection{Dynamical Long-Range Order}
\label{Sec:MysteryChaos}

It is well established experimentally and numerically that many seemingly unrelated sudden processes in astrophysics \cite{Asc11}, geophysics \cite{Gut55}, neurodynamics \cite{Beg04,Chialvo10}, econodynamics \cite{Pre11}, and other branches of modern science exhibit power-law statistics, the very reason why the Richter scale is logarithmic. This is simply one example of the spontaneous long-range dynamical behavior (LRDB) that emerges in many nonlinear dynamical systems (DSs) with no underlying long-range interactions that could potentially explain such behavior. Two other well-known examples of LRDB are the infinitely long memory of perturbations known as the butterfly effect \cite{ButterFly}, and the algebraic power-spectra commonly known as $1/f$ noise or the long-term memory effect \cite{Kog96} found in many existing DSs, including apparently all living objects \cite{BookHeartBrainNoise,Biology1fNoise}.

It was understood that the LRDB must be a signature of some type of spontaneous dynamical long-range order (DLRO). The existence and omnipresence of this DLRO has long been recognized by the scientific community, as evidenced by a myriad of related concepts, including 
chaos \cite{Rue14,Mot14,Shep14},
turbulence \cite{RuelleTurb, Turbulence}, 
dynamical complexity \cite{DynamicalComplexity}, 
self-organization \cite{SelfOrganization},
pattern formation \cite{patternFomration}, and
self-organized criticality \cite{Bak87}. 

Several successful approaches to various realizations of DLRO have been established. For example, the concept of deterministic chaos is a centerpiece of the well-developed dynamical system (DS) theory. 
Nevertheless, there existed no universal theoretical understanding of DLRO. In particular, no rigorous stochastic generalization of the concept of deterministic chaos existed previously, whereas all natural DSs are never completely isolated from their environments and are thus always stochastic.

A class of models with the potential to reveal the mathematical essence of the ubiquitous DLRO is the stochastic (partial) differential equations (SDEs). Indeed, SDEs most likely have the widest applicability in modern science. In physics, for example, SDEs are the effective equations of motion (EoM) for all physical systems above the scale of quantum degeneracy/coherence. In quantum models, SDEs are used in a variety of ways. For example, SDEs play a central role in quantum optics (see, e.g., \cite{QuantumOpt} and the references therein). 
In many-body quantum models, SDEs are used in the investigation of non-equilibrium quantum dynamical phenomena in the form of the effective EoM of the collective quantum modes \cite{SDEQuuantum} and order parameters \cite{GinzburgLandauSDE}. They also represent a useful tool for quantum statistics~\cite{Ringel}. In other scientific disciplines, SDEs are even more fundamental, as they appear at the level of the very formulation of dynamics, unlike the EoM in physics, which descend from least action~principles.

The theory of stochastic dynamics has a long history. Many important insights into stochastic dynamics have been provided so far (see, e.g.,  \cite{Oks10,Kunita1,Baxendale1,Arn03,Watanabe1,Crauel1,Kap90,LaJen1} and the references therein). Nevertheless, the mathematical essence of DLRO remained elusive.

\subsection{Topological Supersymmetry of Continuous Time Dynamics}
\label{Sec:Emergence}

One way of deducing the potential theoretical origin of DLRO is provided by the following qualitative yet solid argument. From the field-theoretic point of view, LRDB is indicative of the presence of a gapless excitation with an infinite correlation length/time. There are only two possible scenarios for such a situation to occur: the accidental or critical scenario and the Goldstone scenario. In~the accidental scenario, the parameters of the model can be fine-tuned to ensure that a certain excitation has zero gap. This is exactly the situation with (structural) phase transitions, where an excitation called the soft mode becomes gapless exactly at the transition temperature (or other parameter). This allows the system to move effortlessly from a previously stable vacuum to a new vacuum. Immediately following the transition, the soft mode ``hardens'' again, \emph{i.e.}, it acquires a finite gap that signifies the dynamical stability of the new vacuum. In other words, only at exactly the transition point the soft mode is gapless and thus has an infinite correlation length/time.

The accidental scenario for DLRO contradicts the fact that DLRO is robust against moderate variations in the parameters of the model. For example, a slight variation in the magnitude of the electric current flowing through a dirty conductor will not destroy the $1/f$ noise. In other words, in phase diagrams, DLRO occupies full-dimensional phases and not the lower dimensional transitions/boundaries between different full-dimensional phases. 
This observation unambiguously suggests that the Goldstone scenario is the only possibility for the field-theoretic explanation of DLRO. More specifically, the Goldstone theorem states that, under the conditions of the spontaneous breakdown of a global continuous symmetry, the ground state is degenerate and that, in spatially extended models, this degeneracy tailors the existence of a~gapless excitation called the Goldstone--Nambu particle. As such, DLRO may be the result of the spontaneous breakdown of some global continuous symmetry. 

It is understood that the symmetry responsible for DLRO cannot be a conventional bosonic symmetry because DSs with no bosonic symmetries can also exhibit DLRO, e.g., be chaotic. In~other words, DLRO must be a result of the spontaneous breakdown of some fermionic symmetry or supersymmetry. It has long been known that supersymmetries are indeed present in some classes of SDEs. The work on supersymmetric theories of SDEs began with the Parisi--Sourlas stochastic quantization procedure \cite{ParSour,ParSour1,CG,CG1,DH,Gozzi0,Olenskoi1,KurchanSpin,Dijkgraaf,KS, ZinnJustin,Bau88,Bau89-1,Bau93}, which leads from a Langevin SDE, \emph{i.e.}, an SDE with a~gradient flow vector field, to a model with $N=2$ supersymmetry. The Parisi--Sourlas quantization procedure was later identified as a realization of the concept of Nicolai maps \cite{Nicolai1,Nicolai2} and ``half'' of the $N=2$ supersymmetry as a corresponding Becchi--Rouet--Stora--Tyutin (BRST) or topological supersymmetry,\cite{Bau88} which is a definitive feature of Witten-type topological or cohomological field theories~\cite{Frenkel2007215, Birmingham1991129, labastida1989, Witten98, Witten981, Wit82, DynSusyBrWitten,Bau89}. Similar~supersymmetries have been studied in classical mechanics \cite{Gozzi1,Gozzi2,Gozzi3,Gozzi4,Gozzi_New,Deotto_1,Niemi1, Niemi2} and its stochastic generalization~\cite{Kurchan}. 

From the perspective of the theory of ubiquitous DLRO, the consideration of specific models is clearly insufficient. In reality, EoM are never exactly Langevin or classical mechanical, and the generalization of the discussion to all or at least most of SDEs is necessary. In other words, the supersymmetry responsible for DLRO must be an attribute of all SDEs to be able to account for omnipresence of DLRO in nature. The remaining question in the Goldstone scenario of the theory of DLRO now is whether such supersymmetry exists.

Traces of this supersymmetry can be found in the literature. In \cite{Gaw86}, for example, the authors considered a non-potential generalization of the Langevin stochastic dynamics and noted that half of the $N=2$ supersymmetry survives a non-potential perturbation. Nevertheless, to the best of the knowledge of the present author, this $N=1$ supersymmetry in the context of all SDEs has not been addressed previously. One possible reason for this is the pseudo-Hermitianity of the stochastic evolution operator of a general SDE. Specifically, the theory of pseudo-Hermitian evolution operators appeared only relatively recently \cite{Mos02,Mos021,Mos022,Mos023,Mos13} as a generalization of the theory of $PT$-symmetric evolution operators \cite{Bend98,Bend981,Fernandez98,Bend99,Mezincescu2000}. 
It was only after the theory of pseudo-Hermitian operators became available that studies on topological supersymmetry in the context of the general SDEs could be resumed. The idea that the spontaneous breakdown of this supersymmetry pertinent to all SDEs may be the mathematical essence of DLRO, or rather of one of its realizations known previously as self-organized criticality, was reported in \cite{Ovc11}. Further work in this direction \cite{Ovc12,Ovc13,Ovc14,Ovc16} resulted in the formulation of what can be called the supersymmetric theory of stochastics (STS). The goal of this paper is to present the current state of the STS in a self-consistent manner. This paper can be viewed as a compilation of a few previous works and as a compilation that corrects several mistakes made during the early stages of the development of the STS and that clarifies a couple of points that were previously swept under the carpet. This paper also presents a few new results, including a discussion of the pseudo-time reversal~symmetry. 

Given the multidisciplinary character of STS, it would take an enormous amount of work to review all the relevant results from DS theory, cohomological field theory, the classical theory of SDEs, and physics. This goal is not pursued in this paper, and references are provided on only the most relevant results that are known to the author and that the material presented here is directly based on. The author would like to apologize in advance if some important related works have escaped his~attention.

\subsection{Relation to Existing Theories}
\label{Intro:Relation}

The topological supersymmetry breaking picture of DLRO aligns well with the previous understanding of the concept of dynamical chaos. For example, the nontrivial connection between chaos and topology is at the heart of the topological theory of chaos \cite{Gil98}. Furthermore, it was also known that, in some cases, the transition into chaos must be a phase transition of some sort, as evident from certain universal features of the onset of chaotic behavior \cite{UniversalityInChaos}. The only unexpected insight from the supersymmetry breaking picture of DLRO is the fact that its mathematical essence is in a sense opposite to the semantics of the word chaos. Indeed, chaos literally means ``absence of order'', whereas the phase with the spontaneously broken supersymmetry is the low-symmetry or ``ordered'' phase. This is why DLRO may be a more accurate identifier for this phenomenon than, say, stochastic chaos. In this paper, both terms will be used interchangeably. 

STS in a nutshell is the following. An SDE defines the noise-configuration-dependent trajectories in the phase space. The collection of all these trajectories can be viewed as a family of noise-configuration-dependent phase space diffeomorphisms. Instead of studying the trajectories, one can equivalently study the actions, called the pullbacks, that these diffeomorphisms induce on the exterior algebra of the phase space. The original trajectories can be reconstructed from these pullbacks so that the later contain all the information on the SDE-defined dynamics. 

The pullbacks have one very important advantage over the trajectories. Unlike trajectories in the general case of a nonlinear phase space, the pullbacks are linear objects and can thus be averaged over the noise configurations. Such a stochastically averaged pullback is the finite-time stochastic evolution operator (SEO). Thus, it becomes clear where the supersymmetry originates from: all the diffeomorphism-induced pullbacks and consequently the finite-time SEO are commutative with the exterior derivative, which is thus a (super-)symmetry of any SDE. In other words, the existence of this supersymmetry in all SDEs is merely the algebraic version of the most fundamental and indisputable statement that continuous dynamics preserves the continuity of the phase space. 

{\em Using the concept of trajectories, the same idea can be explained as follows. Diffeomorphic character and/or topological supersymmetry of continuous time dynamics is the property that (for any noise configuration) close initial points generate close trajectories. When the topological supersymmetry is broken spontaneously, it can be said that this property is violated and close initial points may give rise to trajectories that part in the limit of long propagation (described by the non-supersymmetric ground state). This is nothing else but the famous butterfly effect of chaotic dynamics. In other words, spontaneous topological supersymmetry breaking is a stochastic generalization of deterministic chaos.}

The idea to study pullbacks induced by random maps averaged over noise configurations appeared first, to the best of this author's knowledge, in DS theory, where the analogue of the finite-time SEO is known as the generalized transfer operator \cite{Rue02}. From this perspective, the STS can be viewed as a continuation of DS theory. On the side of the Parisi--Sourlas quantization procedure, the path integral representation of the Witten index of the STS is a member of the cohomological field theories. Furthermore, the SEO of the general SDE is pseudo-Hermitian; thus, the STS is within the domain of applicability of the theory of pseudo-Hermitian operators. In other words, the STS is a~multidisciplinary mathematical construction. It combines a few major mathematical disciplines that are naturally synergetic within the STS. This synergy can ensure fruitful cross-fertilization during  future work on the STS. To date, the STS has already provided a few novel findings, therein making it interesting from several points of view, as discussed below.

For DS theory, an interesting result from the STS is the established equivalence between the so-called sharp-trace of the generalized transfer operator, the stochastic Lefschetz index of the corresponding SDE-defined diffeomorphisms, and the Witten index of the STS. From the perspective of the conventional theory of SDEs, a valuable result from the STS is the demonstrated equivalence between the Stratonovich interpretation of SDEs and the (bi-graded) Weyl symmetrization procedure. For a field theorist, there are two potentially interesting results from the STS. First, the cohomological field theories, or rather the methodology developed within them (e.g., the localization principle and topological invariants as expectation values on instantons), together with the theory of pseudo-Hermitian evolution operators, may find multiple applications in almost all branches of modern science. Second, there are very few known analytical mechanisms that can result in the spontaneous breakdown of supersymmetry \cite{AFewMechanismsForSusybreaking}, which is basically one of the main reasons behind the introduction of the concept of explicit (or soft) supersymmetry breaking \cite{SoftSusyBreaking}. The STS provides yet another such mechanism: the topological supersymmetry in (deterministic) chaotic DSs is spontaneously broken by the non-integrability of the flow vector field. 

From a wider perspective, SDEs find applications in almost all modern scientific disciplines, ranging from social sciences and econodynamics to astrophysics and high-energy physics. Therefore,~the STS in general and this paper in particular may be interesting to specialists working in any of these areas of science.

\subsection{Models of Interest and the Structure of This Paper}
\label{Sec:Models}

The following class of SDEs that covers most of the models in the literature will be of primary~interest:
\begin{eqnarray}
\dot x(t)  = F(x(t)) + (2\Theta)^{1/2}e_a(x(t))\xi^a(t) \equiv {\mathcal F} (t).\label{SDE}
\end{eqnarray}

Here and in the following, summation over repeated indexes is assumed; $x(t):\mathbb{R} \to X$ is a~trajectory of the DS in a $D$-dimensional topological manifold called the phase space, $X$; $F(x) \in TX_x$ is the flow vector field from the tangent space of $X$ at the point $x$; $\xi = \{\xi^a\in \mathbb{R},a=1,2... \}$ are noise variables; and $e_a(x) \in TX_x$ is a set of vector fields. The position-dependent/independent $e$ are often called multiplicative/additive noise. The notation $\mathcal F$ is introduced to separate the flow perturbed by the noise from the deterministic flow, $F$. As will be discussed in Section \ref{ItoStratSubSection}, the SDE in Equation~(\ref{SDE}) is the Stratonovich SDE along the lines of stochastic calculus on manifolds (see, e.g., \cite{ManifoldsSDE} and the references therein). It will also be argued that the STS appears to point to the possibility that the Stratonovich approach is the only correct choice for continuous time models. 

The parameter $\Theta$ represents the temperature or rather the intensity of the noise. As will be made clear below in Section \ref{Chap:StochasticDynamics}, the vector fields $e_a$ define the noise-induced metric on $X$: $g^{ij}(x)=e^i_a(x) e^j_a(x)$. Therefore, in situations wherein the number of vector fields $e_a$ equals the dimensionality of the phase space, these vector fields can be identified as veilbeins (see, e.g., Chapter 7 of  \cite{Nakahara}). In the general case, however, the number of $e$s must not necessarily be equal to the dimensionality of the phase space.

Most of the discussion will be directed toward models with Gaussian white noise. The probability distribution of its configurations is
\begin{eqnarray}
P_\text{Ns}(\xi) = C e^{ - \int dt (\xi^a(t) \xi^a(t) )/2 },\label{GaussianProbability}
\end{eqnarray}
with $C$ being a normalization constant such that
\begin{eqnarray}
\langle 1 \rangle_\text{Ns} \equiv \iint D\xi \cdot 1 \cdot P(\xi) = 1.
\end{eqnarray}

Here, the functional or infinitely dimensional integration is over all the configurations of the noise. The stochastic expectation value of some functional $f(\xi)$ is defined as
\begin{eqnarray}
\langle f(\xi) \rangle_\text{Ns} \equiv \iint D\xi f(\xi) P_\text{Ns}(\xi).\label{DefStochAverage}
\end{eqnarray}

The fundamental correlator of the Gaussian white noise is
\begin{eqnarray}
\langle \xi^{a}(t)\xi^{b}(t')\rangle_\text{Ns} = \delta^{ab}\delta(t-t').\label{GaussianAverage}
\end{eqnarray}


\begin{figure}[h]
\centering
\includegraphics[height=3.9cm, width=8cm]{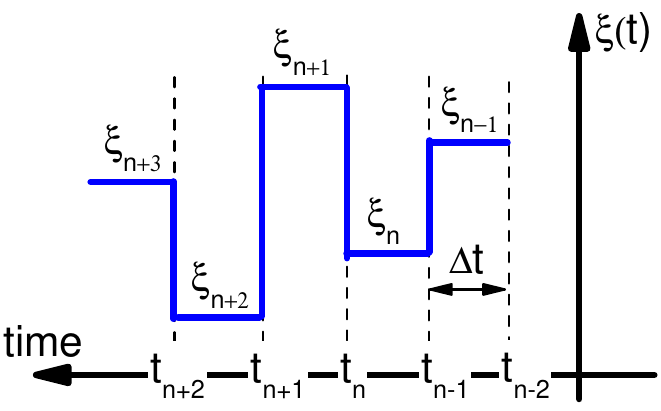}
\caption{\label{Figure_2_1} {Piece-wise constant approximation} for Gaussian white noise. Each $\xi_n\equiv \left.\xi(t)\right|_{t_n<t<t_{n-1}}$ is a~random Gaussian variable. The time in the figure flows from right to left. This is conventional in both quantum theory and the theory under consideration, as discussed at the end of Section \ref{SecLieDerivative}.}
\end{figure}

The infinite-dimensional integrations in Equations (\ref{DefStochAverage}) and (\ref{GaussianAverage}) can be given a more concrete meaning by splitting the time domain into a large number of intervals with infinitesimal duration $\Delta t$ and then taking the continuous time limit, namely, $\Delta t\to0$. Before taking this limit, each noise configuration can be viewed as a piece-wise constant function (see Figure \ref{Figure_2_1}) on each interval, \emph{i.e.}, the value of the noise variable $\xi^a(t)=\xi^a_n$ for $t_{n}>t>t_{n-1}$. The discrete-time version of the probability distribution of the Gaussian white noise in Equation (\ref{GaussianProbability}) is
\begin{eqnarray}
P_\text{Ns}(\xi) \propto e^{- \Delta t\sum_n \xi^a_n \xi^a_n/2},
\end{eqnarray}
and that of the correlator in Equation (\ref{GaussianAverage}) is
\begin{eqnarray}
\langle \xi^a_n \xi ^b_{n'}\rangle_\text{Ns} = \Delta t^{-1} \delta^{ab}\delta_{nn'},\label{AverageNoise}
\end{eqnarray}
whereas all the other (even) order correlators are
\begin{eqnarray}
\langle \xi^{a_1}_{n_1} ... \xi^{a_{2k}}_{n_{2k}}\rangle_\text{Ns}\propto \Delta t^{-k}.\label{AverageNoise1}
\end{eqnarray}

The theory of stochastic dynamics defined by Equation (\ref{SDE}) can be constructed in two steps. The~first step is to understand the deterministic temporal evolution defined by the ordinary differential equation (ODE) obtained from the SDE in Equation (\ref{SDE}) by fixing the noise configuration. This problem will be addressed in Section \ref{Chap:Preliminaries}, where a few concepts closely related to the continuous-time dynamics will also be introduced. The second step is the stochastic generalization of this deterministic evolution, which will be addressed in Section \ref{Chap:StochasticDynamics}.

The realistic noises are more complicated than Gaussian white noise, which is, of course, a~mathematical idealization. In Section \ref{Chap:Pathintegrals}, the path integral representation of the theory will enable the generalization to noise of any form. Further generalization to the spatially extended models with infinite-dimensional phase spaces will also be discussed briefly in Section \ref{SecContinuousSpace}. Having established general technical aspects of the STS, the discussion will concentrate on the analysis of the structure of the ground states in Section \ref{Chap:DynamicsTypes}. The classification of ergodic stochastic models on the most general level related to topological supersymmetry breaking will be proposed. This in particular will help reveal the theoretical picture of the stochastic dynamics on the border of ``ordinary chaos'', known previously under such names as intermittency, complexity, and self-organized criticality. Finally, in Section \ref{Chap:SecConclusion}, the paper will be concluded with a brief discussion of a few potentially fruitful directions for future work.



\section{Continuous-Time Dynamics and Related Concepts}
\label{Chap:Preliminaries}
\vspace{-6pt}
\subsection{Dynamics as Maps}
\label{Pullback}

For a fixed noise configuration, Equation (\ref{SDE}) is an ODE with a time-dependent flow vector field in its Right-Hand Side (R.H.S.). 
This ODE defines a two-parameter family of maps of the phase space onto itself, namely, $M_{tt'}: X \to X$:
\begin{eqnarray}
M_{tt'}: x' \mapsto x = M_{tt'}(x').\label{MapsDefined}
\end{eqnarray}

These maps have straightforward interpretations: $x(t)=M_{tt'}(x')$ is the solution of the ODE with the condition $x(t')=x'$. Clearly,
\begin{eqnarray}
M_{tt} = \text{Id}_X, M_{tt'}\circ M_{t't''} = M_{tt''},\text{ and } M_{t't} = M_{tt'}^{-1}.\label{CompositionMaps}
\end{eqnarray}

The only difference here with the stationary flows described, e.g., in Chapter 5 of  \cite{Nakahara} is that the maps depend on both the initial and final moments of evolution, \emph{i.e.}, $t'$ and $t$, and not only on the duration of the evolution, \emph{i.e.}, $t-t'$. This is the result of the dependence of the noise configuration on time, which breaks the time-translation symmetry. Following stochastic averaging over the Gaussian white noise, which does possess time-translation symmetry, this symmetry of the model will be restored (see Section \ref{Chap:StochasticDynamics}). 

Only physical models in which the maps (for finite time evolution) are invertible and differentiable will be considered. On the mathematical level, this means that $F$ and $e$'s are sufficiently smooth in $X$ such that the Picard-Lindel\"of theorem (see, e.g.,  \cite{Theory_Of_ODE}) on the existence and uniqueness of the solution of an ODE for any initial condition is applicable. In other words, all maps are diffeomorphisms. 

To avoid the necessity of addressing various subtle mathematical aspects not directly related to the subject of interest, the fixed noise configuration will be assumed as a continuous function of time. However, this continuity is not necessary. The noise configuration only needs to be integrable in the sense that there must exist a ${\mathcal W}^a(t)$ such that $d{\mathcal W}^a(t)/dt = \xi^a(t)$. For Gaussian white noise, ${\mathcal W}^a(t)$ is called the Wiener process.  

A physicist's proof of the invertibility of maps defined by Equation (\ref{CompositionMaps}) is as follows. A physical ODE provides only one outcome $x$ at $t$ for each initial condition $x'$ at $t'$. The same must be true for the time-reversed physical ODE, which provides only one $x'$ at $t'$ for any $x$ at $t$. In other words, the map $M_{tt'}$ is a one-to-one map, \emph{i.e.}, it is invertible.

If at time $t'$ the DS is described by a total probability function  $P(x)$, the expectation value of some function $f(x): X\to\mathbb{R}$ is
\begin{eqnarray}
\overline{f}(t') = \int_X f(x)P(x)dx^1...dx^D.
\end{eqnarray}

According to Equation (\ref{MapsDefined}), this expectation value at a later time moment $t>t'$ is
\begin{eqnarray}
\overline{f}(t) = \int_X f(M_{tt'}(x))P(x)dx^1...dx^D.\label{DetermEvolution} \end{eqnarray}

This view on dynamics can be clarified through the following example. Consider $X=\mathbb{R}^D$ and an ODE of the simple form $\dot x=v$, where $v\in\mathbb{R}^D$ is a constant vector field. The corresponding diffeomorphisms are $M_{tt'}(x) = x + v(t-t')$. For $f(x)$ being one of the coordinates, \emph{i.e.}, $f(x)=x^i$, Equation (\ref{DetermEvolution}) states that $\overline{x^i}(t) = \overline{x^i}(t') + v^i (t-t')$, just as it should.


One can now make the transformation of the variable of integration in Equation (\ref{DetermEvolution}), \emph{i.e.}, \linebreak $x \to M_{t't}(x)$,
\begin{eqnarray}
\overline{f}(t) = \int_X f(x) M^*_{t't}(P(x)dx^1...dx^D).\label{Average}
\end{eqnarray}

Here, $M^*_{t't}$ is the operation of the variable transformation applied to the coordinate-free object consisting of $P(x)$ and the collection of all the differentials $dx^1...dx^D$,\index{Pullback}
\begin{eqnarray}
M^*_{t't}(P(x)dx^1...dx^D)=\nonumber P\left(M_{t't}(x)\right) \times \\ \times J(TM_{t't}(x))dx^1...dx^D,\label{PullbackP}
\end{eqnarray}
where $J$ is the Jacobian of the tangent map, $TM_{t't}(x): TX_{x} \to TX_{M_{tt'}(x)}$,
\begin{eqnarray}
TM_{t't}(x): dx^i \mapsto d (M_{t't}(x))^i = TM_{t't}(x)^i_k dx^k,\label{DefinitionOfTangentMap}
\end{eqnarray}
with
\begin{eqnarray}
TM_{t't}(x)^i_k = \partial (M_{t't}(x))^i/\partial x^k\label{TangentMap}
\end{eqnarray}
being the coordinate representation of the tangent map.

Equation (\ref{Average}) suggests that the forward temporal evolution of the variables of the DS is equivalent to the backward temporal evolution of the coordinate-free object representing the total probability distribution (TPD) \begin{eqnarray}
\psi^{(D)} =  P(x) dx^1 ... dx^D \in \Omega^D(X).
\end{eqnarray}

In algebraic topology, this object is known as a top differential form (D-form), the infinite-dimensional linear space of all D-forms is denoted as $\Omega^D(X)$, and the operation $M^{*}_{t't}$ in Equation (\ref{PullbackP}) is called the action or the pullback induced by $M_{t't}$ on $\psi^{(D)}$. 

\begin{figure}[h]
\centerline{\includegraphics[height=3.99cm, width=8cm]{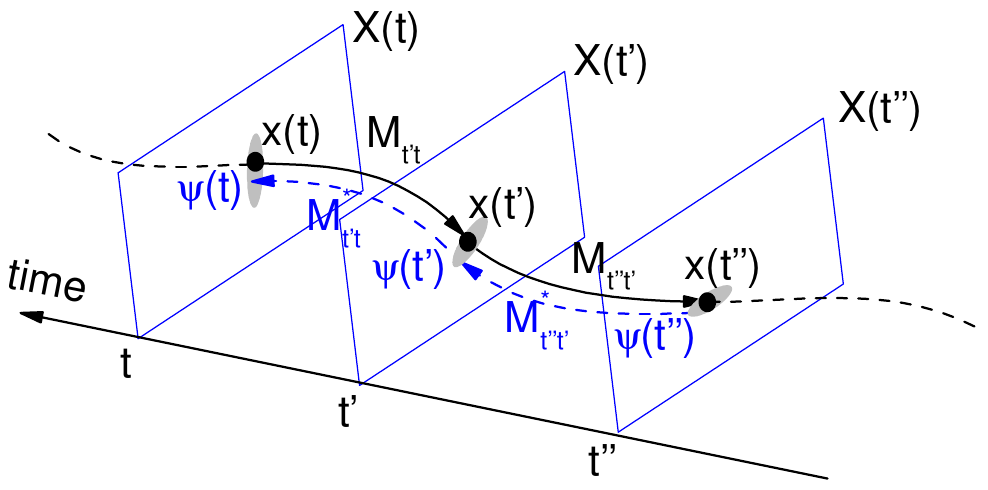}}
\caption{\label{Figure_2_2} Continuous-time deterministic dynamics with a fixed noise configuration can be viewed as a~two-parameter family of diffeomorphisms of the phase space onto itself or between the copies of the phase space: $M_{tt'}:X(t')\to X(t)$. The temporal evolution of a differential form is a pullback induced by the inverse diffeomorphism $M_{t't}^*:\psi(t')\to\psi(t)$.}
\end{figure}

Note that the diffeomorphism in Equation (\ref{Average}) is for the inverse temporal evolution as compared to the time flow in the SDE. This seeming confusion of the time direction can be clarified as follows. The~pullbacks act in the opposite direction  compared to the diffeomorphisms inducing them. This is the reason for the term pullback. The graphical representation of this situation is given in Figure~\ref{Figure_2_2}. There, one introduces an infinite number of copies of the phase space for each time moment, $X(t)$, and dynamics is defined as a two-parameter family of diffeomorphisms between these copies: $M_{tt'}:~X(t')\to X(t)$.

In this path-integral-like picture of dynamics, the pullback in Equation (\ref{PullbackP}) can be given as
\begin{eqnarray}
M^*_{t't}(P(x(t'))dx^1(t')...dx^D(t') = P\left(M_{t't}(x(t))\right) \nonumber \times \\ \times J(TM_{t't}(x))dx^1(t)...dx^D(t).\label{PullbackP1}
\end{eqnarray}
or
\begin{eqnarray}
M^*_{t't}:\Omega^D(X(t'))\to\Omega^D(X(t))
\end{eqnarray}
as opposed to
\begin{eqnarray}
M_{t't}:X(t)\to X(t') .
\end{eqnarray}

The relation between the direction of the flow of time for maps and the corresponding pullbacks can be expressed via the following diagram:
\begin{eqnarray}
\begin{array}{ccccc}
t & \stackrel{\text{flow of time}}{\longleftarrow} &
t' & \stackrel{\text{flow of time}}{\longleftarrow} &
t'' \\
X(t) & \stackrel{M_{t't}}{\longrightarrow} &
X(t') & \stackrel{M_{t''t'}}{\longrightarrow} &
X(t'') \\
\Omega^{D}(X(t)) & \stackrel{M^*_{t't}}{\longleftarrow} &
\Omega^{D}(X(t')) & \stackrel{M^*_{t''t'}}{\longleftarrow} &
\Omega^{D}(X(t'')).
\end{array}
\end{eqnarray}

This diagram particularly suggests that the composition law for pullbacks is
\begin{eqnarray}
M_{t''t}^* = M^*_{t't} M^*_{t''t'}. \label{CompositionPullback}
\end{eqnarray}

\subsection{Differential Forms as Wavefunctions}
\label{SecHilertSpace}

\begin{figure}[h]
\centerline{\includegraphics[width=0.8\linewidth]{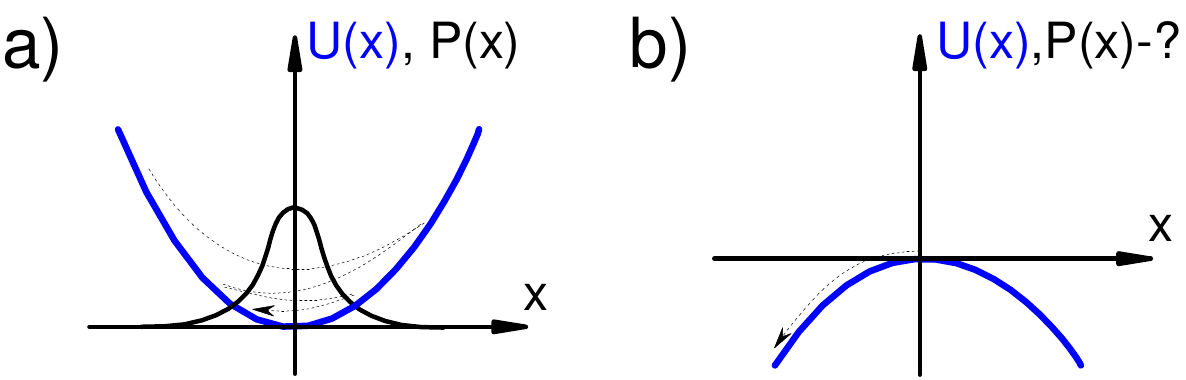}}
\caption{\label{Figure_2_3} (\textbf{a}) A one-variable Langevin stochastic differential equations (SDE) with stable potential, \emph{i.e.}, $U(x)$ (blue parabola oriented up), exhibits dynamics (broken dashed arrow) that can be characterized as the gradual settling to a~steady-state probability distribution $P(x)$ (bell-shaped curve). These dynamics exhibit a loss of the dynamical memory of the initial condition; (\textbf{b}) In the case of an unstable Langevin potential (blue parabola oriented down), the dynamics escape to infinity (dashed arrow pointing left). The dynamics is sensitive to the initial condition. No meaningful steady-state probability distribution can be associated with the ground state in this case. }
\end{figure}

The description of a stochastic model in terms of only TPDs as in the previous subsection is insufficient in the general case. This can be observed from the following qualitative example. Consider~the simplest Langevin SDE with $X=\mathbb{R}$, $F=\partial U(x)/\partial x$, $e=1$. Consider also the case of the stable Langevin potential $U$, as shown in Figure \ref{Figure_2_3}a. It is clear that, after a sufficiently long temporal evolution, this DS will forget its initial condition, and its (only) variable will be distributed according to some steady-state TPD, which is the ground state of this DS (see Section \ref{LangevinSDE} for details). In contrast, when the Langevin potential is unstable, as in Figure \ref{Figure_2_3}b, the DS will never forget its initial condition because a small difference in the initial conditions will grow exponentially. No meaningful steady-state TPD can be prescribed to its unstable variable. This example signifies that the steady-state probability distributions make sense only for stable variables.

The previous example may not look physical because, for any initial condition, the DS escapes to infinity and never returns. Perhaps a better example for the same purpose is a (deterministic) chaotic DS, in which the unstable variables exist even after the infinitely long temporal evolution, \emph{i.e.}, even in the ground state of the DS. In DS theory, the existence of these unstable variables is revealed by positive (global) Lyapunov exponents. Such a chaotic ground state must not be a probability distribution in its unstable variables. That this is indeed so will be observed in  Section \ref{DeterministicChaos} below. The DS theory predecessors of such ground states are the Sinai--Ruelle--Bowen conditional probability functions on the global unstable manifolds \cite{RevModPhys.57.617}.

Section \ref{ChaosTopBreaking} will demonstrate on a more rigorous level that it is a mathematical necessity that the Hilbert space of a stochastic DS be not only the space of the TPDs but rather the entire exterior algebra~of $X$:
\begin{eqnarray}
\Omega(X)=\bigoplus_{k=0}^D \Omega^k(X),
\end{eqnarray}
with the elements being the differential forms of all degrees (see, e.g., Chapter 5 of \cite{Nakahara})
\begin{eqnarray}
\psi^{(k)} = (1/k!) \psi^{(k)}_{i_1...i_k}  dx^{i_1}\wedge ... \wedge dx^{i_k} \in \Omega^k(X).\label{DefinitionOfForm}
\end{eqnarray}

Here, $0\le k\le D$, $\psi^{(k)}_{i_1...i_k}\equiv \psi^{(k)}_{i_1...i_k}(x)$ is an antisymmetric tensor, $\wedge$ is the wedge or antisymmetrized product of differentials, e.g., $dx^1\wedge dx^2 = - dx^2\wedge dx^1 = dx^1\otimes dx^2 - dx^2 \otimes dx^1$, and $\Omega^k(X)$ is the space of all differential forms of degree $k$ ($k$-forms). 

This by no means contradicts the intuitive understanding that it must be possible to associate a~TPD with any wavefunction. As will be clear later, the TPD associated with a wavefunction is not the wavefunction itself but rather, as in quantum theory, is the bra-ket combination, which is a D-form and/or a TPD (see, e.g., Section \ref{SecSpectrum} and the discussion following Equation (\ref{Eigensystem})).

One possible interpretation of the differential forms is the \emph{generalized} (total, conditional, marginal) probability distributions in the coordinate-free setting. The following example demonstrates how the conditional probability distribution can be represented as a differential form (the dimensionality of $X$ is $D=3$):
\begin{eqnarray}
\psi^{(2)} & = &(1/2!) \psi^{(2)}_{i_1i_2} dx^{i_1}\wedge dx^{i_2} \nonumber\\
&=&P(x^2x^3|x^1) dx^2 \wedge dx^3 + P(x^1x^3|x^2) dx^3 \wedge dx^1 \nonumber\\
&&+P(x^1x^2|x^3) dx^1\wedge dx^2\in\Omega^2(X),\nonumber
\end{eqnarray}
where $\psi^{(2)}_{12} = -\psi^{(2)}_{21} = P(x^1x^2|x^3)$, $\psi^{(2)}_{31} = -\psi^{(2)}_{13} = P(x^1x^3|x^2)$, and $\psi^{(2)}_{23}(x) = -\psi^{(2)}_{32} = P(x^2x^3|x^1)$. Similarly, the TPD introduced previously is
\begin{eqnarray}
\psi^{(D)} = (1/D!) \psi^{(D)}_{i_1...i_D} dx^{i_1}\wedge ... \wedge dx^{i_D} \nonumber \\ = P dx^1\wedge ... \wedge dx^D \in \Omega^D(X),\nonumber
\end{eqnarray}
where $\psi^{(D)}_{i_1...i_D} = P \epsilon_{i_1...i_D}$, with $\epsilon_{i_1...i_D} = (-1)^{p(i_1...i_D)}$ being the Levi-Civita antisymmetric tensor and with $p(i_1...i_D)$ being the parity of the permutation of indexes.

The geometrical meaning of a $k$-form is a differential of a $k$-dimensional oriented volume. Therefore, a $k$-form can be integrated over a $k$-dimensional submanifold or a $k$-chain, $c_k$, 
\begin{eqnarray}
\int_{c_k}\psi^{(k)} = p_{c_k} \in \mathbb{R}.\label{PartProb}
\end{eqnarray}
This quantity can be interpreted as follows. If one introduces local coordinates such that the $k$-chain belongs to the $k$-dimensional manifold cut out by $(x^{k+1},...,dx^D) = (\text{Const}^{(k+1)}, ..., \text{Const}^{(D)})$, then Equation (\ref{PartProb}) is the probability of finding variables $(x^1,...,x^k)$ within this $k$-chain given that all the other variables are known with certainty to be equal $(\text{Const}^{(k+1)}, ..., \text{Const}^{(D)})$.

It is worth stressing that the interpretation of the differential forms as the generalized probability distributions is valid only locally in the general case. Only in a neighborhood of a given point and with a properly chosen local coordinates can a differential form be thought of as a conditional probability distribution. Globally, however, this may not be possible because there may not exist global coordinates such that a given differential form is positive everywhere on $X$ and normalizable. Moreover, if it were possible to interpret all differential forms as conditional probability distributions in the global sense, then there would be no reason to consider the extended Hilbert space in the first place. Indeed, a~conditional probability distribution can be constructed from the TPD so that it does not contain any additional information, and it would suffice to describe the DS in terms of the TPD only.  

The exact physical meaning of the wavefunctions in the STS is an open question. At this moment, as a working interpretation of the wavefunction, one can adopt the point of view on the wavefunction from quantum theory. Namely, the ket of the wavefunction at a given moment of time is an abstract object that contains information about the system's past, whereas the bra-ket combination of a wavefunction has the meaning of the TPD.

To finalize the above justification for the use of the extended Hilbert space, it must be stressed that the idea of using the entire exterior algebra as a Hilbert space of a DS is by no means new. This is a well-known method in the supersymmetric theory of Hamilton models in references \cite{Gozzi2,Gozzi_New,Deotto_1}, where it was even demonstrated to a certain degree that the information of chaoticity of a Hamilton model is better represented by differential forms. Moreover, the mathematical object known as the generalized transfer operator that will play a central role in Section \ref{Chap:StochasticDynamics} was designed in the DS theory to probe chaos, and this object was defined on the entire exterior algebra \cite{Rue02}.

To establish the law of the temporal evolution of $k$-forms, one assumes that the DS is described by $\psi^{(k)}(x)$  at time moment $t'$. By analogy with Equation (\ref{Average}), the quantity in Equation (\ref{PartProb}) at a later time moment $t>t'$ is
\begin{eqnarray}
p_{c_k}(t) = \int_{M_{tt'}(c_k)}\psi^{(k)}(x) = \int_{c_k}M_{t't}^*\psi^{(k)}(x).\label{EvolutionOfForm}
\end{eqnarray}

Here, $M^*_{t't}: \Omega^k(X) \to\Omega^k(X)$ is the generalization of the pullback in Equation (\ref{PullbackP}) to pullbacks acting on $\Omega^{(k)}(X)$. Explicitly,\index{Pullback}
\begin{eqnarray}
M^*_{t't} \psi^{(k)}(x) = (1/k!)\psi^{(k)}_{i_1...i_k}(M_{t't}(x)) \times \nonumber \\ \times d(M_{t't}(x))^{i_1}\wedge ... \wedge d(M_{t't}(x))^{i_1},\label{PullBack}
\end{eqnarray}
where the $k$-form is from Equation (\ref{DefinitionOfForm}) and $d(M_{t't}(x))^{i}$ is from Equation (\ref{DefinitionOfTangentMap}).

\subsection{Operator Algebra}
\label{sec:OperAlgebra}
\vspace{-6pt}

\subsubsection{Lie Derivative}
\label{SecLieDerivative}

The infinitesimal pullback is known as the physical or Lie derivative
\begin{eqnarray}
\hat{\mathcal{L}}_{{\mathcal F}} \psi = \lim_{\Delta t\to 0}\frac{\hat 1_{\Omega(x)}-M^*_{(t-\Delta t) t}}{\Delta t} \psi.\label{DefinitionLie}
\end{eqnarray}

The infinitesimal map defined by Equation (\ref{SDE}) can be given as
\begin{eqnarray}
M_{(t-\Delta t)t} (x) \approx x - \Delta t {\mathcal F}(t),\label{MapAddit}
\end{eqnarray}
with ${\mathcal F}$ being the R.H.S. of Equation (\ref{SDE}). Accordingly, the infinitesimal tangent map defined in Equation~(\ref{TangentMap}) is
\begin{eqnarray}
TM_{(t-\Delta t)t}(x)^i_k \approx \delta^i_k - \Delta t T{{\mathcal F}}^i_k(t),\label{DifferentialMatrix}
\end{eqnarray}
with 
\begin{eqnarray}
T{{\mathcal F}}^i_k = \partial {\mathcal F}^i/\partial x^k. 
\end{eqnarray}

Using Equations (\ref{DefinitionLie})--(\ref{DifferentialMatrix}) and the definition of the pullback in Equation (\ref{PullBack}), one arrives at the following expression for the Lie derivative:
\begin{eqnarray}
\hat{\mathcal{L}}_{{\mathcal F}} \psi^{(k)} &=& \frac1{k!} \left({\mathcal F}^i \frac\partial{\partial x^i}\psi_{i_1...i_k}^{(k)} + \sum_{j=1}^k T{{\mathcal F}}^{\tilde i_j}_{i_j} \psi_{i_1... \tilde i_j...i_k}^{(k)} \right)\times \nonumber \\ &&\times  dx^{i_1}\wedge ... \wedge dx^{i_k},
\end{eqnarray}
with $\psi^{(k)}$ being from Equation (\ref{DefinitionOfForm}).

The finite-time pullback satisfies the following equation:
\begin{eqnarray}
\partial_ t M^*_{t't} &=& \lim_{\Delta t\to 0} \frac{ M^*_{t't}-M^*_{t'(t-\Delta t)} }{\Delta t} \nonumber \\ &=& \lim_{\Delta t\to 0} \frac{M^*_{(t-\Delta t)t} M^*_{t'(t-\Delta t)}-M^*_{t' (t-\Delta t) }}{\Delta t}\nonumber\\
&=& \lim_{\Delta t\to 0} \frac{M^*_{(t-\Delta t)t} - \hat 1_{\Omega(X)}}{\Delta t}M^*_{t' (t-\Delta t) } \nonumber \\&=& - \hat {\mathcal L}_{{\mathcal F}(t)}M^*_{t't},
\end{eqnarray}
where $M^*_{t't}=M^*_{(t-\Delta t)t} M^*_{t'(t-\Delta t)}$, which follows from Equation (\ref{CompositionPullback}), has been used together with the definition of the Lie derivative in Equation (\ref{DefinitionLie}). The integration of this equation with the initial condition $M^*_{tt}=\hat 1_{\Omega(X)}$ results in
\begin{eqnarray}
M^*_{t't} = {\mathcal T} e^{- \int_{t'}^t d\tau \hat {\mathcal L}_{{\mathcal F}(\tau)}}, \label{IntegrationFormulaLie}
\end{eqnarray}
where $\mathcal T$ denotes the operator of chronological ordering. This operator is necessary because $\hat{\mathcal L}_{{\mathcal F}(\tau)}$ at different $\tau$s do not commute. Equation (\ref{IntegrationFormulaLie}) can be represented in the form of a Taylor series as
\begin{eqnarray}
M^*_{t't} = \hat 1_{\Omega(X)} - \int_{t'}^t d\tau \hat {\mathcal L}_{{\mathcal F}(\tau)} + \nonumber \\  + \int_{t'}^t d\tau_1 \hat {\mathcal L}_{{\mathcal F}(\tau_1)}\int_{t'}^{\tau_1} d\tau_2 \hat {\mathcal L}_{{\mathcal F}(\tau_2)} - ...\label{Chronological}
\end{eqnarray}

As in quantum theory, in the Taylor series expansion of the finite-time evolution operator, the new infinitesimal evolution operators (the Lie derivatives in this case) accumulate from the left. In other words, operators at later moments of time are always on the left of the operators at earlier moments of time, just like letters in Arabic script. In other words, the time flows from right to left in the operator representation of stochastic evolution. This is exactly the reason why  the arrow of time points left in Figures \ref{Figure_2_1} and \ref{Figure_2_2}, which may appear unconventional.  


\subsubsection{Exterior Derivative}
\label{OperatorAlgebra}

One of the fundamental operators of the exterior algebra is the exterior multiplication $dx^i\wedge:~\Omega^k(X)\to\Omega^{k+1}(X)$. This operator can be defined via its action on a $k$-form from Equation (\ref{DefinitionOfForm}):
\begin{eqnarray}
dx^i\wedge \psi^{(k)} = (1/k!)\psi_{i_1...i_k} dx^i\wedge dx^{i_1}\wedge...\wedge dx^{i_k}.
\end{eqnarray}

Viewing the differentials in the definition of a $k$-form in Equation (\ref{DefinitionOfForm}) as the operators of exterior multiplication, one can also define the operation of the exterior product of differential forms:
\begin{eqnarray}
\psi^{(k)}\wedge \psi^{(n)} \in \Omega^{k+n}(X).
\end{eqnarray}

The other fundamental operator of the exterior algebra is the interior multiplication $\hat \imath_i:\Omega^{k}(X)\to \Omega^{k-1}(X)$, which is defined as
\begin{eqnarray}
\hat {\imath}_{i} \psi^{(k)} = \frac1{k!}\sum\nolimits_{j=1}^k  (-1)^{j+1}\psi_{i_1...i_{j-1} i i_{j+1}...i_k}\nonumber \times \\ \times dx^{i_1}\wedge ...\widehat{dx^{i_j}}...\wedge dx^{i_k},
\end{eqnarray}
where $\widehat{dx^{i_j}}$ denotes a missing element. As can be readily verified, the (anti)commutation relations for these operators are
\begin{eqnarray}
\left[ dx^{i_1} \wedge, dx^{i_2} \wedge\right]        = 0,
\left[ \hat{\imath}_{j_1}, \hat {\imath}_{j_2}\right] = 0,
\left[ dx^i \wedge, \hat{\imath}_j\right]             = \delta^i_j.\label{CommutationRelations}
\end{eqnarray}

Here and in the following, the square brackets denote the bi-graded commutator:
\begin{eqnarray}
[\hat X, \hat Y] = \hat X \hat Y - (-1)^{deg(\hat X)deg(\hat Y)} \hat Y \hat X,\label{BiGRadedCommutator}
\end{eqnarray}
with $deg(\hat X) = \#(dx\wedge) - \#(\hat i)$ being the degree of operator $\hat X$, \emph{i.e.}, the difference between the numbers of exterior and interior multiplication operators in $\hat X$. For example, $deg(dx\wedge )=1$ and $deg(\hat i)=-1$ so~that the bi-graded commutators in Equation (\ref{CommutationRelations}) are actually anticommutators.

The centerpiece of the theory under consideration is the exterior derivative or de Rham operator:
\begin{eqnarray}
\hat d = dx^i\wedge \frac\partial{\partial x^i}.
\end{eqnarray}

The exterior derivative is a bi-graded differentiation, \emph{i.e.}, for any operators $\hat X$ and $\hat Y$,
\begin{eqnarray}
[\hat d, \hat X \hat Y] = [\hat d, \hat X] \hat Y - (-1)^{deg(\hat X)} \hat X [\hat d, \hat Y].\label{BiGraded}
\end{eqnarray}

In the new notations, the Lie derivative can be given via the Cartan formula
\begin{eqnarray}
\hat{\mathcal{L}}_{{\mathcal F}} = {\mathcal F}^i\partial/\partial x^i + T{{\mathcal F}}^i_j dx^j\wedge \hat {\imath}_i = [\hat d, \hat{\imath}_{{\mathcal F}}], \label{CartanFormuola}
\end{eqnarray}
where $\hat{\imath}_{{\mathcal F}} = {\mathcal F}^i \hat{\imath}_{i}$ is the interior multiplication by ${\mathcal F}$, and 
$[\hat d, \hat{\imath}_i] = \partial/\partial x^i$ and $[\hat d, {\mathcal F}^i] = T{{\mathcal F}}^i_jdx^j\wedge$ have been used.

\subsubsection{Hodge Dual}
\label{sec:HodgeStar}

Yet another operation that will be used later is the Hodge star \begin{eqnarray}
\star:\Omega^{k}(X)\to \Omega^{D-k}(X),\label{HodgeStar}
\end{eqnarray}
defined as
\begin{eqnarray}
&\star\psi^{(k)}_{i_1...i_k}dx^{i_1}\wedge...\wedge dx^{i_k}\nonumber \\ &= \frac{g^{1/2}}{(D-k)!}\psi^{(k)}_{\tilde i_1...\tilde i_k} g^{\tilde i_1 i_1}...g^{\tilde i_k i_k}\epsilon_{i_1...i_D}dx^{i_{k+1}}\wedge...\wedge dx^{i_D},\label{DefinitionOfHodge}
\end{eqnarray}
where $\epsilon$ is the Levi-Civita antisymmetric tensor and $g=\det g_{ij}$ is the determinant of the metric on~$X$. As previously mentioned, the natural choice of metric on $X$ is the noise-induced metric $g^{ij}=e^i_ae^j_a$. In~Section \ref{Sec:TimeReversalSymmetry}, it will be noted that, for certain purposes, other metrics on $X$ can be used. The Hodge star has the following property:
\begin{eqnarray}
\star \star = (-1)^{\hat k(D-\hat k)},\label{HodgeSquare}
\end{eqnarray}
where $\hat k$ is the operator of the degree of the differential form
\begin{eqnarray}
\hat k = dx^i\wedge \hat{\imath}_i, \hat k \psi^{(k)} = k \psi^{(k)}, \psi^{(k)} = \Omega^k(X),\label{K_Operator}
\end{eqnarray}
so that
\begin{eqnarray}
\star\star \psi^{(k)} = (-1)^{k(D-k)}\psi^{(k)}, \psi^{(k)}\in\Omega^k(X).
\end{eqnarray}

This can easily be verified using
\begin{eqnarray}
\epsilon_{i_1...i_D}g^{\tilde i_1 i_1}...g^{\tilde i_k i_k}
g^{i_{k+1} \tilde i_{k+1}}...g^{i_D \tilde i_D} = g^{-1} \epsilon_{\tilde i_1...\tilde i_D},
\end{eqnarray}
where the indexes are lowered by the Euclidean metric and
\begin{eqnarray}
&\epsilon_{\tilde i_1...\tilde i_k\tilde i_{k+1}...\tilde i_D}\epsilon_{\tilde i_{k+1}...\tilde i_D i_1... i_k} = (-1)^{k(D-k)} (D-r)!\times \nonumber \\&\times 
\left(\begin{array}{ccc}
  \delta_{\tilde i_1i_1} & \dots & \delta_{\tilde i_1i_k} \\
  \vdots &  \ddots & \vdots \\
  \delta_{\tilde i_k i_1}&\dots&\delta_{\tilde i_kj_k}\\
\end{array}\right).
\end{eqnarray}

In other words, the square of the Hodge star is a unity operator up to a sign. Up to the same sign, the Hodge star is its own inverse:
\begin{eqnarray}
\star^{-1} = (-1)^{\hat k(D-\hat k)}\star,
\end{eqnarray}
and
\begin{eqnarray}
\star \star^{-1} = \star^{-1}\star = \hat 1_{\Omega(X)}.\label{StarSquared}
\end{eqnarray}

The Hodge star naturally defines an internal product on $\Omega(X)$
\begin{eqnarray}
(\phi|\psi) = \int_X \star \phi^*\wedge \psi\label{Inner_Product}
\end{eqnarray}
for $\phi,\psi\in\Omega$. The internal product is Hermitian positive definite, \emph{i.e.},
\begin{eqnarray}
(\phi|\psi) = (\psi|\phi)^*, \text{ and } (\psi|\psi) \ge 0. \label{HermitianMetric}
\end{eqnarray}

Thus, it may serve as a Hermitian metric on $\Omega$. As will be discussed in Section \ref{Sec:TimeReversalSymmetry}, the eigensystem of the pseudo-Hermitian $\hat H$ provides its own non-trivial metric on the Hilbert space. It~is this metric that must be viewed as the fundamental metric of the Hilbert space of the model and for which the standard notation $\langle \cdot | \cdot \rangle$ must be reserved, whereas the round brackets can be used for Equation (\ref{HermitianMetric}).   

Equation (\ref{HermitianMetric}) can also be used for the definition of the concept of the Hermitian conjugate of an~operator
\begin{eqnarray}
(\phi|\hat A \psi) = (\hat A^\dagger\phi|\psi)\label{DefinitionHerConJug}
\end{eqnarray}
for any $\phi,\psi\in \Omega$ and any operator $\hat A:\Omega(X)\to\Omega(X)$. Using this definition, it is straightforward to~derive
\begin{eqnarray}
(dx^i\wedge)^\dagger = g^{ij} \hat \imath_j, (\imath_i)^\dagger =  g_{ij}dx^j\wedge,
\end{eqnarray}
and the explicit expression for the so-called codifferential, which is the Hermitian conjugate of the exterior derivative,
\begin{eqnarray}
\hat d^\dagger = - g^{ij} \hat\imath_j
\left(
\frac{\partial}{\partial x^i} + g_{ik}(g^{lk})_{'m}dx^m\wedge \hat \imath_l \right. \nonumber \\ \left . + (1/2)\frac{\partial log(g)}{\partial x^i}\right).\label{dConjugate}
\end{eqnarray}

Finally, in the forthcoming discussion, the concept of the Hodge Laplacian will also be recalled:
\begin{eqnarray}
\hat \Delta_{H} = [\hat d, \hat d^\dagger].\label{HodgeLaplacian}
\end{eqnarray}

\subsection{Fermionic Variables}
\label{SecFermionVar}

The exterior algebra has an alternative field-theoretic representation in terms of the fermionic variables that will be used in the path integral representation of the theory in Section \ref{SecPathIntegralGTO} as well as at the end of the next section.

Following reference \cite{Wit82}, one notes that the (anti-)commutation relations in  Equation (\ref{CommutationRelations}) are equivalent to those of Grassmann or anticommuting variables, \emph{i.e.}, $\chi^i$, \index{Grassmann numbers} and derivatives over them, \emph{i.e.},~$\partial/\partial\chi^j$:
\begin{eqnarray}
\left[ \chi^{i_1} , \chi^{i_2} \right]_+         = 0,
&\left[ \frac\partial{\partial\chi^{j_1}}, \frac\partial{\partial\chi^{j_2}}\right]_+  = 0,\nonumber \\ &
\left[ \chi^i, \frac\partial{\partial\chi^j}\right]_+ = \delta^i_j,
\end{eqnarray}

Therefore, one can make the formal substitution
\begin{eqnarray}
dx^i\wedge \to \chi^i, \text{ and }\hat{\imath}_j\to \frac\partial{\partial\chi^j},\label{FermionicVariables}
\end{eqnarray}
and a wavefunction in the new notations becomes
\begin{eqnarray}
&\psi_{i_1...i_k}^{(k)}(x)dx^{i_1}\wedge ...\wedge dx^{i_k} \to \nonumber \\ &\to \psi_{i_1...i_k}^{(k)}(x)\chi^{i_1}...\chi^{i_k}\equiv \psi^{(k)}(x\chi),\label{wavefunction2}
\end{eqnarray}
whereas the expression for the exterior derivative, \label{Exterior derivative}
\begin{eqnarray}
\hat d = \chi^i \frac\partial{\partial x^i}, \label{ExteriorDer}
\end{eqnarray}
reveals why $\hat d$ is a ``super'' operator: it destroys a bosonic or commuting variable $x^i$ and creates a~fermionic or anticommuting variable $\chi^i$.

Equation (\ref{wavefunction2}) can be viewed as a $k$-th term of the Taylor expansion of a wavefunction
\begin{eqnarray}
\psi(x\chi) = \sum\nolimits_{k=0}^D \frac 1{k!}\psi_{i_1...i_k}^{(k)}(x)\chi^{i_1}...\chi^{i_k},
\end{eqnarray}
which is now a function of a pair of variables that are the supersymmetric partners with respect to the Operator (\ref{ExteriorDer}).

Some properties of fermionic variables are similar to those of bosonic variables. For example, one~can introduce the fermionic $\delta$-function
\begin{eqnarray}
\int d^D \chi \delta^D(\chi-\chi') f(\chi) = f(\chi').
\end{eqnarray}

Here, $f(\chi)$ is an arbitrary function of a fermionic variable, $\chi'$ is yet another fermionic variable, the differential is $d^D\chi=d\chi\text{}^D...d\chi\text{}^1$, and 
\begin{eqnarray}
\delta^D(\chi-\chi') = (-1)^D(\chi-\chi')^1...(\chi-\chi')^D.
\end{eqnarray}

Note that the definition of the fermionic  $\delta$-function depends on the relative position of the differentials because $\int d^D\chi \delta^D(\chi-\chi')=(-1)^{D} \int \delta^D(\chi-\chi')d^D\chi$, where $(-1)^{D^2}=(-1)^D$ has been~used.

The above property of fermionic variables and their $\delta$-function can be established using\linebreak Berezin rules of integration over Grassmann numbers. The latter include identities such as\linebreak $\int d\chi^1 = 0, \int \chi^1 d\chi^1 = - \int d\chi^1 \chi^1 = 1$.

Another property of fermionic variables that has a straightforward bosonic analogue is the exponential representation of a fermionic delta function
\begin{eqnarray}
\delta^D(\hat A\chi) = \int d^D\bar\chi e^{\bar\chi 
\hat A \chi},\label{ExpFermi}
\end{eqnarray}
where $\bar\chi$ is yet another additional fermionic variable.

Other properties of fermionic variables may be in a sense opposite to their bosonic counterparts,~e.g.,
\begin{eqnarray}
\int d^D\chi \delta^D(\hat A\chi) = \text{det} \hat A,
\end{eqnarray}
whereas for bosonic variables, one would have $\int d^Dx \delta^D (\hat A x) = |\text{det} \hat A|^{-1}$ for $x\in \mathbb{R}^D$. There are many other interesting properties and relations associated with fermionic variables (see, e.g., \cite{GrassmannNumbers}). In the forthcoming discussion, however, only those introduced so far will be used.

\section{Operator Representation}
\label{Chap:StochasticDynamics}
\vspace{-6pt}
\subsection{Stochastic Generalization of Dynamics}
\label{SecGenFPOper}

In the previous section, the noise configuration was assumed to be fixed, and the dynamics was essentially deterministic. The next step is to account for all possible realizations of the noise. This goal can be achieved as follows. 

The stochastic generalization of Equation (\ref{Average}) is
\begin{eqnarray}
\overline{f}(t) &=& \left\langle\int_X f(x)M^*_{t't}P(x)dx^1...dx^D\right\rangle_\text{Ns} \nonumber\\
&=& \int_X f(x) \hat {\mathcal{M}}_{tt'}(P(x)dx^1...dx^D),\label{StochAverage}
\end{eqnarray}
and that of Equation (\ref{EvolutionOfForm}) is
\begin{eqnarray}
p_{c_k}(t) = \left\langle\int_{c_k}M_{t't}^*\psi^{(k)}(x)\right\rangle_\text{Ns} = \int_{c_k}\hat {\mathcal{M}}_{tt'}\psi^{(k)}(x),
\end{eqnarray}
where the notation for the stochastic average is from Equation (\ref{DefStochAverage}) and the new operator $\hat {\mathcal{M}}_{tt'}:~\Omega(X)\to\Omega(X)$ is defined as
\begin{eqnarray}
\hat {\mathcal{M}}_{tt'} = \langle M^*_{t't}\rangle_\text{Ns}.\label{GTO}
\end{eqnarray}

The operation of the stochastic averaging here is legitimate because the pullbacks are linear operators on $\Omega(X)$. Now, the possibly highly nonlinear (stochastic) dynamics is described by linear operators acting on the linear Hilbert space. The price one pays for this ``linearization'' is the infinitely larger dimensionality of $\Omega(X)$ compared to the dimensionality of $X$. One may wonder now if this dramatic increase in the dimensionality of the objects of interest could be an unnecessary complication. It is not. As long as one is interested in the stochastic dynamics, he has to consider the infinite-dimensional space of the probability distributions regardless.  

The finite-time stochastic evolution operator (SEO) in Equation (\ref{GTO}) is known in DS theory as the generalized transfer operator \cite{Rue02}. The only new element in Equation (\ref{GTO}) is that the pullbacks in Equation (\ref{GTO}) are those of the \emph{inverse} maps.

In the case of white (not necessarily Gaussian) noise, the noise variables at different times do not correlate, and thus,
\begin{eqnarray}
\hat {\mathcal{M}}_{tt''} = \langle M^*_{t''t}\rangle_\text{Ns}
=\langle M^*_{t't} M^*_{t''t'}\rangle_\text{Ns} \nonumber \\
=\langle M^*_{t't}\rangle_\text{Ns} \langle M^*_{t''t'} \rangle_\text{Ns}
= \hat {\mathcal{M}}_{tt'} \hat {\mathcal{M}}_{t't''},\label{CompositionGTO}
\end{eqnarray}
where the composition law for pullbacks in Equation (\ref{CompositionPullback}) has been used. Unlike in Equation (\ref{CompositionPullback}), the time now flows from right to left, as it should. This is why the positions of the time arguments of the finite-time SEO are swapped in Equation (\ref{GTO}) compared to those in the corresponding pullbacks.

The quantum mechanical analogue of the finite-time SEO is the finite-time quantum evolution operator, denoted typically as $\hat U=e^{-it\hat H_q}$, with $\hat H_q$ being some Hermitian Hamiltonian. The capitalized $U$ signifies here that the quantum evolution is unitary. The stochastic evolution is not unitary; thus, the current notation for the finite-time SEO borrowed from reference \cite{Rue02} is more suitable.

One can now use the picture of the time intervals as discussed in Section \ref{Sec:Models}. Explicitly, the time domain of the temporal evolution, \emph{i.e.}, $(t,t')$, is a union of a large number, \emph{i.e.}, $N \gg 1$, of elementary intervals, namely, $(t,t') = \bigcup_{n=0}^{N-1}(t_{n+1}, t_n)$, where $t_n = t' + n \Delta t$ and $\Delta t = (t-t')/N$. The law of the infinitesimal stochastic evolution can now be found as
\begin{eqnarray}
\partial_t  \hat{\mathcal{M}}_{tt'} = \lim_{\Delta t \to 0} \frac{\hat{\mathcal{M}}_{tt'}-\hat{\mathcal{M}}_{t_{N-1}t'}}{\Delta t} \nonumber \\  = \lim_{\Delta t \to 0} \frac{\hat{\mathcal{M}}_{t_N t_{N-1}}-\hat 1_{\Omega(X)} } {\Delta t} \hat{\mathcal{M}}_{t_{N-1}t'},\label{FPEqM0}
\end{eqnarray}
where the following equality has been used:
\begin{eqnarray}
\hat{\mathcal{M}}_{tt'} = \hat{\mathcal{M}}_{t_{N}t_{N-1}} \hat{\mathcal{M}}_{t_{N-1}t'}.
\end{eqnarray}

If one now introduces the ``infinitesimal'' SEO as
\begin{eqnarray}
\hat H = \lim_{\Delta t\to0}
\frac{\hat 1_{\Omega(X)}- \hat{\mathcal{M}}_{t_N t_{N-1}}}{\Delta t},\label{FPOp0}
\end{eqnarray}
one obtains
\begin{eqnarray}
\partial_t \hat{\mathcal{M}}_{tt'} = - \hat H \hat{\mathcal{M}}_{tt'}.\label{FPEqM}
\end{eqnarray}

The integration of this differential equation with the condition $\hat{\mathcal{M}}_{tt}=\hat 1_{\Omega(X)}$ leads to
\begin{eqnarray}
\hat{\mathcal{M}}_{tt'} = e^{- \hat H (t-t')}.
\end{eqnarray}

One can now introduce the time-dependent wavefunction $\psi(t)=\hat{\mathcal M}_{tt'}\psi(t')$ by analogy with the Schr\"odinger representation of quantum theory. Equation (\ref{FPEqM}) then takes the familiar form of the stochastic evolution equation or the \emph{generalized} Fokker--Planck (FP) equation:
\begin{eqnarray}
\partial_t\psi(t) = - \hat H \psi(t).\label{FPEq}
\end{eqnarray}

To establish the explicit expression for the SEO in Equation (\ref{FPOp0}), one recalls that the noise variable and consequently the R.H.S. of Equation (\ref{SDE}) are constant within all time intervals, including the last~one
\begin{eqnarray}
\left.{\mathcal F}(xt)\right|_{t_{N}>t>t_{N-1}} = {\mathcal F}_{N}(x) \equiv F(x) + (2\Theta)^{1/2}e_a(x) \xi^a_{N}, 
\end{eqnarray}
and that
\begin{eqnarray}
\hat{\mathcal{M}}_{t_Nt_{N-1}} = \langle M^*_{t_{N-1} t_N}\rangle_\text{Ns} = \langle e^{-\Delta t \hat{\mathcal{L}}_{{\mathcal F}_N}}\rangle_\text{Ns},\label{Pullback1}
\end{eqnarray}
as follows from the definition of the Lie derivative in Equation (\ref{DefinitionLie}). Using the linearity of the Lie derivative in its vector field,
\begin{eqnarray}
\hat{\mathcal{L}}_{{\mathcal F}_N} = \hat{\mathcal{L}}_{F} + (2\Theta)^{1/2}\xi^a_N \hat{\mathcal{L}}_{e_a},\label{Linearity}
\end{eqnarray}

Equations (\ref{AverageNoise}), (\ref{AverageNoise1}), (\ref{FPOp0}) and (\ref{Pullback1}), one arrives at
\begin{eqnarray}
\hat H = \hat{\mathcal{L}}_{F} - \Theta \hat{\mathcal{L}}_{e_a}\hat{\mathcal{L}}_{e_a}.\label{FPOp}
\end{eqnarray}

The physical meaning of the two terms in the SEO (\ref{FPOp}) is clear. The first term is the deterministic flow along $F$, and the second term represents the noise-induced diffusion. The operator
\begin{eqnarray}
&\hat{\mathcal{L}}_{e_a}\hat{\mathcal{L}}_{e_a} = g^{ij}(x)\frac\partial{\partial x^i}\frac\partial{\partial x^j} + g_1^i(x,dx\wedge, \hat {\imath})\frac\partial{\partial x^i} \nonumber \\& + g_0(x,dx\wedge, \hat {\imath}),
\end{eqnarray}
with $g_{0,1}$ being some functions of its arguments, can be called the diffusion Laplacian. This operator is a member of the family of Laplace operators. In the general case, however, this operator is neither Hodge (or de Rham) Laplacian (\ref{HodgeLaplacian}) nor Bochner (or Beltrami) Laplacian. Nevertheless, just as the Hodge Laplacian, the diffusion Laplacian has the important property of being $\hat d$-exact, \emph{i.e.}, $\hat{\mathcal{L}}_{e_a}\hat{\mathcal{L}}_{e_a} = [\hat d, \hat{\imath}_{i}e^i_a\hat{\mathcal{L}}_{e_a}]$.

The time-interval picture and the piece-wise constant noise may not appear sufficiently convincing in the context of the Ito--Stratonovich dilemma discussed in Appendix \ref{ItoStratonovichDilemma} and in the next subsection. As such, it is worth re-deriving Equation (\ref{FPOp}) directly for the white noise picture. This can be performed using yet another equivalent form of Equation (\ref{FPOp0}):
\begin{eqnarray}
\hat H = \left\langle\lim_{\Delta t\to0}
\frac{\hat 1_{\Omega(X)}- M^*_{t,t+\Delta t}}{\Delta t}\right\rangle_\text{Ns}.\label{FPOp0_1}
\end{eqnarray}

Using Equation (\ref{Chronological}), the above equation results in
\begin{eqnarray}
\hat H &=& \lim_{\Delta t\to0} \Delta t^{-1}\left\langle
\int_t^{t+\Delta t} \hat {\mathcal L}_{{\mathcal F}(\tau)}d\tau \right . \nonumber \\ &&\left. - \int_t^{t+\Delta t} \int_t^{\tau_1} \hat {\mathcal L}_{{\mathcal F}(\tau_1)}\hat {\mathcal L}_{{\mathcal F}(\tau_2)}d\tau_1 d\tau_2 + ...
\right\rangle_\text{Ns}.\label{FPOp0_2}
\end{eqnarray}

Now, using the continuous-time analogue of Equation (\ref{Linearity})
\begin{eqnarray}
\hat{\mathcal{L}}_{{\mathcal F}(\tau)} = \hat{\mathcal{L}}_{F} + (2\Theta)^{1/2}\xi^a(\tau) \hat{\mathcal{L}}_{e_a},\label{Linearity_1}
\end{eqnarray}
together with $\langle\xi^a(t)\rangle_\text{Ns}=0$ and  Equation (\ref{GaussianAverage}), one readily arrives at
\begin{eqnarray}
\hat H = \hat {\mathcal L}_{F} - 
C \Theta \hat{\mathcal{L}}_{e_a} \hat{\mathcal{L}}_{e_a},\label{H_Op_Fin}
\end{eqnarray}
where
\begin{eqnarray}\label{Coefficient}
C = 2 \lim_{\Delta t\to0} \Delta t^{-1}\int_t^{t+\Delta t} \int_t^{\tau_1} \delta(\tau_1-\tau_2) d\tau_1 d\tau_2.
\end{eqnarray}

The subtle point here is that the upper limit of the integration over $\tau_2$ is exactly at the peak of the $\delta$-function, \emph{i.e.}, $\tau_2=\tau_1$. The $\delta$-function, no matter how narrow, is a symmetric function of its argument, \emph{i.e.}, $\delta(\tau_1-\tau_2)=\delta(\tau_2-\tau_1)$, and its integral over its entire domain is unity. Therefore, $\int_t^{\tau_1} \delta(\tau_1-\tau_2) d\tau_2$ must be interpreted as $1/2$. Consequently, $C=1$, and the SEO in Equation ( \ref{H_Op_Fin}) is the same operator obtained earlier in Equation (\ref{FPOp}) within the piece-wise constant picture of the noise.


The conventional Fokker--Planck (FP) equation for the TPD is simply Equation (\ref{FPEq}) for the D-forms. Its explicit expression can be readily found. Using the Cartan Formula (\ref{CartanFormuola}) and noting that $\hat d\psi^{(D)}=0$ for any $\psi^{(D)}\in\Omega^D(X)$ so that $\hat{\mathcal L}_{G} \psi^{(D)} = \partial/\partial x^i G^i \psi^{(D)}$ for any vector field $G\in TX$, one finds that the FP equation is
\begin{eqnarray}
&\partial_t \psi^{(D)}(x,t) = - \left(\frac\partial{\partial x^i} F^i(x)\right. \nonumber \\ &\left.- \Theta \frac\partial{\partial x^i} e^i_a(x)\frac\partial{\partial x^j} e^j_a(x) \right)\psi^{(D)}(x,t).\label{StratonovichFP}
\end{eqnarray}

This is the well-known FP equation in the so-called Stratonovich interpretation of SDEs. This~brings the discussion to the Ito--Stratonovich dilemma addressed next.


\subsection{Ito--Stratonovich Dilemma}
\label{ItoStratSubSection}


The Ito--Stratonovich dilemma \cite{Ito, Stratonovich, Kampen, Wong, Shreve} is a well-known ambiguity in the exact form of the FP operator that appears when an SDE is looked upon as a continuous time limit of a related stochastic \emph{difference} equation. A discussion of this issue can begin by rewriting the original SDE~(\ref{SDE}) in an~equivalent~form
\begin{eqnarray}
d x(t) = F(x(t))dt + (2\Theta)^{1/2} e_a(x(t)) d {\mathcal W}^a(t), \label{WienerSDE}
\end{eqnarray}
where ${\mathcal W}^a(t)=d\xi^a(t)/dt$ is introduced to emphasize that $\xi$ is ``integrable'' in a certain mathematical sense so that the SDE is well defined. 

The classical view on SDEs is through the continuous-time limit of the related stochastic \emph{difference} equations (SdE):
\begin{eqnarray}
&\Delta x_n = F(x_{n-1} + \alpha \Delta x_n)\Delta t \nonumber \\ & + (2\Theta)^{1/2} e_a(x_{n-1} + \alpha \Delta x_n) \Delta {\mathcal W}^a_n.\label{WienerSdE}
\end{eqnarray}

Here, $\Delta x_n=x_n-x_{n-1}$, with $x_n$ and $x_{n-1}$ being the DS variables at two consecutive time moments; $t_n=t_{n-1}+\Delta t$; and $\Delta {\mathcal W}^a_n = \xi^a_n \Delta t$, with $\xi_n^a$ being the noise variable acting between $t_n$ and $t_{n-1}$; 
moreover, the parameter $\alpha\in(0,1)$ controls at which point of the elementary time step the R.H.S. of the SdE is evaluated. There are three major choices for $\alpha$ in the literature: the Ito choice, \emph{i.e.}, $\alpha =0$, of the starting point $x_{n-1}$; the Stratonovich choice, \emph{i.e.}, $\alpha=1/2$, of the mid-point $(x_{n}+x_{n-1})/2$; and the Kolmogorov or ``isothermal'' choice, \emph{i.e.}, $\alpha=1$ \cite{Isothermal}, of the final point $x_{n}$.

As shown in Appendix \ref{ItoStratonovichDilemma}, the FP equation of the continuous-time limit of the SdE is Equation~(\ref{StratonovichFP}) with the shifted flow vector field:
\begin{eqnarray}
F^i \to F^i_\alpha = F^i + 2 \Theta (\alpha-1/2)(e^i_a)_{'j}e^j_a.\label{Falpha}
\end{eqnarray}

It is also shown in Appendix \ref{ItoStratonovichDilemma} that if, instead of observing the SdE as a formal equation (implicitly) defining the increment $\Delta x$, one considers a continuous-time flow with the piece-wise constant noise from Figure \ref{Figure_2_1}, the freedom in choosing $\alpha$ disappears, and one must always use the Stratonovich choice of $\alpha=1/2$. This explains why the approximation-free derivation in the previous subsection led to the Stratonovich FP Operator (\ref{StratonovichFP}). The point is that the SDE was believed to define a~continuous-time flow on $X$. The action induced by this flow on the exterior algebra can be given by what can be called the stochastic flow equation (SFE)
\begin{eqnarray}
d \psi(t) = -\left(\hat{\mathcal L}_F dt + (2\Theta)^{1/2} \hat {\mathcal L}_{e_a} d {\mathcal W}^a(t) \right)\psi(t).\label{WienerSFE}
\end{eqnarray}

This equation is just as well defined as the original SDE~(\ref{WienerSDE}). Indeed, the transition from Equation~(\ref{WienerSDE}) to (\ref{WienerSFE}) uses only the two following fundamental properties: diffeomorphisms defined as $\dot x = {\mathcal F}$ induce the (time-reversed) action on $\Omega$ defined by $\partial_t \psi = - \hat {\mathcal L}_{\mathcal F}\psi $, and the Lie derivative is linear in its argument, $\hat {\mathcal L}_{C_1 F_1+C_2 F_2} = C_1\hat {\mathcal L}_{F_1} + C_2 \hat {\mathcal L}_{F_2}$, where $F_{1,2}$ are some (differentiable) vector fields on $X$ and $C_{1,2}$ are constants. Therefore, if the SDE together with $d{\mathcal W}^a$ is well defined, then the SFE is also well defined.

One way to understand why the FP operator for Equation (\ref{WienerSFE}) does not depend on the interpretation of the noise is as follows. As is clear from Equation (\ref{Falpha}), the FP operator for Equation~(\ref{WienerSDE}) is independent of the interpretation of the SDE when the noise is additive, \emph{i.e.}, when  the $e$ are independent of the position of $X$ so that the factor $(2\Theta)^{1/2}e_a$, to which the noise is coupled in Equation~(\ref{WienerSDE}), is independent of time. Similarly, the factor $(2\Theta)^{1/2}\hat {\mathcal L}_{e_a}$, to which the noise is coupled in Equation (\ref{WienerSFE}), is always independent of time even for the position-dependent $e$, which renders the FP operator derived from Equation (\ref{FPOp0_2}) noise-interpretation independent.

Within the classical view on SDEs as a continuous-time limit of SdEs, no interpretation can in principle have a qualitative mathematical advantage over the others. This follows from the mere fact that different interpretations can be transformed among themselves by a mere shift of the flow vector field in accordance with Equation (\ref{Falpha}). Indeed, consider, for example, an Ito SDE and a Stratonovich SDE with the flow appropriately shifted such that the FP operators of the two models are the same. These two models define the same stochastic model, and there is simply no room to accommodate any mathematical advantage of one model over the other.

\begin{figure}[t]
\centerline{\includegraphics[width=0.99\linewidth]{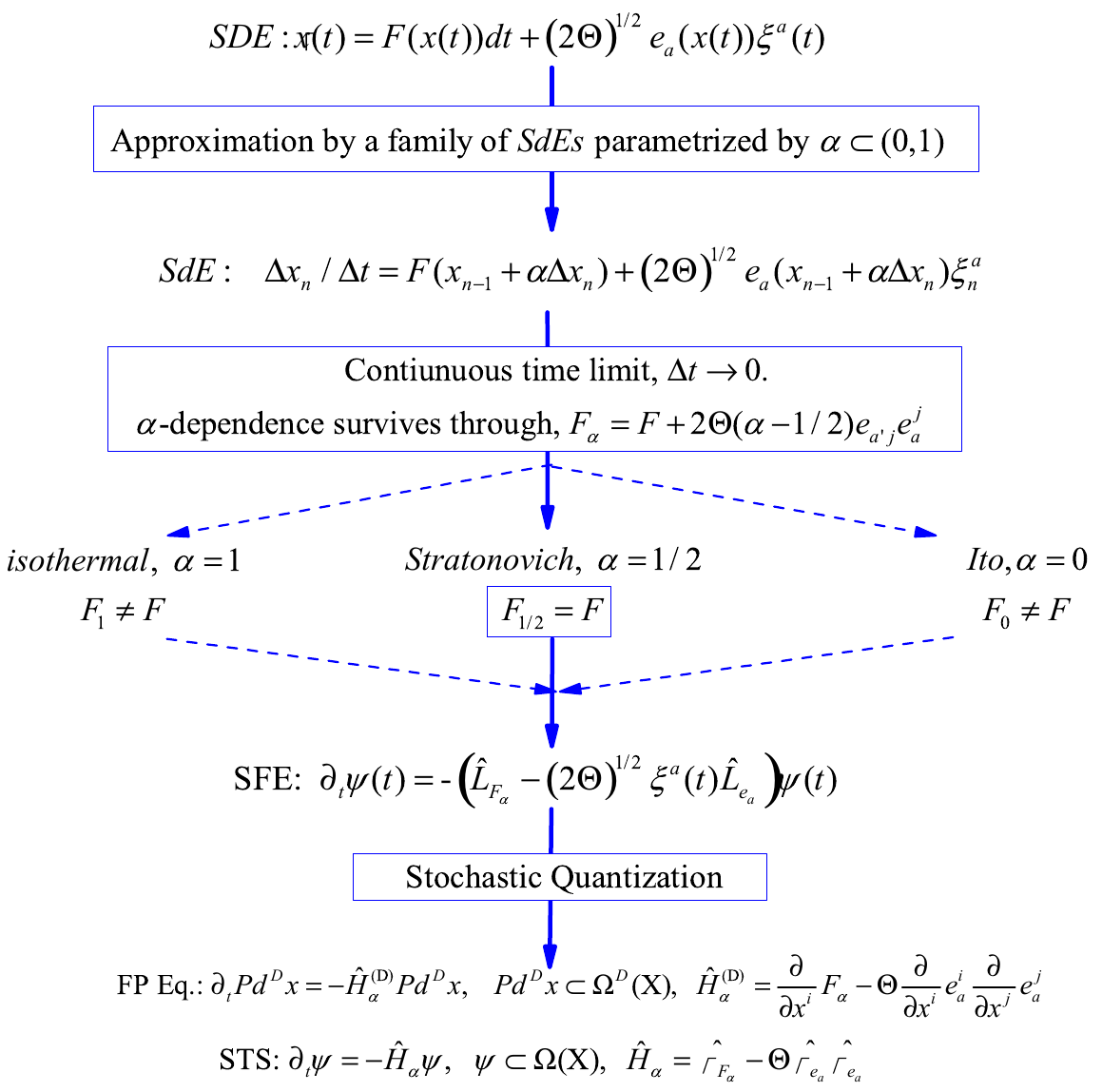}}
\caption{\label{Figure_3_1} A diagram summarizing the relations among SDEs, SdEs, and SFEs. An SDE can be interpreted as a continuous-time limit of an $\alpha$-family of a corresponding SdE. In the continuous time limit, the SdE corresponds to an SFE with the $\alpha$-dependent flow vector field. The Ito ($\alpha=0$), Stratonovich ($\alpha=1/2$), and "isothermal" ($\alpha=1$) \cite{Isothermal} choices of $\alpha$ are explicitly given. The Stratonovich interpretation of SDEs has one distinct advantage in that the flow vector field of the original SDE is the same as that of the corresponding SFE.}
\end{figure}

In other words, the classical theory of SDEs has an intrinsic redundancy in the sense that each stochastic model has infinitely many representatives corresponding to different interpretations of the SDE. This redundancy can only be removed if there existed a reference point outside the classical view of SDEs as a continuous-time limit of $\alpha$-parameterized SdEs. The SFEs introduced above may serve as this external reference point. If this point of view is adopted, it must be said that \emph{the Ito--Stratonovich dilemma is resolved in favor of the Stratonovich interpretation of SDEs} because Stratonovich SDEs provide the same FP operator as the corresponding SFEs. Within this framework, one would state that the continuous-time limit of two SdEs with the same $F$ and $e$ but with different $\alpha$ is two different SDEs with shifted flow vector fields. This point of view is summarized in Figure \ref{Figure_3_1}. 

Supporting this proposition, it can be mentioned that the Stratonovich approach is more natural from a mathematical point of view (see, e.g., \cite{Moon14} and the references therein). On the side of physics, the Stratonovich interpretation is superior because all noise sources are never white; rather, they have finite correlation times, and in the white-noise limit, the colored-noise SDEs become Stratonovich SDEs~\cite{Wong}. Furthermore, it is experimentally established that the Stratonovich approach is more accurate in regard to the numerical simulations of physical models \cite{West}. Simultaneously, the only known advantage of the Ito interpretation of being ``respective'' of the Markovian property is a~misinterpretation of the fact that the SdEs with $\alpha\ne0$ define the increment as a function of the initial point implicitly, as discussed at the end of Appendix \ref{ItoStratonovichDilemma}. 

The resolution of the Ito--Stratonovich dilemma, however, is a purely mathematical problem, and the above picture should only be viewed as one of the scenarios of its possible solution. Fortunately, it~is not important for further discussion whether the Ito--Stratonovich dilemma is resolved at this point. To liberate the STS from the burden of the Ito--Stratonovich dilemma, one can always think that the STS is a theory of SFEs and not of SDEs. If one is interested in the STS of the Ito or any other interpretation of the SDE, all he has to do is to shift the flow vector field accordingly. 

The Ito--Stratonovich dilemma will be readdressed in Section \ref{SecStratonovichWeyl}, where it will be shown that the Stratonovich interpretation of SDEs is equivalent to the Weyl symmetrization rule, whereas the so-called Martingale property of the Ito interpretation is equivalent to the unphysical convention of placing all the momentum operators after all the position operators. It will also be reestablished that different interpretations of SDEs can be transformed among themselves by a shift of the flow vector field and that this result is correct for the entire SEO and not only for the FP operator as discussed here.      

\subsection{Properties of the Stochastic Evolution Operator}
\label{SecSpectrum}

In this subsection, some of the most important properties of the eigensystem of the SEO are discussed. For simplicity, it is assumed that the phase space is closed, the noise-induced metric $g^{ij}$ is positive definite everywhere on $X$, $\Theta>0$ so that the SEO is elliptic, and $\hat H$ is diagonalizable, with a~discrete spectrum bounded from below. It is natural to believe that most of the claims here hold true or are at least transformative to more general classes of models.

\subsubsection{Fermion Number Conservation}

The SEO is of ``zeroth'' degree, \emph{i.e.}, $deg \hat H = 0$ (see the definition of the degree of an operator in Equation (\ref{BiGRadedCommutator})). This means that it does not mix wavefunctions of different degrees. As a result, the operator of the degree of the differential form in Equation (\ref{K_Operator}) is commutative with the SEO:
\begin{eqnarray}
[\hat k, \hat H] = 0.\label{FermionNumberConservation}
\end{eqnarray}

As noted in Section \ref{SecFermionVar}, the differentials of wavefunctions can be viewed as fermions. Therefore, $\hat k$ can be interpreted as the number of fermions, and Equation (\ref{FermionNumberConservation}) reflects the conservation of this~quantity.

Yet another method of expressing the idea of the conservation of the number of fermions is to note the block-diagonal structure of $\hat H$:
\begin{eqnarray}
\hat H = \text{diag} (\hat H^{(D)},\hat H^{(D-1)},...,\hat H^{(0)}), \end{eqnarray}
where 
\begin{eqnarray}
\hat H^{(k)}:\Omega^{k}(X)\to\Omega^{k}(X),
\end{eqnarray}
is the projection of $\hat H$ onto $\Omega^{k}(X)$.

\subsubsection{Completeness}

The SEO is a real operator. Therefore, its  spectrum consists of real eigenvalues and pairs of complex conjugate eigenvalues that are called Ruelle--Pollicott resonances  in DS theory. This form of the spectrum is a sufficient condition for $\hat H$ to be pseudo-Hermitian \cite{Mos02}. As a pseudo-Hermitian operator, $\hat H$ has a complete bi-orthogonal eigensystem
\begin{subequations}
\label{Eigensystem}
\begin{eqnarray}
\hat H \psi_n &=& \mathcal{E}_n \psi_n, \\ \bar \psi_n \hat H  &=& \bar \psi_n \mathcal{E}_n , \\
\sum\nolimits_n |\psi_n\rangle \langle \psi_n | &=& \hat 1_{\Omega(X)}, \\ \langle \psi_n | \psi_m \rangle  &=&  \int_X \bar \psi_n\wedge \psi_m  =  \delta_{nm},\label{Orthogonality}
\end{eqnarray}
\end{subequations}
where $\psi$ and $\bar\psi$ are the right and left eigenfunctions (or rather eigenforms), and the bra-ket notation is 
\begin{eqnarray}
|\psi_n \rangle \equiv \psi_n\in\Omega^{k_n}(X) \text{ and } \langle \psi_n | \equiv \bar \psi_n \in\Omega^{D-k_n}(X).\label{BraKetEig}
\end{eqnarray}

Here, $k_n$ is the eigenvalue of the operator of the degree of a wavefunction:
\begin{eqnarray}
\hat k |\psi_n \rangle = k_n |\psi_n \rangle.
\end{eqnarray}

As noted in the previous subsection, the degree of a wavefunction is a ``good quantum number'' because $\hat k$ is commutative with $H$ and consequently the both operators can be simultaneously diagonalized. 

The eigenstates of a given degree, say, $k$, provide a complete bi-orthogonal eigensystem on $\Omega^k(X)$ such that the resolution of unity on $\Omega^{k}(X)$ is
\begin{eqnarray}
\sum\nolimits_{n, k_n = k} |\psi_n\rangle \langle \psi_n | = \hat 1_{\Omega^k(X)}.
\end{eqnarray}

The bra-ket combination of an eigenstate of any degree is a D-form,
\begin{eqnarray}
\bar \psi_n \wedge \psi_n = P_n \in \Omega^D(X),
\end{eqnarray}
which has the meaning of the TPD associated with this eigenstate.

\subsubsection{Pseudo-Time-Reversal Symmetry}
\label{Sec:TimeReversalSymmetry}

The pseudo-Hermitianity of the SEO is closely related to the pseudo-time-reversal symmetry of the model. To begin the discussion of this symmetry, one needs the explicit expression for $\hat H^\dagger$ that can be established as follows. The~defining property of $\hat H^\dagger$ is given by Equation (\ref{DefinitionHerConJug}):
\begin{eqnarray}
\int_X \star \phi^*\wedge \hat H \psi = \int_X \star(\hat H^\dagger \phi)^*\wedge \psi.\label{Conjugate}
\end{eqnarray}

One notes now that the Lie derivative is a differentiation, \emph{i.e.},
\begin{eqnarray}
\hat{\mathcal L}_G \left(\psi_1\wedge\psi_2\right) = (\hat{\mathcal L}_G \psi_1)\wedge\psi_2 + \psi_1\wedge(\hat{\mathcal L}_G \psi_2)
\end{eqnarray}
for any $\psi_{1,2}\in\Omega(X)$ and any vector field $G\in TX$. In addition, 
\begin{eqnarray}
\int_X \hat{\mathcal L}_G \psi^{(D)} = \int_X (\hat d \hat\imath_G + \hat\imath_G\hat d) \psi^{(D)} \nonumber  \\ = \int_X \hat d (\hat\imath_G \psi^{(D)}) = 0
\end{eqnarray}
because $\hat d\psi^{(D)}=0$ for any $\psi^{(D)}\in\Omega^D(X)$. The last two formulas lead to the conclusion that, for any $\psi_1\wedge \psi_2\in\Omega^D$,
\begin{eqnarray}
\int_X \psi_1\wedge(\hat{\mathcal L}_G \psi_2) =  \int_X(-\hat{\mathcal L}_G \psi_1)\wedge \psi_2.\label{DiffByParts}
\end{eqnarray}

This equality can now be used to rewrite the Left Hand Side (L.H.S.) of Equation (\ref{Conjugate}) as
\begin{eqnarray}
\int_X \star \phi ^*\wedge \hat H \psi = \int_X \star ( \star^{-1} \hat H_{\mathfrak{T}} \star \phi)^*\wedge \psi,\label{EqualAdditional_1}
\end{eqnarray}
where
\begin{eqnarray}
\hat H_{\mathfrak{T}} = - \hat{\mathcal L}_F - \Theta \hat{\mathcal L}_{e_a}\hat{\mathcal L}_{e_a}.\label{FPOpRev}
\end{eqnarray}

Here, $\star^{-1}$ is the inverse of the Hodge star defined in Section \ref{sec:HodgeStar}, and the following has been~used:
\begin{eqnarray}
\star^{-1} \hat H_{\mathfrak{T}} \star \phi^* = ( \star^{-1} \hat H_{\mathfrak{T}} \star \phi)^*\label{ComplexCommute}
\end{eqnarray}
because both $\hat H_{\mathfrak{T}}$ and $\star$ are real.  Equations  (\ref{Conjugate}) and (\ref{EqualAdditional_1})  give
\begin{eqnarray}
\hat H^\dagger = \star^{-1} \hat H_{\mathfrak{T}} \star.\label{HDaggerExpression}
\end{eqnarray}

The next goal is to examine the properties of the model with respect to the time-reversal operation. One first notes that Equation (\ref{FPOpRev}) is the SEO of the SDE obtained from Equation (\ref{SDE}) by reversing the flow of time:
\begin{eqnarray}
\dot x = - \mathcal{F} (x).\label{SDEReversed}
\end{eqnarray}

In other words, the ``naive''reversal of time
\begin{eqnarray}
\mathfrak{T}: F \to -F, e_a \to -e_a\label{MathfrakT}
\end{eqnarray}
has the following effect on the SEO:
\begin{eqnarray}
\mathfrak{T}: \hat H \to \hat H_{\mathfrak{T}}\label{HMathfrak}
\end{eqnarray}
in the original stochastic evolution in Equation (\ref{FPEq}). 

Let us recall now that the time reversal  in quantum mechanics is also accompanied by swapping bras and kets. This is needed because the information of the system's past and future are stored in the kets and bras, respectively, whereas the past and future are interchanged by the time reversal. This~bra-ket swapping in the STS is accomplished using the Hodge star
\begin{eqnarray}
T:\hat H \to \hat H^\dagger=\star^{-1} \hat H_{\mathfrak{T}} \star,
\end{eqnarray}
where $T=\star \mathfrak{T}$ denotes the composition of the operations.

In Section \ref{Chap:DynamicsTypes}, it will be discussed that the structure of the wavefunction of the supersymmetric ground states is such that the coordinate directions with/without differentials correspond to the stable/unstable (local) variables. On the other hand, the time reversal makes the stable variables unstable and vice versa. Therefore, this operation must act on a wavefunction in such a way that, in the directions with/without differentials, the wavefunction loses/acquires differentials. This~understanding strengthens the above relation between the time-reversal operation and the Hodge star operation, which acts on a wavefunction in exactly this manner.

Clearly, an operation $T$ does not seem to be a symmetry of the model because $\hat H\ne \hat H^\dagger$ for pseudo-Hermitian $\hat H$. Nevertheless, the stochastic evolution defined by $\hat H^\dagger$ may turn out to be physically equivalent to that defined by $\hat H$. On a mathematical level, this ``equivalency'' means that there exist
\begin{eqnarray}
\eta: \Omega^{k}(X) \to \Omega^{k}(X)  
\end{eqnarray}
such that
\begin{eqnarray}
\eta: \hat H^\dagger \to \hat H = \eta^{-1}\hat H^\dagger \eta.
\end{eqnarray}

The existence of such $\eta$ is the definitive property of all pseudo-Hermitian operators. Therefore,~the model does possess the so-called $\eta T$-symmetry:
\begin{eqnarray}
\eta T: \hat H \to \hat H = (\star\eta)^{-1} \hat H_{\mathfrak{T}} \star\eta.\label{EtaStarT}
\end{eqnarray}

The operator $\eta$ can be called the Hilbert space metric. It relates the bras and the Hodge duals of the eigenstates as
\begin{eqnarray}
\langle \psi_n| = (\psi_k|\eta_{kn},
\end{eqnarray}
where the notation $( \psi_k | \equiv \star \psi_k^*$ was previously introduced in Equation (\ref{Inner_Product}). From the orthogonality property in {Equation} (\ref{Orthogonality}), 
one finds
\begin{eqnarray}
\langle \psi_n|\psi_m\rangle = (\psi_k|\psi_m)\eta_{kn} = \delta_{mn}.
\end{eqnarray}

In other words, $\eta$ is the inverse of the ``overlap'' matrix, \emph{i.e.}, of (the transpose of) the matrix of the inner products of the kets of the eigenstates.

The $\eta T$ operation acts on a wavefunction as
\begin{eqnarray}
\eta T:\psi \mapsto \eta^{-1}\star^{-1}\psi^*.
\end{eqnarray}

If $\psi_{T,n}$ is an eigenstate of the time-reversed SDE (\ref{SDEReversed}),
\begin{eqnarray}
\hat H_{\mathfrak{T}} \psi_{T,n} = \mathcal{E}_n\psi_{T,n},
\end{eqnarray}
then $\eta T(\psi_{T,n})$ is an eigenstate of the original SDE (\ref{SDE}) although with a complex conjugate~eigenvalue,
\begin{eqnarray}
\hat H \eta T(\psi_{T,n}) = \mathcal{E}_n^* \eta T(\psi_{T,n}).
\end{eqnarray}

It can be said that the eigenstates with complex eigenvalues (the Ruelle--Pollicott resonances) break the $\eta T$ symmetry. If one such eigenstate is a ground state of the model, the $\eta T$ symmetry can be said to be spontaneously broken because the ground state of the time-reversed SDE has a different~eigenvalue.

Equation (\ref{EtaStarT}) can also be rewritten as 
\begin{eqnarray}
\hat H^{(k)} = (\star\eta)^{-1}\hat H_{\mathfrak{T}}^{(D-k)}\star\eta.
\end{eqnarray}

That is, $\hat H^{(k)}$ and $\hat H_{\mathfrak{T}}^{(D-k)}$ are related by the similarity transformation $\star\eta$. This immediately suggests that the two operators are isospectral:
\begin{eqnarray}
spec(\hat H^{(k)}) = spec(\hat H_{\mathfrak{T}}^{(D-k)}).\label{IsoSpectralTimeReversed}
\end{eqnarray}

This result will be used later in the discussion of the possible forms of the SEO spectra in Section~\ref{SubSec:TermEquil}.

Up to this moment, it was not specified what phase-space metric $g$ is being used. This phase space metric, or $X$-metric for short, enters the above formulas through the definition of the Hodge dual in Equation (\ref{DefinitionOfHodge}). What has been said so far in this subsection is correct for any ``good enough'' $X$-metric. In other words, one has a freedom in choosing $g$. This freedom can in principle be used for the simplification of the explicit expressions for $g$-dependent objects such as $\hat H^\dagger$ and $\eta$. For $g$-independent objects, such as $\hat H$, $\hat H_{\mathfrak{T}}$, the Eigensystem (\ref{Eigensystem}), and the composition $\star \eta$, the choice of $g$ is unimportant.

\subsubsection{Topological Supersymmetry}

The SEO is $\hat d$-exact, \emph{i.e.}, it has the form of a bi-graded commutator
\begin{eqnarray}
\hat H = [\hat d, \hat{\bar d}],\label{DExactFPOperator}
\end{eqnarray}
where
\begin{eqnarray}
\hat{\bar d} = F^i\hat{\imath}_i - \Theta e^i_a\hat{\imath}_i\hat {\mathcal L}_{e_a}.\label{dbar}
\end{eqnarray}

The exterior derivative is commutative with the SEO:
\begin{eqnarray}
[\hat d, \hat H] = 0.
\end{eqnarray}

This can be observed from the nilpotency property of the exterior derivative, \emph{i.e.}, $\hat d^2=0$, leading to the conclusion that $\hat d$ commutes with any $\hat d$-exact operator: $[\hat d, [\hat d, \hat X]] = 0, \forall \hat X$.

The commutativity of an operator with the SEO indicates that this operator is a symmetry of the model. The reason why $\hat d$ is a symmetry can be explained as follows. The finite-time SEO is a stochastically averaged pullback induced by the SDE-defined diffeomorphisms. Therefore, the finite-time SEO commutes with $\hat d$ because any pullback induced by a diffeomorphism is commutative with $\hat d$. In other words, this symmetry is a consequence of the fact that continuous(-time) dynamics preserves the continuity of the phase space, as previously mentioned in Section \ref{Intro:Relation}. 

Note also that not all possible evolution operators that commute with $\hat d$ are necessarily $\hat d$-exact as in Equation (\ref{DExactFPOperator}). A $\hat d$-exact evolution operator implies more than simply the commutativity with $\hat d$. As will be discussed below, the additional implication of a $\hat d$-exact evolution operator is that all the $\hat d$-symmetric eigenstates have a zero eigenvalue.

In terms of the fermionic variables of Section \ref{SecFermionVar}, the exterior derivative substitutes commuting or bosonic variables with anticommuting or fermionic ones. Therefore, it can be identified as a~supersymmetry. 

\subsubsection{Topological Supersymmetry \emph{vs.} $N=2$ Supersymmetry}
\label{Top_vs_Susy}

The topological supersymmetry operator is the same for all SDEs. The operator does not contain any information on the specifics of dynamics, which in turn are solely encoded in the other fermionic operator, \emph{i.e.}, $\hat {\bar d}$, in Equation (\ref{dbar}). One way to look at this operator is as the operator of the current of the probability density. This point of view is at least partially correct, as is evident from the stochastic evolution equation for top differential forms
\begin{eqnarray}
\partial_t \psi^{(D)} = - \hat d (\hat {\bar d}\psi^{(D)}) = - \hat d j,
\end{eqnarray}
which can be recognized as a continuity equation for the total probability density, and the R.H.S. is the divergence of the current of the probability density: $j=\hat {\bar d}\psi^{(D)}\in\Omega^{(D-1)}$.

The current operator is not a supersymmetry of the model. The operator lacks the important property of being nilpotent, \emph{i.e.}, $\hat {\bar d}^2\ne 0$, and thus is not commutative with the SEO:
\begin{eqnarray}
[\hat H, \hat {\bar d}] = [\hat d, \hat {\bar d}^2]\ne 0.\label{Commutativity}
\end{eqnarray}

Only for a very special class of models is this operator nilpotent and consequently a~supersymmetry of the model. The best known examples from this class of models are the Langevin SDEs and the Hamilton models, \emph{i.e.}, models that have been studied in the literature almost exclusively in the context of the relation between supersymmetry and stochastics. The reason why these models have received most of the scientific attention is that only when $\hat{\bar d}^2=0$ can the model be said to be $N=2$ supersymmetric. In particular, the evolution operator is a square of $N=2$ mixed-degree fermionic~operators,
\begin{eqnarray}
\hat {q}_1=\hat d + \hat {\bar d}, \text { and } \hat {q}_2 = i(\hat d - \hat {\bar d}),
\label{SuperchargesQ}
\end{eqnarray}
and
\begin{eqnarray}
\hat H = \hat q_1^2 = \hat q_2^2 \text{ only if } \hat {\bar d}^2=0. \label{N_2_Susy}
\end{eqnarray}

In the general case, however, Equation (\ref{N_2_Susy}) is incorrect, and the model can only be identified as a~member of the family of the cohomological field theories for which the evolution operator is a~bi-graded commutator with the topological supersymmetry operator, as in Equation (\ref{DExactFPOperator}). In other words, the difference between general form SDEs and Langevin SDEs and/or Hamilton models is the same as the difference between topological quantum mechanics (see, e.g., \cite{TQM}) and $N=2$ supersymmetric quantum mechanics (see, e.g., \cite{SQM}).

In addition, note that, to identify a model as a cohomological field theory, one must also require that the ground states of the model be supersymmetric. In the STS, however, the ground states are not supersymmetric in the most interesting situations with the spontaneously broken topological supersymmetry. Therefore, the identification of the STS as a cohomological field theory is technically~inaccurate.

It is well known that $N=2$ supersymmetry leads to the pairing of the non-supersymmetric eigenstates into boson-fermion doublets, whereas all the supersymmetric eigenstates are singlets with exactly zero eigenvalues. In the next Subsection \ref{Sec:BosonFermionPairing}, it will be shown that the topological supersymmetry tailors the same structure of the eigensystem.

\subsubsection{Boson-Fermion Pairing of Eigenstates}
\label{Sec:BosonFermionPairing}

From the group-theoretic point of view, the topological supersymmetry is a continuous one-parameter group of transformations
\begin{eqnarray}
\hat G_s = (\hat G_{-s})^{-1} = e^{s\hat d} = 1 + s \hat d, s \in \mathbb{R},
\end{eqnarray}
of which the SEO is invariant:
\begin{eqnarray}
\hat G_s \hat H \hat G_{-s} = \hat H.
\end{eqnarray}

As in the case of any other symmetry, the eigenstates must be irreducible representations of this group. There are only two types of irreducible representations of this symmetry: most of the eigenstates are non-$\hat d$-symmetric ``bosonic-fermionic'' doublets or pairs of eigenstates, and some of the eigenstates are $\hat d$-symmetric singlets.

Each pair of non-$\hat d$-symmetric eigenstates, which will be denoted as $|\vartheta\rangle$ and $|\vartheta'\rangle$, can be defined via a single bra-ket pair, \emph{i.e.}, $|{\underline{\vartheta}}_n\rangle$ and $\langle{\underline{\vartheta}}_n|$, such that 
\begin{eqnarray}
\langle{\underline{\vartheta}}_n|\hat d |{\underline{\vartheta}}_n \rangle = 1,\label{Reduced_Orthodonality}
\end{eqnarray}
so the bra-ket pairs of the non-$\hat d$-symmetric pairs of eigenstates can be given as
\begin{eqnarray}
|\vartheta_n \rangle = |{\underline{\vartheta}}_n\rangle, \text{ } \langle\vartheta_n| = \langle{\underline{\vartheta}}_n|\hat d, \label{FirstType}
\end{eqnarray}
and
\begin{eqnarray}
|\vartheta'_n \rangle = \hat d |{\underline{\vartheta}}_n\rangle, \text{ } \langle\vartheta'_n| = \langle{\underline{\vartheta}}_n|.\label{SecondType}
\end{eqnarray}

Here, the expression $\langle{\underline{\vartheta}}_n|\hat d$ must be understood as a differential form $\bar \vartheta_n$ such that $\int_X \bar{{\underline{\vartheta}}}_n\wedge \hat d \gamma = \int_X \bar{\vartheta}_n\wedge \gamma$ for any $\gamma$, where the barred notation for the bras in {Equation} (\ref{Orthogonality}) has been used. Using the standard relation $\int \hat d (\gamma_1\wedge\gamma_2) = 0$, valid for all $\gamma_{1,2}$, it can be easily established that, up to a sign, the differential form $\bar \vartheta_n = \hat d\bar{{\underline{\vartheta}}}_n$.

The orthogonality relations for the $\underline{\vartheta}$ are
\begin{eqnarray}
\langle{\underline{\vartheta}}_n|\hat d|{\underline{\vartheta}}_k\rangle =\delta_{nk},
\langle{\underline{\vartheta}}_n|{\underline{\vartheta}}_k\rangle =0.\label{Overlap}
\end{eqnarray}

These relations can be derived from the structure of the non-$\hat d$-symmetric eigenstates in Equations~(\ref{FirstType}) and (\ref{SecondType}) as follows. Consider an eigen-bra $\langle{\underline{\vartheta}}_n|\in\Omega^{D-k}$. There are two types of eigen-kets of degree $k$ that can potentially  overlap with $\langle{\underline{\vartheta}}_n|$ non-trivially: $|{\underline{\vartheta}}_{n_1}\rangle\in\Omega^{k}$ and the $\hat d$-exact eigen-kets $\hat d|{\underline{\vartheta}}_{n_2}\rangle\in\Omega^{k}$ with $|{\underline{\vartheta}}_k\rangle\in\Omega^{k-1}$. The eigensystem is bi-orthogonal; thus, only one eigen-ket can provide a non-zero overlap with $\langle{\underline{\vartheta}}_n|$. As observed from Equation (\ref{Reduced_Orthodonality}), this eigen-ket is $\hat d$-exact. All the other kets must have zero overlap with $\langle{\underline{\vartheta}}_n|$ because they correspond to different eigenvalues of the SEO, which  has no spectral degeneracy in the most general situation. As long as this argument is valid in the most general case of no spectral degeneracy, it is always valid even in the case where there is spectral degeneracy due to some additional symmetry of the model or accidentally.

The pairing of the non-supersymmetric eigenstates into the boson-fermion doublets can be demonstrated as follows. Consider the complete set of the eigenstates of $\hat H^{(0)}$, \emph{i.e.}, $|{\underline{\vartheta}}_{n_0}\rangle\in\Omega^0$, where $n_0$ is the label running over all these eigenstates. Consider also the set of $\hat d$-exact states, \emph{i.e.}, $|{\underline{\vartheta}}'_{n_0}\rangle=\hat d |{\underline{\vartheta}}_{n_0}\rangle\in \Omega^1$. These are the eigenstates of $\hat H^{(1)}$ with the same eigenvalues as $|{\underline{\vartheta}}_{n_0}\rangle$ because $\hat d$ commutes with $\hat H$. This set is incomplete in $\Omega^1$, and there exists another set of eigenstates of $\hat H^{(1)}$, \emph{i.e.}, $|{\underline{\vartheta}}_{n_1}\rangle\in\Omega^1$, with $n_1$ being yet another label running over this new set of eigenstates. Unlike $|{\underline{\vartheta}}'_{n_0}\rangle$, $|{\underline{\vartheta}}_{n_1}\rangle$ are not $\hat d$-closed, \emph{i.e.}, the operator $\hat d$ does not annihilate them in general. Once again, one considers the set of the $\hat d$-exact eigenstates of $\hat H^{(2)}$, $|{\underline{\vartheta}}'_{n_1}\rangle = \hat d |{\underline{\vartheta}}_{n_1}\rangle\in\Omega^2$. This recurrent procedure terminates at $\Omega^D$, the eigenstates of which are all $\hat d$-closed. At the end of this procedure, almost all (see below) the eigenstates are non-supersymmetric pairs related by $\hat d$. 

A similar procedure exists for bras. The only difference is that the procedure runs in the opposite direction, \emph{i.e.}, from $\langle {\underline{\vartheta}}_{n_D}|$ to $\langle {\underline{\vartheta}}_{n_0}|$, because $\bar{{\underline{\vartheta}}}_{n_k}\in\Omega^{D-k}$, as observed from {Equation} (\ref{Orthogonality}), and consequently $\langle {\underline{\vartheta}}_{n_k}|\hat d$ is an eigen-bra of $\hat H^{(k-1)}$ and not of $\hat H^{(k+1)}$, as could be incorrectly expected. 

The boson-fermion pairing procedure does not count eigenstates that are non-trivial in the de~Rham cohomology, in other words, the eigenstates that are $\hat d$-closed but that are not $\hat d$-exact, \emph{i.e.}, 
\begin{eqnarray}
\hat d|\theta_k\rangle=0, \text{ but }|\theta_k\rangle \ne \hat d|\cdot\rangle.\label{Singlets}
\end{eqnarray}

These eigenstates are the supersymmetric or $\hat d$-symmetric singlets. Their bras are also non-trivial in the de Rham cohomology:
\begin{eqnarray}
\langle\theta_k|\hat d=0, \text{ but }\langle\theta_k| \ne \langle \cdot |\hat d.
\end{eqnarray}

An important property of the supersymmetric eigenstates is that the expectation value of any $\hat d$-exact operator vanishes on these eigenstates:
\begin{eqnarray}
\langle\theta_k| [\hat d, \hat X] |\theta_l\rangle = 0.\label{ZeroExactOperators}
\end{eqnarray}

Note now that the SEO is a $\hat d$-exact operator. Therefore, all the $\hat d$-symmetric eigenstates have zero eigenvalues: $0=\langle\theta_k| \hat H|\theta_k\rangle = \mathcal{E}_{\theta_k}\langle\theta_k| \theta_k\rangle = \mathcal{E}_{\theta_k}$.

Each de Rham cohomology class must provide one $\hat d$-symmetric eigenstate of the form
\begin{eqnarray}
|\theta_k\rangle = |h_k\rangle + \hat d|\cdot\rangle,\label{Harmonic}
\end{eqnarray}
where $|h_k\rangle$ is the harmonic differential form from this particular class of the de Rham cohomology. That~this is true can be demonstrated using perturbation theory, as done in Appendix \ref{PertubativeStates}. The~statement that each de Rham cohomology class provides one $\hat d$-symmetric eigenstate must be correct even outside the domain of the applicability of the perturbation theory. This follows from the ``completeness argument'': if it is not so, then the eigensystem of the pseudo-Hermitian SEO is incomplete, which contradicts the theory of pseudo-Hermitian operators.

\subsubsection{Topological Supersymmetry and $N=2$ Pseudo-Supersymmetry}

In Section \ref{Top_vs_Susy}, it was discussed that, in the general case, $\hat {\bar d}$ is not a supersymmetry of the model. Nevertheless, the second supercharge does exist. Although its explicit form may not be easy to establish, the second supercharge can be easily constructed out of the eigensystem of the model. Using~the notations in Equations (\ref{FirstType}) and (\ref{SecondType}), the SEO can be given as
\begin{eqnarray}
\hat H = \sum\nolimits_{n}\left(|{\underline{\vartheta}}_n\rangle\mathcal{E}_{n}\langle{\underline{\vartheta}}_n|\hat d + \hat d |{\underline{\vartheta}}_n\rangle\mathcal{E}_{n}\langle{\underline{\vartheta}}_n|\right) = [\hat d, \hat {d}^\ddagger],\label{DExactFPOperator1}
\end{eqnarray}
where the second supercharge of the model is
\begin{eqnarray}
\hat {d}^\ddagger = \sum\nolimits_{n}|{\underline{\vartheta}}_n\rangle\mathcal{E}_{n}\langle{\underline{\vartheta}}_n|.
\end{eqnarray}

The two operators $\hat {d}^\ddagger$ and $\hat {\bar d}$ must differ by a $\hat d$-closed piece, \emph{i.e.}, 
\begin{eqnarray}
[\hat d, \hat {d}^\ddagger-\hat {\bar d}] =0,
\end{eqnarray}
because Equations (\ref{DExactFPOperator}) and (\ref{DExactFPOperator1}) define the same SEO. 

Just like $\hat{\bar d}$, this operator has fermionic degree $-1$. Unlike $\hat {\bar d}$, the second supercharge is nilpotent, \emph{i.e.}, $(\hat { d^\ddagger})^2=0$, as can be verified using Equation (\ref{Overlap}), and commutative with the SEO, \emph{i.e.},  $[\hat H,\hat d^\ddagger]=[\hat d, (\hat {d}^\ddagger)^2]=0$, as follows from Equation (\ref{DExactFPOperator1}) and the nilpotency of this operator. Moreover, with the  introduction of the generalization of the two operators in Equation (\ref{SuperchargesQ}), $\hat {\tilde q}_1=\hat d + \hat d ^\ddagger$ and $\hat {\tilde q}_2 = i(\hat d - \hat d ^\ddagger)$, the SEO can be given a form similar to that in Equation (\ref{N_2_Susy}): $\hat H = \hat {\tilde q}_1^2 = \hat {\tilde q}_2^2$.

\begin{figure}[h]
\centerline{\includegraphics[height=7.5cm, width=8.5cm]{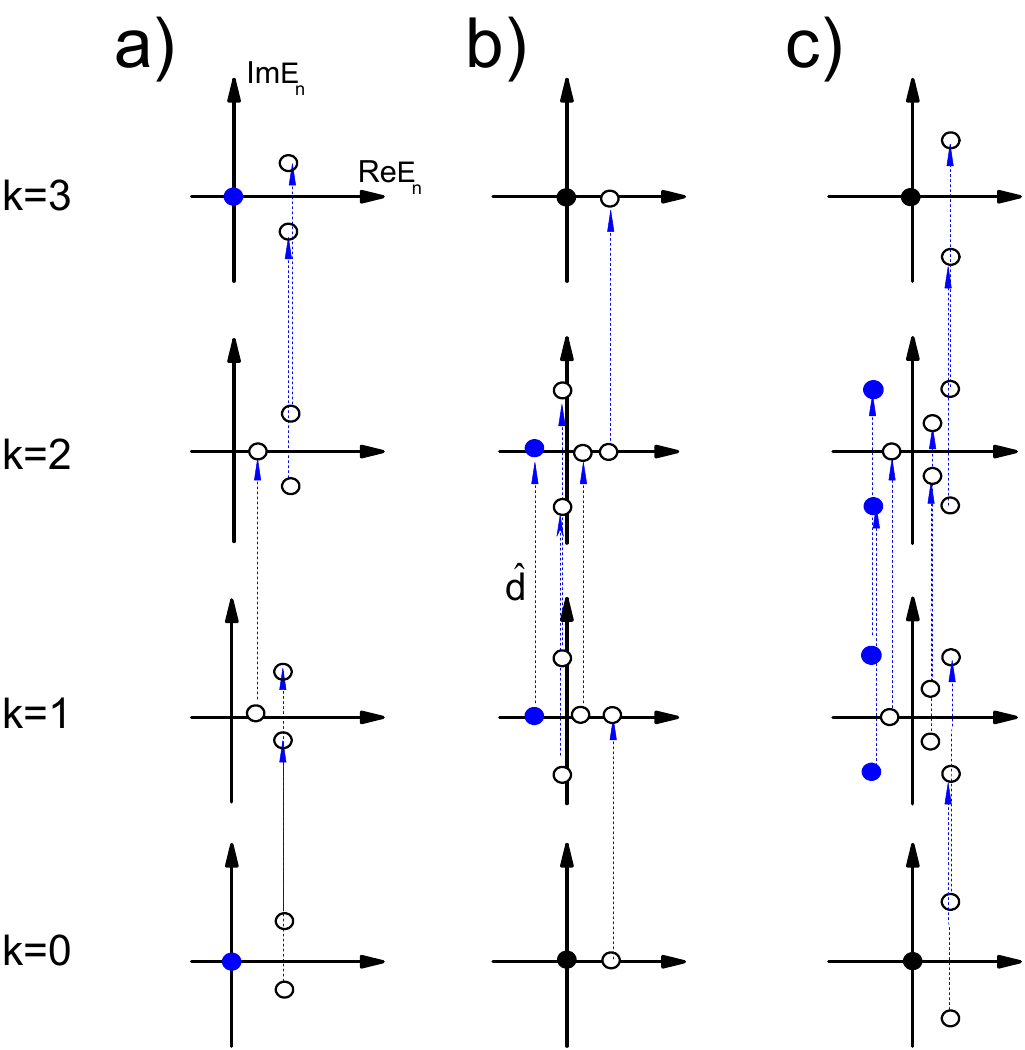}}
\caption{\label{Figure_3_2} The three possible types of spectra of the SEO for an SDE on a three-dimensional sphere. The~spectra are given separately for the four degrees indicated by the parameter $k$ on the left. The~zeroth- and third-degree cohomology classes of the 3-sphere provide two supersymmetric eigenstates, indicated as thick dots at the origin for the $k=0$ and $k=3$ spectra. The ground states are represented as the leftmost filled (blue or black) dots. (\textbf{a}) The case of thermodynamic equilibrium when the topological supersymmetry is unbroken because the ground states are supersymmetric; (\textbf{b},\textbf{c})~The~cases of spontaneously broken supersymmetry when the ground states have nonzero eigenvalues and are thus non-supersymmetric; The ground state is ambiguous in case (\textbf{c}) because there are two~Ruelle--Pollicott resonances with the same lowest real part of its eigenvalue. In Section \ref{Sec:OneWayForUnique}, it is discussed that it is possible to view only one of these eigenstates as the ground state. Thin dotted arrows represent the action of $\hat d$ that couples all the non-$\hat d$-symmetric eigenstates into boson-fermion~pairs.}
\end{figure}

The second supercharge is responsible for the same boson-fermion pairing of the non-supersymmetric eigenstates. Its effect on the eigenstates is in essence opposite that of $\hat d$:
\begin{eqnarray}
\hat {d}^\ddagger |{\vartheta'}_n\rangle = {\mathcal E}_n|\vartheta_n\rangle, \hat {d}^\ddagger|{\vartheta}_n\rangle = 0, \text{ and } \hat d^\ddagger|\theta_k\rangle = 0, 
\end{eqnarray}
where the notations from Equations (\ref{FirstType}) and (\ref{SecondType}) have been used. This operator can be visualized in Figure \ref{Figure_3_2} as reverse arrows representing $\hat d$. In other words, the second supercharge does not contain any additional information on the supersymmetric structure of the SEO, and for this reason, it will not be considered in this paper any further. It must be noted, however, that in situations in which there exists an operator $\underline{\eta}$ such that $\hat d^\ddagger = \underline{\eta}^{-1}\hat d^\dagger \underline{\eta}$ and $\hat d = \underline{\eta}^{-1}\hat (d^\ddagger)^\dagger \underline{\eta}$, the model can be said to possess $N=2$ pseudo-supersymmetry \cite{Mos02}. It is an open question under what conditions the $N=2$ pseudo-supersymmetry is present and what its relation is with the pseudo-time reversal symmetry discussed in Section \ref{Sec:TimeReversalSymmetry}.

\subsubsection{Thermodynamic Equilibrium and Stochastic Poincar\'e--Bendixson Theorem}
\label{SubSec:TermEquil}

In Section \ref{Sec:BosonFermionPairing} above, it was argued that each de Rham cohomology class must provide one~supersymmetric eigenstate. Whether this is true is not important for further discussion. What is important is the existence of the supersymmetric state of thermodynamic equilibrium (TE), \emph{i.e.}, the steady-state (zero-eigenvalue) TPD: $\psi_{TE}\in\Omega^{(D)}$ (in DS theory the TE state is known as the invariant measure). Its presence can be established through the physical version of the completeness argument. Indeed, all the non-$\hat d$-symmetric eigenstates from $\Omega^D(X)$ are $\hat d$-exact, \emph{i.e.}, they are of Type (\ref{SecondType}). This means that the integral of all such eigenstates over $X$ is zero: $\int_{X} \hat d {\underline{\vartheta}} = 0$. On the other hand, a wavefunction from $\Omega^D(X)$ has the meaning of the TPD. The integral of a meaningful TPD over $X$ must not vanish. This suggests that at least one $\hat d$-symmetric eigenstate from $\Omega^D(X)$ must exist for physical models.\label{Thermodynamic Equilibrium} It can also be shown that the bra of the TE state is a~constant function: $\bar \psi_{TE}=const\in\Omega^0(X)$.  

It is also an important piece of understanding that the TE state is always the ``ground state'' for $\Omega^{D}$. In other words, among all the eigenstates from $\Omega^{D}$, it has the smallest real part of its eigenvalue, which is zero, of course. Indeed, imagine that this is not true and that there exists an eigenstate from $\Omega^{D}(X)$ such that its eigenvalue is real and negative. As mentioned in the previous paragraph, the integral of the ket of such a non-$\hat d$-symmetric eigenstate over $X$ vanishes. Thus, the ket of this eigenstate must be negative somewhere on $X$. One can now take a random TPD and evolve it in time sufficiently long. This will eventually lead to the situation whereby the TPD will become negative somewhere on $X$ due to the dominant contribution from this presumably existing non-$\hat d$-symmetric eigenstate. The negative TPD is not physical. Thus, it can be concluded that, for physical models, this situation is not realizable, and there are no eigenstates in $\Omega^{D}$ with real negative eigenvalues. In a similar manner, one can rule out the possibility that there is a pair of Ruelle--Pollicott resonances in $\Omega^{D}$ with a negative real part of their eigenvalues. Thus, for physical models, the TE state is the ``ground state'' in $\Omega^{D}$.     

The same reasoning applies to the SEO (\ref{FPOpRev}) of the time-reversed SDE (\ref{SDEReversed}). Combined with the fact that $spec \hat H^{(0)} = spec \hat H_{\mathfrak{T}}^{(D)}$, which follows from Equation (\ref{IsoSpectralTimeReversed}), this observation suggests that $\hat H^{(0)}$ also never breaks the supersymmetry, \emph{i.e.}, the supersymmetric zero-eigenvalue eigenstate is the ground state in $\Omega^0$. The ket of this ground state of $\Omega^0$ is a constant function on $X$. 

This brings the discussion to the Poincar\'e--Bendixson theorem stating that smooth deterministic flows can be chaotic only in three-plus dimensions (see, e.g., \cite{Teschl}). 
The above analysis of the SEO spectra seemingly leads to the stochastic version of this theorem. Indeed, as long as $\hat H^{(D)}$ and $\hat H^{(0)}$ do not break the topological supersymmetry as discussed in the two previous paragraphs, the overall supersymmetry cannot be spontaneously broken unless the dimensionality of the phase space is three or higher. This can be straightforwardly deduced from Figure \ref{Figure_3_2}. Clearly, if the dimensionality of the phase space is less than three, at least one eigenstate of a pair of the non-supersymmetric eigenstates (the degrees of which differ by one) with a negative real part of their eigenvalue must be either in $\Omega^D$ or $\Omega^0$, which contradicts the above properties of the spectra of $\hat H^{(D)}$ and $\hat H^{(0)}$.

\subsubsection{Realizable Spectra}

The properties of the SEO discussed previously limit its possible spectra to only the three types given in Figure \ref{Figure_3_2}. A natural question that may arise at this point is whether there exist other general limitations on the possible forms of the SEO spectra. For example, the spectra of the type in Figure \ref{Figure_3_2}c do appear somewhat suspicious because the pair of the Ruelle--Pollicott resonances are two equally good candidates for the title of the ground state. Furthermore, in DS theory, there are theorems stating that, for a certain class of models that mimic chaotic behavior, namely, the so-called expanding models, the eigenvalues of the ground states must be real \cite{Rue02}.

There are no other limitations. To convince oneself, one needs at least one example for each of the two types of SEO spectra with spontaneously broken supersymmetry (Figure \ref{Figure_3_2}b,c). These~examples must not necessarily be analytical. Well-established numerical examples are sufficient. Such examples exist in the theory of the magnetohydrodynamical phenomenon of kinetic dynamo (KD), as was very recently found in \cite{Torsten} and discussed briefly in Appendix \ref{sec:Appendix_KD}. Thus, both types of the supersymmetry breaking spectra in Figure \ref{Figure_3_2} are realizable.

\subsection{Witten Index}
\label{SecWittenPartFuncIntroduced}

One of the fundamental partition-function-like objects is the ''sharp trace'' of the finite-time SEO known in supersymmetric quantum theory as the Witten index:\index{Witten index}
\begin{eqnarray}
W_{tt'} &=& \text{Tr} (-1)^{\hat k} \hat {\mathcal M}_{tt'}  = \text{Tr} (-1)^{\hat k} e^{-\hat H(t-t')} \nonumber \\&=& \sum\nolimits_{n} (-1)^{k_n} e^{-\mathcal{E}_n(t-t')}.\label{WittenIndex}
\end{eqnarray}

It was previously established that all the eigenstates with non-zero eigenvalues are non-$\hat d$-symmetric. They come in pairs of even and odd degrees. Thus, their contributions cancel out from the Witten index. Only $\hat d$-symmetric eigenstates with zero eigenvalue contribute to the Witten index, which is thus independent of the duration of the time evolution:
\begin{eqnarray}
W_{tt'} \equiv W = \sum\nolimits_{k=0}^D (-1)^k b_k,
\end{eqnarray}
where $b_k$ is the number of $\hat d$-symmetric states of degree $k$. If one believes that each de Rham cohomology class provides one $\hat d$-symmetric state, then $b_k$ are the Betti numbers, and $W$ equals the Euler characteristic of $X$, $Eu(X)$. 


The next goal is to discuss the physical meaning of Equations (\ref{WittenIndex}) and (\ref{PartitionFunction}) and provide an~alternative proof of $W=Eu(X)$ by identifying it with the stochastic Lefschetz index. This can be performed using the fermionic variables in Section \ref{SecFermionVar}. The pullback in Equation (\ref{PullBack}) can be given as
\begin{eqnarray}
M^*_{t't} \psi(x\chi) &=& \int d^Dx' d^D\chi' M^*_{t't}(x\chi,x'\chi')\psi(x'\chi'),\\
M^*_{t't}(x\chi,x'\chi') &=& \delta^D(x'-M_{t't}(x)) \delta^D(\chi'- TM_{t't}(x)\chi), \label{PullBackRepresentation}
\end{eqnarray}
where the bosonic and fermionic $\delta$-functions substitute the arguments $x'$ and $\chi'$ by $M_{t't}(x)$ and $TM_{t't}(x)\chi$, respectively, with $TM_{tt'}(x)$ being the tangent map in Equation (\ref{TangentMap}).

Following stochastic averaging, one arrives at the finite-time SEO in the following representation:
\begin{eqnarray}
\hat{\mathcal{M}}_{tt'}(x\chi,x'\chi') = \langle M^*_{t't}(x\chi,x'\chi') \rangle_\text{Ns}.
\end{eqnarray}

The Witten index in Equation (\ref{WittenIndex}) takes the form of the trace of the finite-time SEO with periodic boundary conditions for both the commuting and anticommuting variables:
\begin{widetext}
\begin{eqnarray}
W &=& \int d^Dxd^D\chi \hat{\mathcal{M}}_{tt'}(x\chi,x\chi) = \left\langle \sum_{x=M_{t't}(x)} \frac{\text{det} (\hat 1_{TX} - TM_{t't}(x))}{|\text{det}(\hat 1_{TX} - TM_{t't}(x))|} \right\rangle_\text{Ns}
\nonumber\\
&=&
\sum\nolimits_{k=0}^D (-1)^k \left\langle \sum_{x=M_{t't}(x)} \frac{m_k(x)}{|\text{det}(\hat 1_{TX} - TM_{t't}(x))|} \right\rangle_\text{Ns},\label{WittenIndex2}
\end{eqnarray}
where the characteristic polynomial formula $\text{det} (\hat 1 + \lambda TM_{t't}(x)) = \sum\nolimits_{k=0}^D \lambda^k m_k(x)$ has been utilized,~with
\begin{eqnarray}
m_k(x) = \sum_{i_1<i_2<...<i_k}
\text{det}
\left(
\begin{array}{ccc}
TM_{t't}(x)^{i_1}_{i_1} &\dots& TM_{t't}(x)^{i_1}_{i_k}\\
\vdots &\ddots& \vdots\\
TM_{t't}(x)^{i_k}_{i_1} &\dots& TM_{t't}(x)^{i_k}_{i_k}
\end{array}
\right).\label{ms}
\end{eqnarray}

The denominator in Equation (\ref{WittenIndex2}) originates from the integration over the bosonic variables, whereas $m_k$ can be viewed as a fermionic trace over $\Omega^k(X)$. Indeed, the basis of the differentials in $\Omega^k(X)$, which the fermionic variables represent, is given by the $C^k_D$ ordered combinations of the differentials: $dx^{i_1}\wedge ... \wedge dx^{i_k}, i_1<...<i_k$. Thus, the trace of the fermionic variables over $\Omega^k(X)$ is
\begin{eqnarray}
\text{Tr}^{ferm}_{\Omega^{k}(X)} M^*_{t't} = \sum_{i_1<...<i_k}\hat {\imath}_{i_k}...\hat {\imath}_{i_1}d(M_{t't}(x))^{i_1}\wedge ... \wedge d(M_{t't}(x))^{i_k} \nonumber\\
= \sum_{i_1<...<i_k}\frac{\partial}{\partial \chi^{i_k}}...\frac{\partial}{\partial \chi^{i_1}} TM_{t't}(x)^{i_1}_{\tilde i_1}\chi^{\tilde i_1} ... \wedge TM_{t't}(x)^{i_k}_{\tilde i_k}\chi^{\tilde i_k} = m_k(x),
\end{eqnarray}
\end{widetext}
where $d(M_{tt'}(x))$ is defined in Equation (\ref{DefinitionOfTangentMap}). One arrives now at
\begin{eqnarray}
W = \sum\nolimits_{k=0}^D (-1)^k \text{Tr}_{\Omega^k(X)}\langle M^*_{t't}\rangle_\text{Ns},
\end{eqnarray}
which is yet another version of Equation (\ref{WittenIndex}).

The Witten index in Equation (\ref{WittenIndex2}) can be given as
\begin{subequations}
\label{WittenIndex3}
\begin{eqnarray}
W = \left\langle I_L \right\rangle_{\text{Ns}},
\end{eqnarray}
where
\begin{eqnarray}
I_{L} = \sum_{x=M_{tt'}(x)} \text{sign }\text{det} (\hat 1_{TX} - TM_{t't}(x))\label{Lefschetz1}
\end{eqnarray}
\end{subequations}
is known as the Lefschetz index of the map $M_{t't}$. The Lefschetz--Hopf theorem states that, under some general conditions, 
\begin{eqnarray}
I_L = \sum\nolimits_{k=0}^D (-1)^D \text{Tr}_{H^k(X)} M^*_{t't},
\end{eqnarray}
where the trace is over the de Rham cohomology $H^k(X)$. In the limit $t'\to t$, when $M_{tt'}\to\text{Id}_X$, the Lefschetz index reduces to the signed sum of the Betti numbers, \emph{i.e.}, to the Euler characteristics of $X$. On~the other hand, it was previously established that $W$ is independent of $t$. This leads to the conclusion that the Witten index equals the Euler characteristic of $X$ for any duration of temporal~evolution.

\begin{figure}[h]
\centerline{\includegraphics[height=3cm, width=6cm]{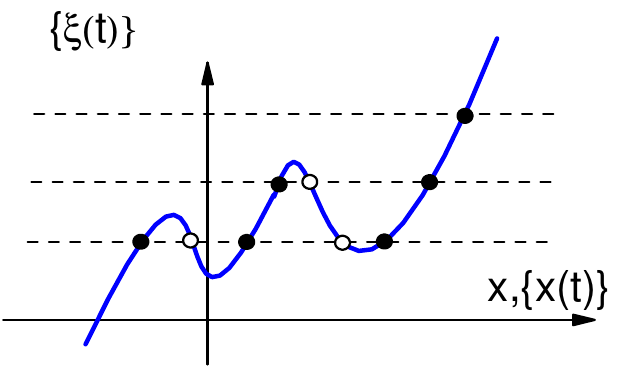}}
\caption{\label{Figure_3_3} Schematic representation of the topological character of the Witten index. The vertical axis represents noise configurations. The horizontal axis represents the phase space $X$ for the Lefschetz index interpretation of the Witten index in Section \ref{SecFermionVar} or the space of all the closed paths in $X$ for the Mathai--Quillen interpretation in Section \ref{SecPathIntegralGTO}. As one gradually varies the noise configuration, the fixed points of the SDE-defined diffeomorphism appear and disappear in pairs with positive (filled~dots) and negative (hollow dots) determinants. As a result, the sum of the signs of the determinants is independent of the noise configuration.}
\end{figure}

The topological character of $W$ can be qualitatively understood in the following manner (see~Figure~\ref{Figure_3_3}). For each noise configuration, there may exist many periodic solutions of the SDE, \emph{i.e.}, fixed points of $M_{t't}$. As one gradually varies the noise configuration, periodic solutions appear and disappear in pairs with the positive and negative determinants of the matrix $\hat 1_{TX}- TM_{t't}$. The constant $I_L$, however, remains the same. Up to this constant, $W$ represents the (normalized) partition function of the stochastic noise.

The interpretation of the Witten index as the partition function of the noise is important for the following reason. The noise partition function is a very fundamental object of a stochastic model, and it must certainly have its representative in the theory. Clearly, this representative exists only if one views the differential forms of all degrees as the Hilbert space and not simply the TPD (top differential forms), as in the conventional approach to SDEs. Without viewing the differential forms of all degrees as the rightful wavefunctions of the model, the partition function of the noise would not have its representative in the theory, and such a situation is clearly somewhat suspicious because, as previously mentioned, the partition function of the noise is one of the fundamental objects of the model, \emph{i.e.}, the object that appears at the level of the very formulation of stochastic dynamics.

\subsection{Dynamical Partition Function}
\label{SecPartFunc}

Yet another fundamental object is the \emph{dynamical} partition function (DPF) and/or the ''counting trace' of the finite-time SEO:\index{Partition function}
\begin{eqnarray}
Z_{tt'} &=& \text{Tr} \hat {\mathcal M}_{tt'} = \text{Tr} e^{-\hat H(t-t')} = \sum\nolimits_{n} e^{-\mathcal{E}_n(t'-t)}.\label{PartitionFunction}
\end{eqnarray}

By analogy with Equation (\ref{WittenIndex2}), the DPF is the trace of the finite-time SEO with periodic/anti-periodic boundary conditions for the bosonic/fermionic variables:
\begin{eqnarray}
Z_{tt'} &=& \int d^D xd^D \chi \hat{\mathcal{M}}_{tt'}(x(-\chi),x\chi) \nonumber \\ &=& \left\langle \sum_{x=M_{t't}(x)} \frac{\text{det} (\hat 1_{TX} + TM_{t't}(x))}{|\text{det}(\hat 1_{TX} - TM_{t't}(x))|} \right\rangle_\text{Ns} \nonumber \\
&=& \sum\nolimits_{k=0}^D \left\langle \sum_{x=M_{t't}(x)} \frac{m_k(x)}{|\text{det}(\hat 1_{TX} - TM_{t't}(x))|} \right\rangle_\text{Ns}\nonumber \\ &=& \text{Tr} \langle M^*_{t't}\rangle_\text{Ns},
\end{eqnarray}
where the $m$ are defined in Equation (\ref{ms}).

The physical meaning of the DPF is observed in the limit of the infinitely long temporal evolution. Consider models in which the absolute values of the eigenvalues of the tangent map \linebreak $Spec(TM_{t't})=(\mu_1(t,t'),...,\mu_D(t,t'))$, in the long-time limit $t-t'\to\infty$, are such that $|\mu_i(t,t')|\approx e^{\lambda_i (t'-t)}$, with $\lambda$ being the stochastic versions of the (global) Lyapunov exponents. The class of models that satisfy this condition must certainly exist in the deterministic limit as follows from the classical DS~theory. 

In the assumption that none of the $\lambda$ vanish, one has, in the limit of $t-t'\to\infty$,
\begin{eqnarray*}
Z_{tt'} &=& \left\langle \sum_{x=M_{t't}(x)} \frac{\text{det} (\hat 1_{TX} + TM_{t't}(x))}{|\text{det}(\hat 1_{TX} - TM_{t't}(x))|} \right\rangle_\text{Ns} \\&=& 
\left\langle \sum_{x=M_{t't}(x)} \frac{\prod_{i=1}^D(1 + \mu_i(t,t'))}{|\prod_{i=1}^D(1 - \mu_i(t,t'))|} \right\rangle_\text{Ns}
\\&\approx& \left\langle \sum_{x=M_{t't}(x)} \frac{\prod_{i, \lambda_i<0}\mu_i(t,t')}{|\prod_{i, \lambda_i<0}\mu_i(t,t')|} \right\rangle_\text{Ns}\\
&\le& \left\langle \sum_{x=M_{t't}(x)} \frac{|\prod_{i, \lambda_i<0} \mu_i(t,t')|}{|\prod_{i, \lambda_i<0} \mu_i(t,t')|} \right\rangle_\text{Ns} =\left\langle \sum_{x=M_{t't}(x)} 1 \right\rangle_\text{Ns} \\&=& \left\langle \text{\# of fixed points of }M_{t't} \right\rangle_\text{Ns}.\label{PartFunc1}
\end{eqnarray*}

In other words, in this class of models, the DPF grows slower than the stochastically averaged number of fixed points of the SDE-induced diffeomorphisms or, equivalently, of the number of periodic solutions of the SDE. 

At this point, it must be stressed that, in this and previous subsections, the summation over the fixed points of the SDE-defined diffeomorphisms (see, e.g., Equation (\ref{WittenIndex2})) has been used as if these fixed points were isolated in $X$. This is not true in general. The fixed points of the diffeomorphisms may appear in submanifolds of $X$, with the Morse--Bott flow vector fields being one example of this situation. How to count the fixed points in this general situation using the standard methodology of DS theory used in this subsection is not clear. This problem, however, does not exist in the operator representation of the theory considered previously as well as in its path integral representation in the next section.

It is also important to discuss the fundamental difference between the \emph{dynamical} partition function of the STS and the \emph{thermodynamic} partition function in statistical (quantum) physics. The latter is defined as $Tr e^{-\beta \hat H_q}$, where $\hat H_q$ is a Hermitian Hamiltonian of a quantum model and $\beta$ is the inverse temperature. Equation (\ref{PartitionFunction}) has a very similar appearance. Furthermore, in the literature on, e.g., $N=2$ supersymmetric quantum mechanics, it is often said Equation (\ref{PartitionFunction}) is the result of the Wick rotation of the ``real'' time of evolution, \emph{i.e.}, $t\to i\times t \sim \beta$. This is not so from the point of view of the STS. The~time $t$ in Equation (\ref{PartitionFunction}) is the original time of the stochastic evolution and not that of the Schr\"odinger evolution. This explains the absence of the imaginary unity in the exponent. The direct quantum analogue of Equation (\ref{PartitionFunction}) is the generating functional $Tr e^{-i t \hat H_q}$.

\subsection{Topological Supersymmetry Breaking, Chaos, and Dynamical Entropy}
\label{ChaosTopBreaking}
\index{Chaos} 

In models with the type of SEO spectra given in Figure \ref{Figure_3_2}b, the DPF grows exponentially in the long-time limit:
\begin{eqnarray}
\left.Z_{t0}\right|_{t\to\infty} \approx 2 e^{|\text{Re}{\mathcal E}_g|t},\label{ExponentialGrowth}
\end{eqnarray}
where the factor of $2$ comes from the $\hat d$-degeneracy of the non-$\hat d$-symmetric ground state and ${\mathcal E}_g$ is the ground state's eigenvalue, \emph{i.e.}, the eigenvalue with the smallest and negative real part. For the type of spectra in Figure \ref{Figure_3_2}c, one has
\begin{eqnarray}
\left.Z_{t0}\right|_{t\to\infty} \approx 4 \cos ( \text{Im}{\mathcal E}_g t) e^{|\text{Re}{\mathcal E}_g|t},
\end{eqnarray}
where the pair of the Ruelle--Pollicott resonances with the least real part of their eigenvalues provides the dominant contribution in the long time limit.

It can be recalled that, in deterministic chaotic DSs, the number of periodic solutions grows exponentially with time in the long time limit. The rate of this exponential growth is related to the concept of dynamical \emph{entropy} (see, e.g.,  \cite{Teschl}). This exponential growth is provided by the infinite number of unstable periodic orbits with arbitrary large periods, which constitute strange or fractal attractors \cite{Gil98}. This exponential growth is basically the reason why chaotic dynamics is sometimes identified as complex dynamics. This term is borrowed from information theory. There,~a~problem is identified as complex if the number of elementary operations needed to obtain its solution grows exponentially with the ''size'' of the problem.

As demonstrated in Section \ref{SecPartFunc}, for a wide class of models, the number of periodic solutions grows faster than the DPF. For spectra given in Figure \ref{Figure_3_2}b,c, the stochastically averaged number of periodic solution grows at least exponentially and the supersymmetry is spontaneously broken because the ground states are non-$\hat d$-symmetric as they have non-zero eigenvalues. Thus, one concludes that the stochastic generalization of the concept of deterministic chaos is the spontaneous breakdown of the topological supersymmetry. Even more convincing evidence that this is indeed so will be provided in Section \ref{Butterfly}, where it will be shown that the spontaneous breakdown of topological supersymmetry must always be accompanied by the emergence of the long-term memory of perturbations that must be associated with the famous butterfly effect.

In DS theory, there exists the so-called Shub conjecture (see, e.g., \cite{BookEntropy} and the references therein), stating that, for a sufficiently smooth map $M:X\to X$, the spectral radius of $M_*: H_*(X)\to H_*(X)$, where $H_*$ denotes the homology group, provides a lower bound for the topological entropy, \emph{i.e.}, the central measure of chaos (see, e.g., \cite{TopEntropy1,TopEntropy2} and the references therein). The spectral radius of the finite-time stochastic evolution operator $\langle M^*_{0t}\rangle_\text{Ns} = e^{-t\hat H}$ is (up to a sign) the real part of its ground-state eigenvalue, which can therefore be recognized as the stochastic generalization of the lower bound for the topological entropy in the Shub conjecture. Thus, if the real part of the ground-state eigenvalue is negative, the topological entropy is positive, and the model must be identified as chaotic.   

As a conclusion of this section, it can be stressed that the stochastic chaos is the opposite or rather complementary concept of that of the thermodynamic equilibrium, \emph{i.e.}, to the situation with unbroken supersymmetry when the supersymmetric state of the thermodynamic equilibrium discussed in Section \ref{SubSec:TermEquil} is among the ground states of the model, as in Figure \ref{Figure_3_2}a. Indeed, the supersymmetry can be spontaneously broken or unbroken but not both at the same time.

%
%
%

\section{Path Integral Representation}
\label{Chap:Pathintegrals}

\begin{figure}[htb]
\centerline{\includegraphics[width=0.9\linewidth]{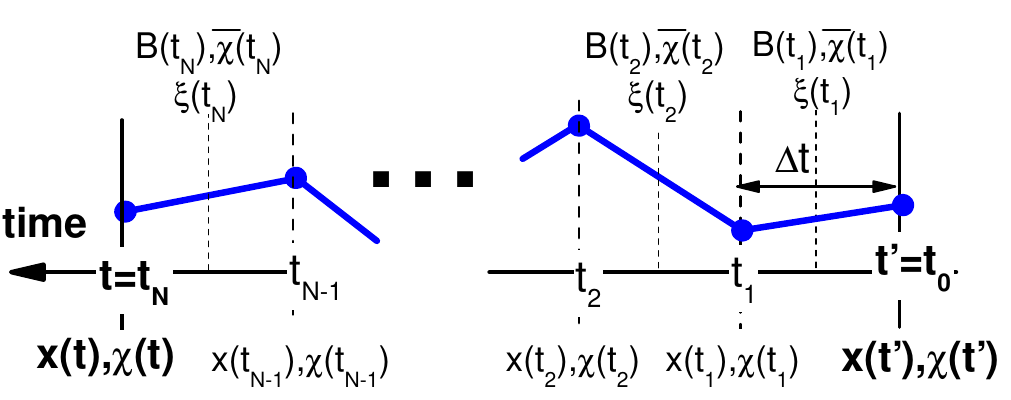}}
\caption{\label{Figure_4_1} Path integral representation of the finite-time stochastic evolution operator (SEO). Each time slice $t_n$ hosts a boson-fermion pair of variables $x(t_n)\in X$ and $\chi(t_n)\in TX_x(t_n)$. In between the time slices, there are pairs of Lagrange multipliers and fermionic momenta from the cotangent space $B(t_n),\bar\chi(t_n)\in TX^*_{x(t_n)}$ or $TX^*_{x(t_{n-1})}$ (in the continuous time limit, this choice makes no difference), as well as the noise variables $\xi(t_{n})$. The time flows from right to left, as explained at the end of Section \ref{SecLieDerivative}. The finite-time SEO is obtained by integrating out all the variables except $x(t),\chi(t)$ and $x(t'),\chi(t')$. Further integration over $x(t), \chi(t)$ with the periodic boundary conditions $x(t)=x(t'), \chi(t)=\chi(t')$ results in the Witten index $W$, whereas the integration with the anti-periodic boundary conditions for the fermionic variables $x(t)=x(t'), \chi(t)=-\chi(t')$ results in the dynamical partition function $Z$.}
\end{figure}

Path integrals are a powerful analytical tool that can greatly simplify various tasks that otherwise would be tedious. They represent a component of the mathematical foundation of quantum theory and have also been used in the studies of stochastic dynamics (see, e.g., \cite{PathStochastics} and the references therein). In the case of the STS, the path integral representation of the theory addressed in this section (see Figure \ref{Figure_4_1}) allows particularly for the generalization of the theory to models with noise of any form, not simply Gaussian white noise.

\begin{widetext}

\subsection{Finite-Time Stochastic Evolution Operator}
\label{SecPathIntegralGTO}

In the discrete-time picture introduced in Section \ref{SecGenFPOper}, the domain of the temporal evolution is split into $N\gg1$ segments with boundaries at $t_n = t' + n \Delta t$, $\Delta t = (t-t')/N$, $t_N\equiv t$ and $t_0\equiv t'$. The finite-time SEO can be given as the stochastically averaged composition of pullbacks at each time segment:
\begin{eqnarray}
\hat {\mathcal{M}}_{tt'} =  \langle M^*_{t't}\rangle_\text{Ns} = \langle M^*_{t_{N-1}t_N} M^*_{t_{N-2}t_{N-1}} \dots M^*_{t_{0}t_1} \rangle_\text{Ns},\label{GTORepresentation1}
\end{eqnarray}
where the composition law for the pullbacks from Equation (\ref{CompositionPullback}) has been used multiple times. 

At each time moment, one can now introduce a copy of the phase space and a pair of bosonic and fermionic variables $x(t_n),\chi(t_n)$ so that Equation (\ref{GTORepresentation1}) can be given as
\begin{eqnarray}
\hat {\mathcal{M}}_{tt'}(x(t)\chi(t),x(t')\chi(t')) = \bigg\langle M^*_{t_{N-1}t}(x(t)\chi(t),x(t_{N-1})
\chi(t_{N-1}))\times \nonumber
\\
\times \prod_{n=1}^{N-1}d^Dx(t_{n})d^D\chi(t_{n})
M^*_{t_{n-1}t_n}(x(t_{n})\chi(t_{n}),x(t_{n-1})\chi(t_{n-1}))
\bigg\rangle_\text{Ns},
\end{eqnarray}
where the pullbacks on the R.H.S. are combinations of the bosonic and fermionic $\delta$-functions, as in Equation (\ref{PullBackRepresentation}).

To exponentiate the bosonic $\delta$-function, one introduces an additional bosonic variable called the Lagrange multiplier or bosonic momentum from the cotangent space of X $B(t_n)\in TX^*_{x(t_n)}$:
\begin{subequations}
\label{Deltas}
\begin{eqnarray*}
\delta^D(x(t_{n-1})-M_{t_{n-1}t_{n}}(x(t_{n}))) = \int \frac{d^D B(t_n)}{(2\pi)^D} e^{-iB_i(t_n)(x(t_{n-1}) - M_{t_{n-1}t_{n}}(x(t_{n})))^i}.
\end{eqnarray*}

In a similar manner, one can exponentiate the fermionic $\delta$-function using Identity (\ref{ExpFermi}) and introducing the fermionic momentum from the cotangent space $\bar \chi\in TX^*_x$:
\begin{eqnarray*}
\delta^D(\chi(t_{n-1})-\hat M_{t_{n-1}t_{n}}\chi(t_{n})) = \int d^D(i\bar\chi(t_n)) e^{i\bar\chi_i(t_n)(\chi(t_{n-1}) - \hat M_{t_{n-1}t_{n}}(x(t_n)) \chi(t_{n}))^i}.
\end{eqnarray*}
\end{subequations}

Here, the imaginary unity is needed to bring the model to the form conventional in the literature on cohomological field theories.

In the continuous-time limit, \emph{i.e.}, $N\to\infty,\Delta t\to0$, one has
\begin{subequations}
\label{ApproxFlow}
\begin{eqnarray}
M_{t_{n-1} t_{n}}(x(t_{n}))^i &\approx& x^i(t_{n}) - \Delta t {\mathcal F}^i(x(t_n)),\\
(TM_{t_{n-1}t_{n}} (x(t_n)) \chi(t_{n}))^i &\approx& \chi^i(t_{n}) - \Delta t T{{\mathcal F}}^i_j (x(t_n)) \chi^j(t_n),
\end{eqnarray}
\end{subequations}
where $T{{\mathcal F}}(x)$ is introduced in Equation (\ref{DifferentialMatrix}). Combining the above representation of the $\delta$-functions and Equation (\ref{ApproxFlow}), one arrives at
\begin{eqnarray}
\hat {\mathcal{M}}_{tt'}(x(t)\chi(t),x(t')\chi(t')) = \langle \iint
D'\Phi e^{\tilde S(x(t)\chi(t)\dots x(t')\chi(t'))} \rangle_\text{Ns}.\label{PathIntegralGTO}
\end{eqnarray}

Here,
\begin{eqnarray}
\tilde S(\Phi) &=& \lim_{N\to\infty}i \sum\nolimits_{n=1}^{N} \Delta t \left( B_i(t_{n}) \left(\frac{x(t_n)-x(t_{n-1})}{\Delta t} - {\mathcal F} (x(t_n))\right )^i
- \bar\chi_i(t_{n}) \left(\frac{\chi(t_n)-\chi(t_{n-1})}{\Delta t} - T{{\mathcal F}} (x(t_n))\chi(t_n)\right )^i\right)\nonumber\\
&=& i\int_{t'}^{t} d\tau \left(B_i(\tau)(\dot x(\tau) - {\mathcal F}(x(\tau)))^i 
-\bar\chi_i(\tau)(\dot \chi(\tau) - T{{\mathcal F}}(x(\tau)) \chi(\tau))^i\right)
\end{eqnarray}
is the action of the model, $\Phi=(x,\chi,B,\bar\chi)$ denotes the collection of the original and additional fields, and the dots in Equation (\ref{PathIntegralGTO}) denote all the intermediate variables over which the path integration occurs with the differential
\begin{eqnarray}
D'\Phi = \lim_{N\to\infty}\frac{d^D B(t_N)}{(2\pi)^D}d(i\bar\chi^D(t_N))
\prod\nolimits_{n=1}^{N-1}d^{4D}\Phi(t_n),
\label{IncompletePathDifferential}
\end{eqnarray}
where
\begin{eqnarray}
d^{4D}\Phi(t_n) = d^Dx(t_n)d^D\chi(t_n)
\frac{d^DB(t_n)}{(2\pi)^D}d^D(i\bar\chi(t_n)).
\end{eqnarray}

The action can be expressed in the so-called $\mathcal Q$-exact form
\begin{eqnarray}
\tilde S(\Phi) &=& \{ \mathcal{Q},  \tilde \Psi (\Phi)\}, \label{QexactAction}
\end{eqnarray}
where
\begin{eqnarray}
\tilde \Psi(\Phi) &=&
\lim_{N\to\infty}\sum\nolimits_{n=1}^{N} \Delta t \left(i \bar\chi_i (t_{n}) \left(\frac{x(t_n)-x(t_{n-1})}{\Delta t} - {\mathcal F} (x(t_n))\right )^i\right),
=i\int_{t'}^{t} d\tau \bar\chi_i(\tau)(\dot x(\tau) - {\mathcal F}(x(\tau)))^i,\label{GaugeFermion}
\end{eqnarray}
is known as the gauge fermion and the curly brackets denote the operator of the topological~supersymmetry:
\begin{eqnarray}
\{\mathcal{Q},\tilde\Psi\} &=& \lim_{N\to\infty}\left(\sum\nolimits_{n=0}^{N} \chi^i(t_n)\frac{\partial}{\partial x^i(t_n)}  + \sum\nolimits_{n=1}^{N} B_i(t_n)\frac{\partial}{\partial \bar\chi_i(t_n)}\right)\Psi
=\int_{t'}^{t} d\tau \left(\chi^i(\tau)\frac{\delta}{\delta x^i(\tau)} + B_i(\tau)\frac{\delta}{\delta \bar\chi_i(\tau)}\right)\Psi.\label{QOperator}
\end{eqnarray}
\end{widetext}
This operator is the path integral version of the exterior derivative. In particular, it has similar properties: it is nilpotent, \emph{i.e.}, $\{ \mathcal{Q},  \{ \mathcal{Q},  X(\Phi) \} \} = 0$ for any $X(\Phi)$, and it is a bi-graded differentiation,~\emph{i.e.},
\begin{eqnarray}
\{ \mathcal{Q},  X Y \}  = \{ \mathcal{Q},  X \} Y + (-1)^{deg(X)} X \{ \mathcal{Q},  Y \}, \label{BeGradedPath}
\end{eqnarray}
where $X$ and $Y$ are some functionals of $\Phi$, and $deg(X)$ is the degree of $X$ defined as the difference between the numbers of $\chi$s and $\bar\chi$s in $X$. Equation (\ref{BeGradedPath}) is the path integral version of Equation (\ref{BiGraded}).



To perform the stochastic averaging, one can first separate the noise term in the action as
\begin{eqnarray}
\tilde S(\Phi) &=& S_0(\Phi) + \int_{t'}^{t} d\tau y_a(\tau) \xi^a(\tau),\text{ } S_0(\Phi)\nonumber \\ &=& \{\mathcal{Q}, \Psi_0(\Phi)\},
\end{eqnarray}
with
\begin{eqnarray}
\Psi_0(\Phi) = i \int_{t'}^{t} d\tau \bar\chi_i(\tau) (\dot x(\tau) - F(x(\tau)))^i,\label{PsiZero}
\end{eqnarray}
and
\begin{eqnarray}
y_a(\tau) = \{ \mathcal{Q}, -i(2\Theta)^{1/2}\bar\chi_i(\tau) e^i_a(x(\tau)) \}.\label{ys}
\end{eqnarray}

As a next step, one can integrate out the noise field, which is no longer assumed to be Gaussian white. The noise, however, remains assumed physical so that the path integration over all the noise configurations remains well defined. 

Integrating out the noise transforms  Equation (\ref{PathIntegralGTO}) into the following {form}:
\begin{subequations}
\label{PathAfterStoch}
\begin{eqnarray}
\hat {\mathcal{M}}_{tt'} = \iint D'\Phi e^{S(\Phi)}, \label{PathIntegralGTO1}
\end{eqnarray}
where the arguments of the finite-time SEO are dropped for brevity and the new action
\begin{eqnarray}
S(\Phi) &=& \text{log}\langle e^{\tilde S(\Phi)} \rangle_\text{Ns} = S_0(\Phi) + \text{log}\langle e^{\int_{t'}^{t} d\tau y_a(\tau)\xi^a(\tau)} \rangle_\text{Ns} \nonumber\\
&=& S_0(\Phi) + \sum\nolimits_{k=1}^{\infty} \frac{1}{k!}\int \left(\prod_{i=1}^kd\tau_i y_{a_i}(\tau_i)\right) c_{(k)}^{a_1...a_k}(\tau_1...\tau_k),\nonumber
\end{eqnarray}
with $c_{(k)}$ being the irreducible correlators of the noise. The zeroth-order term in the Taylor series vanishes because the partition function of the noise is assumed normalized, \emph{i.e.}, for $y=0$, one has $\text{log}\langle e^0 \rangle_\text{Ns} = \text{log}1= 0$.

Now, using the property of the nilpotency of $\mathcal{Q}$, the differentiation rule in Equation (\ref{BeGradedPath}) and the fact that all $y$ in Equation (\ref{ys}) are $\mathcal Q$-exact, one {arrives at}
\begin{eqnarray}
S(\Phi) = \{ \mathcal{Q}, \Psi(\Phi) \},\label{ActionGeneral}
\end{eqnarray}
where the new gauge fermion
\begin{eqnarray}
\Psi(\Phi) = \sum\nolimits_{k=0}^\infty\Psi_{k}(\Phi),
\end{eqnarray}
with $\Psi_0$ defined in Equation (\ref{PsiZero}) and
\begin{eqnarray}
\Psi_{k}(\Phi) &=& - i \sum_{k=1}^\infty \frac{(2\Theta)^{\frac12}}{k!}\int (\prod_{i=1}^kd\tau_i) \bar\chi_i(\tau_1)e^i_{a_1}(x(\tau_1)) \nonumber\\ && \times y_{a_2}(\tau_2) ... y_{a_k}(\tau_k)c_{(k)}^{a_1...a_k}(\tau_1...\tau_k)\label{PsiGeneral}
\end{eqnarray}
\end{subequations}
for $k\ge1$. In other words, even after stochastic averaging, the action is $\mathcal Q$-exact.

\subsection{Interpretations of Stochastic Quantization}

A $\mathcal Q$-exact action as in Equation (\ref{PathAfterStoch}) is a definitive feature of cohomological field theories~\cite{Birmingham1991129, labastida1989, Witten98, Witten981, Frenkel2007215}. Their standard path integral representations include periodic boundary conditions (PBC) for the fermionic variables. This is related to the fact that the PBC are consistent with the $\mathcal Q$-operator:
\begin{eqnarray}
&\{ \mathcal{Q},  \Psi(\Phi) \}_{PBC}  \equiv \left.\{ \mathcal{Q},  \Psi(\Phi) \}\right|_{x(t)=x(t'), \chi(t)=\chi(t')} \nonumber \\ &= \{ \mathcal{Q}, \Psi(\Phi)_{PBC}\}.
\end{eqnarray}

This equality and Equations (\ref{WittenIndex2}) and\index{Witten index} (\ref{PathAfterStoch}) lead to
\begin{eqnarray}
W  = \iint_{PBC} D\Phi e^{\{\mathcal{Q}, \Psi(\Phi)\}} = \iint D\Phi e^{\{\mathcal{Q}, \Psi(\Phi)_{PBC}\}},\label{WittenIndex4}
\end{eqnarray}
where the path integration is over
\begin{eqnarray}
D\Phi = d^Dx(t)d^D\chi(t) D'\Phi,
\end{eqnarray}
with $D'\Phi$ defined in Equation (\ref{IncompletePathDifferential}).

The integrand in Equation (\ref{WittenIndex4}), \emph{i.e.}, $e^{\{{\mathcal Q}, \Psi\}}$, belongs to a class of mathematical objects known as Mathai--Quillen forms. Its integral is of a topological character that can be clarified by rewriting Equation (\ref{WittenIndex4}) as
\begin{widetext}
\begin{eqnarray}
W &=& \left\langle \iint_{PBC} D\Phi e^{\tilde S(\Phi)}\right\rangle_\text{Ns} 
=\left\langle\iint_\text{closed paths} Dx \left(\prod_{\tau}\delta(\dot x(\tau) - {\mathcal F}(x(\tau)))\right) \text{Det}(\partial_\tau - T{{\mathcal F}}(x(\tau)))\right\rangle_\text{Ns}\nonumber\\
&=&\left\langle \sum_\text{closed solutions of SDE} \text{sign Det}(\partial_\tau - T{{\mathcal F}}(x(\tau)))\right\rangle_\text{Ns}. \label{ParisiSourlas}
\end{eqnarray}
\end{widetext}

Here, the functional $\delta$-function in the second line, which emerges from integrating out the field $B$, limits the path integration only to the periodic solutions of the SDE with a fixed noise configuration. The functional determinant of the infinite-dimensional matrix of the functional derivatives of the SDE appears from integrating out the fermionic fields $\chi,\bar\chi$. The third line is the path integral analogue of the Lefschetz index in Equation (\ref{WittenIndex3}).

Equation (\ref{ParisiSourlas}) has the meaning of the infinite-dimensional generalization of the Poincar\'e--Hopf theorem. The later states that, under certain and general conditions and for a vector field with isolated critical points, the sum of the indices of the critical points, \emph{i.e.}, the signs of the determinant of the matrix of the derivatives of the vector field, equals the Euler characteristic of the manifold. In the case of Equation (\ref{ParisiSourlas}), the objects in the Poincar\'e--Hopf theorem are recognized as follows. The manifold is the space of all the closed paths (periodic boundary conditions). The vector field is the SDE with a~fixed noise configuration. The critical points are the periodic solutions of this SDE. 

The constant resulting from the summation in the last line of Equation (\ref{ParisiSourlas}) is independent of the configuration of the noise. In other words, as one varies the noise configuration, the periodic solutions of the SDE appear/disappear in pairs with the opposite signs of their determinants. This~situation can be illustrated graphically similarly to the interpretation of $W$ in the previous section as the stochastically averaged Lefschetz index (see Figure \ref{Figure_3_2}). 

Equation (\ref{ParisiSourlas}) is the functional of the Parisi--Sourlas \index{Parisi--Sourlas quantization} stochastic quantization \cite{ParSour} generalized to SDEs of any form. One way to look at the stochastic quantization is as at the gauge-fixing procedure. In other words, one starts with an empty theory, or rather with a theory with the trivial action $S_{cl}=0$, and then ``fixes the gauge'' by adding a $\mathcal Q$-exact piece to it. This reduces the integration over all possible closed paths only to the closed solutions of the SDE. From this point of view, the fermionic fields of the model must be recognized as the Fadeev--Popov ghosts, \index{Fadeev--Popov ghosts} the $\mathcal Q$-operator as the Becchi--Rouet--Stora--Tyutin (BRST) (super-)symmetry, the closed solutions of the SDE that contribute to Equation (\ref{ParisiSourlas}) as the Gribov copies, and $\Psi$ as the gauge fermion, \emph{i.e.}, the term that has been introduced previously. Note also that, while the topological symmetry is equivalent to the BRST symmetry in stochastic quantization, this is not so in cohomological gauge field theories where $\mathcal Q$ is said to be a BRST-like symmetry. 

It is worth stressing that it is a typical mistake in the literature to treat the functional of the Parisi--Sourlas stochastic quantization (\ref{ParisiSourlas}), \emph{i.e.}, the Witten index, as the generating functional, with the help of which various expectation values and correlators can be calculated. The actual generating functional and/or the DPF corresponds to the anti-periodic boundary conditions (APBC) for the fermionic fields, as discussed in Section \ref{SecPartFunc}. These boundary conditions are not consistent with the~$\mathcal Q$-operator
\begin{eqnarray*}
&\{ \mathcal{Q}, \Psi(\Phi) \}_{APBC}  \equiv \left.\{ \mathcal{Q},  \Psi(\Phi) \}\right|_{x(t)=x(t'), \chi(t)=-\chi(t')}\nonumber \\ & \ne \{ \mathcal{Q}, \Psi(\Phi)_{APBC}\}.
\end{eqnarray*}

As a result, the topological character is lost for the DPF
\begin{eqnarray}
&Z_{t't} = \iint_{APBC} D\Phi e^{\left\{ \mathcal{Q},  \Psi(\Phi) \right\}} \nonumber\\ &\ne \iint D\Phi e^{\left\{ \mathcal{Q}, \Psi(\Phi)_{APBC}  \right\}}.\label{PartFunction3}
\end{eqnarray}

Note that this does not mean that the model described by the DPF no longer has the topological supersymmetry. The topological supersymmetry is not a property of the DPF but rather of the most fundamental object in the theory, the finite-time SEO. Equation \ref{PartFunction3} simply means that this particular object, \emph{i.e.}, the DPF, that we construct from the finite-time SEO is not of topological character. 

The fundamental difference between the DPF and the Witten index can be best revealed in the limit of the infinitely long temporal evolution. There (see Section \ref{GeneratFunctional} below), only the ground states contribute to the DPF, whereas only the $\hat d$-symmetric states contribute to $W$. Thus, the difference between $W$ and $Z$ is particularly pronounced under the conditions of the spontaneously broken topological supersymmetry when the ground states of the model are non-$\hat d$-symmetric.

\subsection{Generalization to Spatially Extended Models}
\label{SecContinuousSpace}

The class of models under consideration can be generalized further to spatially extended models. These models are defined by the following stochastic (partial) (integro-)differential equations:
\begin{eqnarray}
&\dot x(rt) = F(x,rt) + (2\Theta)^{1/2}e(x,rt)\xi(rt) \nonumber \\ &= {\mathcal F}(x,\xi,rt),\label{SDEGen}
\end{eqnarray}
where $r$ is the spatial coordinate of the ``base-space''. In the general case, the flow vector field and the veilbeins are temporarily and spatially non-local functionals of $x(rt)$ that may also have explicit dependences on the base-space coordinates $rt$. The relation between the spatially extended models defined by Equation (\ref{SDEGen}) and the previously discussed models with time being the only base-space coordinate is the same as the relation between quantum nonlinear sigma models (or field theories) and quantum mechanics. In~other words, Equation~(\ref{SDEGen}) is the infinite-dimensional version of Equation~(\ref{SDE}). The phase space now is the infinite-dimensional space of all possible configurations $x(r)$.

The  stochastic quantization procedure of Equation (\ref{SDEGen}) is along the same lines, the action is $\mathcal Q$-exact with the topological supersymmetry operator
\begin{eqnarray}
{\mathcal Q} = \int dr d\tau \left(\chi^i(r\tau)\frac\delta{\delta x^i(r\tau)}+B_i(r\tau)\frac\delta{\delta \bar\chi_i(r\tau)}\right),
\end{eqnarray}
and the gauge fermion before integrating away the noise variables is
\begin{eqnarray}
\tilde \Psi(\Phi) = i \int dr d\tau \bar\chi_i(r\tau)\left(\dot x(r\tau) - {\mathcal F}(x, \xi, r\tau)\right)^i.
\end{eqnarray}

After integrating out the noise, one arrives at a model with a $\mathcal Q$-exact action. 

\subsection{Weyl--Stratonovich Symmetrization and Martingale}
\label{SecStratonovichWeyl}

The story of stochastic quantization would not be complete without a discussion on how the path integral representation can be turned back into the operator representation. The exercise to be conducted in this subsection will reveal a close relation between the Ito--Stratonovich dilemma (see Appendix \ref{ItoStratonovichDilemma} and the end of Section \ref{SecGenFPOper}) and the Weyl symmetrization rule of quantum theory.

Models with Gaussian white noise are of interest herein. The gauge fermion can be acquired from {Equation} (\ref{PsiGeneral}), and recalling that the only irreducible correlator of the Gaussian white noise is the one given in Equation (\ref{GaussianAverage}),
\begin{eqnarray}
\Psi = \int d\tau \left(i \bar\chi_i(\tau)  \dot x^i(\tau)  - \bar d(\Phi(\tau))\right),
\end{eqnarray}
where
\begin{eqnarray}
\bar d(\Phi)  &=& i \bar\chi_iF^i(x) - \Theta i \bar\chi_i e^i_a(x) \{\mathcal{Q}, i\bar\chi_j e^j_a(x)\}\nonumber \\
&=& i \bar\chi_i \left(F^i(x) - \Theta e^i_a(x) \right. \times \nonumber \\&&\times \left.\left(e^j_a(x)(iB_j) + e^j_{a'l}(x)\chi^l (i\bar\chi_j)\right)\right) \label{bard}
\end{eqnarray}
is the path integral version of Operator (\ref{dbar}).

Accordingly, the action is
\begin{eqnarray}
S &=& \{ \mathcal{Q},\Psi\} = \int d\tau \left( iB_i(\tau)\dot x^i(\tau) - i \bar\chi_i(\tau)\dot \chi^i(\tau)\right. \nonumber \\ &&\left. - H(\Phi(\tau))\right),
\end{eqnarray}
where
\begin{eqnarray}
H(\Phi) &=& \{\mathcal{Q}, \bar d(\Phi) \}.\label{FPHamiltonFunction}
\end{eqnarray}

This function is the stochastic analogue of the Hamilton function in the path integral representation of quantum mechanics and   can thus be called the stochastic Hamilton function (SHF). The SHF (\ref{FPHamiltonFunction}) is the path integral version of the SEO (\ref{FPOp}). In particular,  Equations (\ref{CartanFormuola}), (\ref{FPOp}) and (\ref{FPHamiltonFunction}) reveal that $\mathcal Q$ is the path integral version of the commutator with the exterior derivative, whereas $i\bar\chi_i$ is that of the interior multiplication and/or the fermionic momentum operator in Equation (\ref{FermionicVariables}).

Consider now infinitesimal temporal evolution between  $t_{n-1}$ and $t_{n}=t_{n-1}+\Delta t$. In the continuous-time limit $\Delta t\to 0$, the path integral representation of the infinitesimal stochastic evolution~is
\begin{widetext}
\begin{eqnarray}
\psi(x\chi t_n) &=& \int \frac{d^DB}{(2\pi)^D}d^Di\bar\chi d^D y d^D \varphi e^{iB_i(x-y)^i - i \bar\chi_i(\chi-\varphi)^i - \Delta t H(B\bar\chi x_\alpha\chi_\alpha)}
\psi(y \varphi t).
\label{TediousFormula}
\end{eqnarray}
\end{widetext}

Here, for the sake of brevity, the notations are different from those in Figure \ref{Figure_4_1}. The relation with the notations in Figure \ref{Figure_4_1} is as follows: $x,\chi \equiv x(t_{n}),\chi(t_n)$, $B, \bar\chi \equiv B(t_n), \bar\chi(t_n)$ and $y, \varphi \equiv x(t_{n-1}),\chi(t_{n-1})$. Other notations introduced in Equation (\ref{TediousFormula}) are $x_\alpha = (1-\alpha) y + \alpha x$ and $\chi_\alpha = (1-\alpha) \varphi + \alpha \chi$, with the parameter $\alpha$ being from Appendix \ref{ItoStratonovichDilemma} and Figure \ref{Figure_3_1}. This parameter determines at which point of the elementary evolution the function $H$ is evaluated: $\alpha=0$ and $\alpha=1/2$ correspond to the Ito and Stratonovich choices of the very beginning and the mid-point, respectively. The reason why the different possible choices of $\alpha$ have not been discussed before in the context of the path integral representation of the STS is because, in obtaining the path integral representation, the choice of $\alpha$ makes no difference. It is only when going from the path integral representation back to the operator representation that different choices of $\alpha$ lead to different SEOs, as will be observed below. 

Using Equation (\ref{TediousFormula}), the infinitesimal evolution of the wavefunction can be given as
\begin{eqnarray}
\partial_t \psi(x\chi t)&=&\lim_{\Delta t\to 0}(\Delta t)^{-1}(\psi(x\chi t_1)-\psi(x\chi t)).
\end{eqnarray}

The Taylor expansion of the exponent in the R.H.S. of Equation (\ref{TediousFormula}) in $\Delta t$ results in
\begin{eqnarray}
\partial_t \psi(x\chi t)&=& -\int \frac{d^DB}{(2\pi)^D}d^D(i\bar\chi) d^D y d^D\varphi e^{iB_i (x-y)^i - i \bar\chi_i(\chi-\varphi)^i} \nonumber \\&&\times H(B\bar\chi x_\alpha\chi_\alpha) \psi(y\varphi t).\label{FPEqPathIntegral}
\end{eqnarray}

This representation of the stochastic evolution equation highlights the roles of the variables involved: the first integration over $y$ and $\varphi$ transforms the wavefunction into the Fourier space, where $B$ and $\bar\chi$ are diagonal, whereas the consequent integration over $B$ and $\bar\chi$ transforms the wavefunction back into the real space where $x$ and $\chi$ are diagonal. The straightforward conclusion from this observation is that, in the real space, $B$ and $\bar\chi$ are the operators of the bosonic and fermionic momenta
\begin{eqnarray}
i\hat B_i = \partial/\partial x^i, i\hat{\bar\chi}_i = \partial/\partial\chi^i.\label{SignAmbiguity}
\end{eqnarray}

The sign in front of the bosonic momentum is unambiguous because this operator acts on the ordinary commuting variables. In contrary, the sign in front of the fermionic momentum operator is ambiguous or rather can be established unambiguously only by a tedious exercise of carefully tracking all the signs associated with the relative positions of the anticommuting variables and their differentials. There is a simpler way, however, of accomplishing this task. One can demand that the bi-graded commutator of the exterior derivative in the operator representation act in the same way as the operator of the $\mathcal Q$-differentiation in the path integral representation. In other words, because
\begin{eqnarray}
\{\mathcal{Q}, (x^i,\chi^i,B_i,\hat {\bar\chi}_i)\} = (\chi^i,0,0,B_i),
\label{ActionByQ}
\end{eqnarray}
which is just another version of Equation (\ref{QOperator}),
\begin{eqnarray}
[\hat d, (\hat x^i,\hat \chi^i,\hat B_i, \hat {\bar\chi}_i)] = (\hat \chi^i,0,0,\hat B_i). \label{ActionByd}
\end{eqnarray}

It can be readily verified that the choice of sign in Equation (\ref{SignAmbiguity}) satisfies this requirement.

Another important observation from Equation (\ref{FPEqPathIntegral}) is related to the order of operators. Both the variables $x$ and $y$ (and $\chi$ and $\varphi$) in the path integral representation correspond to the operator $\hat x$ (and $\hat \chi$) in the operator representation of the theory. The difference between, say, $x$ and $y$ is that $y$ acts on the wavefunction before the momentum operators $B$, whereas $x$ acts after $B$. In particular, if $H$ included a~term $B_i x^j_\alpha$, this term in the operator representation would become
\begin{subequations}
\label{BiGradedSymmerizationRule}
\begin{eqnarray}
B_ix^j_\alpha \stackrel{B\to \hat B}{\longrightarrow} (1-\alpha) \hat B_i \hat x^j + \alpha \hat x^j \hat B_i.
\end{eqnarray}

The same can be said about fermionic operators, with the only correction being that the symmetrization must be bi-graded, \emph{i.e.},
\begin{eqnarray}
\bar \chi_i \chi^j_\alpha \stackrel{\bar\chi \to \hat {\bar\chi} }{\longrightarrow} (1-\alpha) \hat {\bar\chi}_i \hat \chi^j-\alpha \hat \chi^j \hat {\bar\chi}_i.
\end{eqnarray}
\end{subequations}

This reveals that if, in the Ito case ($\alpha=0$), all the momentum operators must act after all the position operators, in the Stratonovich case ($\alpha=1/2$), the operators must be symmetrized in the bi-graded manner. This symmetrization is the well-known Weyl quantization rule in quantum theory. There, it guarantees that any real Hamilton function in the path integral representation results in a~Hermitian Hamiltonian in the operator representation of the theory. If one could now establish that the Weyl--Stratonovich bi-graded symmetrization of the stochastic Hamilton Function (\ref{FPHamiltonFunction}) results in the SEO in Equation (\ref{FPOp}), it would mean that the Weyl--Stratonovich quantization is a correct choice because Equation (\ref{FPOp}) was obtained without approximations and outside the path integral formulation of the theory (see Section \ref{SecGenFPOper}).  

This is indeed true, as will be demonstrated next. To facilitate the procedure of establishing the expression for the SEO, one can utilize the ``commutativity'' of the operation of the bi-graded symmetrization and the substitution of the momenta fields by their corresponding operators:
\begin{eqnarray}
\left[\left.H(\Phi)\right|_{B\bar\chi\to\hat B \hat{\bar\chi}}\right]_{sym} = \left.\left[H(\Phi)\right]_{sym}\right|_{B\bar\chi\to\hat B \hat{\bar\chi}},\label{FirstFacil}
\end{eqnarray}
where the bi-graded symmetrization of $H(\Phi)$ follows the same rules described above, with the only difference being that the fields $\Phi$ are not operators; rather, they are c-numbers. Two other useful observations are that
\begin{eqnarray}
\left[\{{\mathcal Q}, X_1 X_2...\}\right]_{sym} = \left\{{\mathcal Q}, \left[ X_1X_2...\right]_{sym}\right\},
\end{eqnarray}
where $X$ are some arbitrary functions of $\Phi$, and
\begin{eqnarray}
\left. \{{\mathcal Q}, X \}\right|_{B\bar\chi\to\hat B \hat{\bar\chi}}
=\left[\hat d, \left. X \right|_{B\bar\chi\to\hat B \hat{\bar\chi}}\right],
\end{eqnarray}
as follows from Equations (\ref{ActionByQ}) and (\ref{ActionByd}). Using these properties and Equation (\ref{FPHamiltonFunction}), one obtains
\begin{eqnarray}
\hat H = \left[\left.H(\Phi)\right|_{B\bar\chi\to\hat B \hat{\bar\chi}}\right]_{sym}= \left[\hat d, \hat {\bar d} \right],\label{HamiltAdditional}
\end{eqnarray}
where
\begin{eqnarray}
\hat {\bar d} &=& \left[\left.\bar d(\Phi)\right|_{B\bar\chi\to\hat B \hat{\bar\chi}}\right]_{sym},\label{LastFacil}
\end{eqnarray}
with $\bar d(\Phi)$ given in Equation (\ref{bard}). 

One can now proceed straightforwardly starting with the expression,
\begin{eqnarray}
\hat {\bar d} &=& \left[\frac{\partial}{\partial \chi^i} 
\left(F^i(x) - \Theta e^i_a(x) \left(e^j_a(x)\frac{\partial}{\partial x^j} + e^j_{a'l}(x)\chi^l \frac{\partial}{\partial \chi^j}\right)\right) \right]_{sym}.\nonumber
\end{eqnarray}

The symmetrization in the first term related to the flow vector field is trivial because all the operators commute so that
\begin{eqnarray}
\hat {\bar d} &=& \frac{\partial}{\partial \chi^i} F^i(x) - \Theta \hat d^*,\label{dbar1}
\end{eqnarray}
where 
\begin{widetext}
\begin{eqnarray}
\hat d^* = \left[\frac{\partial}{\partial \chi^i}  e^i_a(x)\hat {\mathcal{L}}_{e_a} \right]_{sym} = \frac12\left(\frac\partial{\partial \chi^i}e^i_a(x) \hat {\mathcal{L}}_{e_a} + \hat{\mathcal{L}}_{e_a} \frac\partial{\partial \chi^i}e^i_a(x)\right),\label{dstar}
\end{eqnarray}
with
\begin{eqnarray}
\hat {\mathcal{L}}_{e_a} &=& \left[e^j_a(x)\frac{\partial}{\partial x^j} + e^j_{a'l}(x)\chi^l \frac{\partial}{\partial \chi^j}\right]_{sym} = \frac12
\left(
\left(
e^j_a(x)\frac{\partial}{\partial x^j} + \frac{\partial}{\partial x^j}e^j_a(x)
\right)
+ e^j_{a'l}(x)
\left(
\chi^l \frac{\partial}{\partial \chi^j} - \frac{\partial}{\partial \chi^j}\chi^l
\right)
\right)\nonumber\\
&=&  e^j_a(x)\frac{\partial}{\partial x^j} + e^j_{a'l}(x)\chi^l \frac{\partial}{\partial \chi^j} = \left[\hat d, e^i_a \frac\partial{\partial\chi^i}\right]
\end{eqnarray}
\end{widetext}
being the Lie derivative along $e_a$ (see Equation (\ref{CartanFormuola})). As can be straightforwardly verified, $\hat{\mathcal{L}}_{e_a}$ and $\frac\partial{\partial \chi^i}e^i_a(x)$ commute in Equation (\ref{dstar}) so that
\begin{eqnarray}
\hat d^* = \frac{\partial}{\partial \chi^i}  e^i_a(x)  \hat {\mathcal{L}}_{e_a}.\label{dbarstar}
\end{eqnarray}

This operator is the stochastic analogue of the codifferential $\hat d^\dagger$, through which the Hodge Laplacian is defined as $\hat \triangle_H = [\hat d, \hat d^\dagger]$. In the stochastic quantization case, the diffusion Laplacian is defined similarly as $\hat \triangle = [\hat d, \hat d^*]$. 

Using Equations (\ref{dbar1}) and (\ref{HamiltAdditional}), the Cartan Formula (\ref{CartanFormuola}), and the exterior differentiation Rule~(\ref{BiGraded}), one arrives at
\begin{eqnarray}
\hat H &=& \hat {\mathcal{L}}_{F} - \Theta \hat {\mathcal{L}}_{e_a}\hat {\mathcal{L}}_{e_a},
\end{eqnarray}
in agreement with Equation (\ref{FPOp}). This result shows that the Stratonovich approach to SDEs is equivalent to the bi-graded Weyl symmetrization rule.

The Ito interpretation of SDEs ($\alpha=0$) in turn corresponds to what is known as the martingale property. This property is often formalized by such formulas as $\langle f(y)(x-y)\rangle=0$, where the averaging is assumed over the noise variable $\xi_n$. From the point of view of the STS, expectation values such as the one above make no sense unless one specifies the bra and ket of the expectation value. Nevertheless,~the martingale property does find its realization in the STS through the unphysical rule for the operator ordering that all the momenta operators are on the left of all the position operators, as  seen from Equation (\ref{BiGradedSymmerizationRule}). In quantum mechanics, such an ordering rule will lead to a non-Hermitian Hamiltonian in the general case.

\begin{widetext}
Using this martingale operator ordering rule, one can now find the SEO for the Ito interpretation of SDEs. Again, to facilitate the derivation, one notes that Equation (\ref{FirstFacil}) are also correct for the martingale ordering. Thus, the Ito SEO is also $\hat d$-exact, \emph{i.e.},
\begin{eqnarray}
\hat H_{Ito} = [\hat d, \hat {\bar d}_{Ito}],\label{ItoSEO}
\end{eqnarray}
and the Ito version of Operator (\ref{LastFacil}) is
\begin{eqnarray}
\hat {\bar d}_{Ito} &=& \left[\left.\bar d(\Phi)\right|_{B\bar\chi\to\hat B \hat{\bar\chi}}\right]_{mart},
\end{eqnarray}
with the subscript ``mart'' denoting the martingale ordering. One can now proceed straightforwardly as follows:

\begin{eqnarray}
\hat {\bar d}_{Ito} &=& \left[\frac{\partial}{\partial \chi^i} 
\left(F^i(x) - \Theta e^i_a(x) \left(e^j_a(x)\frac{\partial}{\partial x^j} + e^j_{a'l}(x)\chi^l \frac{\partial}{\partial \chi^j}\right)\right) \right]_{mart}= \frac{\partial}{\partial \chi^i} 
\left(F^i(x) - \Theta \frac{\partial}{\partial x^j} e^i_a(x) e^j_a(x) + \Theta e^j_{a'l}(x)\frac{\partial}{\partial \chi^j}\chi^l \right)\nonumber\\
&=& \frac{\partial}{\partial \chi^i} 
\left(F^i(x) - \Theta (e^i_a(x))_{'j} e^j_a(x) - \Theta e^i_a(x) \left(e^j_a(x) \frac{\partial}{\partial x^j} + e^j_{a'l}(x)\chi^l\frac{\partial}{\partial \chi^j} \right)\right)= \frac{\partial}{\partial \chi^i} 
\left(F^i_0(x) - \Theta e^i_a(x) \hat{\mathcal L}_{e_a}\right),\nonumber
\end{eqnarray}
\end{widetext}
where $F_0$ is the shifted flow vector field defined in Equation (\ref{Falpha}). 

The previous equation and Equations (\ref{dbar1}) and (\ref{dbarstar}) clearly show that the only difference between the Weyl--Stratonovich SEO (\ref{HamiltAdditional}) and the Ito SEO (\ref{ItoSEO}) is the shifted flow vector field. This finding is in accordance with the discussion in Section \ref{ItoStratSubSection}, which stated that different interpretations of SDEs can be transformed between each other by a shift of $F$. The importance of this result is that it was obtained for the entire SEO and not only for the FP operator acting on only top differential forms, as  done in Appendix \ref{ItoStratonovichDilemma}.

\subsection{Generating Functional and Correlators}
\label{GeneratFunctional}

Various correlators and expectation values in the theory can be established through the introduction of the generating functional
\begin{eqnarray}
Z_{tt'}(J) = \iint_{APBC} D\Phi e^{\{{\mathcal Q}, \Psi\} + \int d\tau J_\alpha(\tau) O^\alpha(\Phi(\tau))} \label{GeneratingFunctional1}.
\end{eqnarray}

Here, the periodic/anti-periodic boundary conditions for the bosonic/fermionic fields are used, and $J_\alpha$ is a set of external ``probing'' fields coupled to the system via a set of operators $O^\alpha$, sometimes called observables. We now present a few examples of observables with the corresponding probing~fields:
\begin{eqnarray}
J_f(\tau)f(x(\tau)), J_{B_i}(\tau) B_i(\tau), J_{\chi^i}(\tau) \chi^i(\tau), J_{\bar\chi_i}(\tau) \bar\chi_i(\tau), ... \label{examples}
\end{eqnarray}
where $f(x)$ is a function on $X$. In models with linear phase spaces, an observable of the form $J_{x^i}x^i$, with $J_{x^i}\in TX^*$, can also be used. The generating Functional (\ref{GeneratingFunctional1}) can be thought of as the DPF perturbed by probing fields. In particular, $Z_{tt'}(0) = Z_{tt'}$.

In practice, what one is interested in is the limit of the infinitely long evolution
\begin{eqnarray}
Z(J) = \lim_{t^{\pm} = \pm\infty} Z_{t^+ t^-}(J).
\end{eqnarray}
\begin{widetext}
The limit here does not exist for interesting cases of the spontaneously broken supersymmetry because the DPF grows exponentially in this limit (see Equation (\ref{ExponentialGrowth})). Taking this limit must always be the very last operation after other manipulations are already performed. For example, the following notation for the family of correlators
\begin{eqnarray}
\langle O^{\alpha_k}(t_k)...O^{\alpha_1}(t_1) \rangle&=& Z(0)^{-1} \left.\frac{\delta^k Z(J)}{\delta J_{\alpha_1}(t_1)...\delta J_{\alpha_k}(t_k)}\right|_{J=0} 
=\frac{\iint_{APBC} D\Phi\text{ } O^{\alpha_k}(t_k)...O^{\alpha_1}(t_1) e^{\{{\mathcal Q}, \Psi(\Phi)\}}}
{\iint D\Phi e^{\{{\mathcal Q}, \Psi(\Phi)\}}}
\end{eqnarray}
must be understood as
\begin{eqnarray}
\langle O^{\alpha_k}(t_k)...O^{\alpha_1}(t_1) \rangle &=& \lim_{t^{\pm} = \pm\infty} Z_{t^+ t^-}(0)^{-1} \left.\frac{\delta ^k Z_{t^+ t^-}(J)}{\delta J_{\alpha_1}(t_1)...\delta J_{\alpha_k}(t_k)}\right|_{J=0}. \label{correlators}
\end{eqnarray}
\end{widetext}

The next goal now is to pass to the operator representation of the theory. Again, models with the Gaussian white noise will be  considered.

Previously, the following relations were established for the DPF:
\begin{eqnarray}
Z_{tt'}(0) = \iint_{APBC} e^{\{{\mathcal Q},\Psi\}} = Tr \hat{\mathcal{M}}_{tt'} = Tr e^{-\hat H(t-t')}. 
\end{eqnarray}

Their analogue for the generating functional is
\begin{eqnarray}
Z_{tt'}(J) &=& \iint_{APBC} D\Phi e^{\{{\mathcal Q}, \Psi\} + \int d\tau J_\alpha(\tau) O^\alpha(\Phi(\tau))} \nonumber \\ &=&  Tr {\mathcal T} e^{ - \int_{t'}^{t} \hat H(\tau) d\tau},\label{GeneratingFunctionalOperator}
\end{eqnarray}
where 
\begin{eqnarray}
\hat H(\tau) = \hat H - J_\alpha(\tau)\hat O^\alpha,
\end{eqnarray}

$\mathcal T$ denotes chronological ordering, and $\hat O^\alpha \equiv [O^\alpha(\hat \Phi)]_{sym}$, \emph{i.e.}, the operator version of $O(\Phi)$ symmetrized in accordance with the Weyl--Sratonovich bi-graded symmetrization rule discussed in the previous subsection. The chronological ordering in Equation (\ref{GeneratingFunctionalOperator}) is needed because $\hat H(\tau)$ and different $\tau$ do not commute so that the order of operators is important. The chronologically ordered exponent in Equation (\ref{GeneratingFunctionalOperator}) can be given in the form of the formal Taylor series similar to Equation (\ref{Chronological}),
\begin{widetext}
\begin{eqnarray}
{\mathcal T} e^{ - \int_{t'}^{t} \hat H(\tau) d\tau} &=& \hat 1_{\Omega(X)} - \int_{t'}^{t}d\tau_1 \hat H(\tau_1) + \int_{t'}^{t}d\tau_1\int_{t'}^{\tau_1}d\tau_2\hat H(\tau_1)\hat H(\tau_2) + ... \nonumber
\end{eqnarray}

To establish the operator representation of Correlators (\ref{correlators}), one first performs the chronological ordering in the denominator of Equation (\ref{correlators}):  
\begin{eqnarray}
\left.\frac{\delta^k Z_{t^+ t^-}(J)}{\delta J_{\alpha_1}(t_1)...\delta J_{\alpha_k}(t_k)}\right|_{J=0} &=& \iint_{APBC} D\Phi\text{ } O^{\alpha_k}(t_k)...O^{\alpha_1}(t_1) e^{\{{\mathcal Q}, \Psi\}} 
=(-1)^p \iint_{APBC} D\Phi\text{ } O^{\alpha_k'}(t_k')...O^{\alpha_1'}(t_1') 
e^{\{{\mathcal Q}, \Psi \}},\nonumber 
\end{eqnarray}
where $t_k'>t_{k-1}'...>t_1'$ is the chronologically ordered permutation of $t_1...t_k$. Accordingly, $\alpha_1'...\alpha_k'$ is the same permutation of $\alpha_1...\alpha_k$, and $(-1)^P$
is the sign that may appear if this permutation is odd for fermionic operators. Further, 
\begin{eqnarray}
&=&(-1)^p\iint_{APBC} D\Phi\text{ } O^{\alpha_k'}(t_k')...O^{\alpha_1'}(t_1') 
e^{\{{\mathcal Q}, \Psi\}} = (-1)^p Tr \hat{\mathcal M}_{t^+t_k'} \hat O^{\alpha_k'}\hat {\mathcal M}_{t_k' t_{k-1}'}...\hat {\mathcal M}_{t_2' t_1'} \hat O^{\alpha_1'} \hat {\mathcal M}_{t_1' t^-}\nonumber\\
&=& (-1)^p \sum_n \langle \psi_n| e^{-(t^+-t_k') {\mathcal E}_n} \hat O^{\alpha_k'}\hat {\mathcal M}_{t_k't_{k-1}'}...\hat {\mathcal M}_{t_2't_1'} \hat O^{\alpha_1'} e^{-(t_1' - t^-) {\mathcal E}_n}  |\psi_n\rangle, \nonumber
\end{eqnarray}

where $\langle \psi_n| \hat{\mathcal M}_{t^+t_k'} = \langle \psi_n| e^{-(t^+-t_1') {\mathcal E}_n}$ and $\hat {\mathcal M}_{t_1' t^-} |\psi_n\rangle = e^{-(t_1' - t^-) {\mathcal E}_n}  |\psi_n\rangle$ have been used. The~contribution from each eigenstate includes a factor $e^{-(t^+-t^-){\mathcal E}_n}$. In the limit of the infinitely long temporal evolution $t^\pm\to\pm\infty$, only the contribution from the ''ground'' states with the least real part of their eigenvalue, \emph{i.e.}, $Re{\mathcal E}_n = \Gamma_g = min_{n} Re {\mathcal E}_n$, survive. All other eigenstates provide exponentially vanishing contributions that can be neglected. Therefore,
\begin{eqnarray}
= (-1)^p \sum\nolimits_g \langle \psi_g| e^{-(t^+-t_k') {\mathcal E}_g} \hat O^{\alpha_k'}\hat {\mathcal M}_{t_k't_{k-1}'}...\hat {\mathcal M}_{t_2't_1'} \hat O^{\alpha_1'} e^{-(t_1' - t^-) {\mathcal E}_g}  |\psi_g\rangle.\label{Intermed1}
\end{eqnarray}

For the spectra presented in Figures \ref{Figure_3_2}a and \ref{Figure_3_2}b, the eigenvalue of the ground states is unique, and the situation is relatively simple. For example, the partition function takes the following~form:
\begin{eqnarray}
&Z_{t^+t^-} (0) = \sum_g \langle \psi_g| e^{-(t^+-t^-){\mathcal E}_g}|\psi_g\rangle = N_g e^{-(t^+-t^-){\mathcal E}_g},\label{Intermed2}
\end{eqnarray}
where $N_g$ is the number of the groundstates. When ${\mathcal E}_g\ne0$, $N_g=2$ because the ground states are the bosonic-fermionic pair. In the case ${\mathcal E}_g=0$, \emph{i.e.}, the situation of unbroken supersymmetry, the ground states are $\hat d$-symmetric, and $N_g$ must be the sum of the Betti numbers because each de Rham cohomology class must provide one $\hat d$-symmetric eigenstate.  

Using Equations (\ref{Intermed1}) and (\ref{Intermed2}), the Correlators (\ref{correlators}) take the following form:
\begin{eqnarray}
\langle O^{\alpha_k}(t_k)...O^{\alpha_1}(t_1) \rangle &=& N_g^{-1} (-1)^p \sum_g  \langle \psi_g| e^{-(t^--t_k')} \hat O^{\alpha_k'} ... \hat O^{\alpha_1'} e^{-(t_k'-t^-)}|\psi_g\rangle
=N_g^{-1} (-1)^p \sum_g  \langle \psi_g| \hat {\mathcal M}_{t_-t_k'} \hat O^{\alpha_k'} ... \hat O^{\alpha_1'} \hat {\mathcal M}_{t_1't^-} |\psi_g\rangle
\nonumber\\&=& 
N_g^{-1} \sum_g  \langle \psi_g| \hat {\mathcal M}_{t^-t^+} {\mathcal T} \left(\hat O^{\alpha_k}(t_k)...\hat O^{\alpha_1}(t_1)\hat {\mathcal M}_{t^+t^-}\right) |\psi_g\rangle, \label{CorrelatorsSchoedinger}
\end{eqnarray}
where the time arguments of the operators $\hat O(t)$ do not suggest that these operators have explicit dependence on time but rather indicate the moments of time that these operators act on the~wavefunction.

In the Heisenberg representation, the wavefunctions are viewed independent of time, and the temporal evolution is passed onto the operators that now have explicit time dependences:
\begin{eqnarray}
\hat O \to \hat O_{H}(t) = e^{(t-t^*)\hat H} \hat O e^{-(t-t^*)\hat H},\label{Heisenberg}
\end{eqnarray}
with $t^*$ being some reference time moment. In this  representation, Equation (\ref{CorrelatorsSchoedinger}) is even simpler:
\begin{eqnarray}
&\langle O^{\alpha_k}(t_k)...O^{\alpha_1}(t_1) \rangle = 
N_g^{-1} \sum_g  \langle \psi_g| {\mathcal T} \left(\hat O_H^{\alpha_k}(t_k)...\hat O_H^{\alpha_1}(t_1)\right) |\psi_g\rangle. \label{CorrelatorsHeisenberg}
\end{eqnarray}
\end{widetext}

Note that the correlators are independent of the choice of the reference time $t^*$ in Equation~(\ref{Heisenberg}) because the model is time-translation invariant. This suggests, in particular, that the expectation values of all operators that have no explicit dependence on time are time independent. Indeed, one can chose $t^*=t$ so that $\hat O_H(t) = \hat O = O(\hat \Phi)$. Equation (\ref{CorrelatorsHeisenberg}) then gives
\begin{eqnarray}
&\langle O(t) \rangle = N_g^{-1} \sum_g  \langle \psi_g| \hat O_H(t) |\psi_g\rangle \nonumber \\ & = N_g^{-1} \sum_g  \langle \psi_g| \hat O |\psi_g\rangle.
\end{eqnarray}

The time independence of these expectation values can be interpreted as the ergodicity of the model. Thus, the ergodicity in STS is the property that the stochastic expectation values in the limit of infinitely long temporal evolution are those over the ground state(s). Models with the spectra in Figure \ref{Figure_3_2}a,b are automatically ergodic.

For operators that are functions on $X$, \emph{i.e.}, $O(t) = f(x(t))\in\Omega^0(X)$, the expectation value has the following form:
\begin{eqnarray}
\langle f(x(t)) \rangle = \sum\nolimits_g \int_X f(x)\bar P_g(x),\label{ExpectationFunction}
\end{eqnarray}
where $P_g(x) = \bar\psi_g(x)\wedge\psi_g(x)/N_g \in \Omega^{D}(X)$ is the TPD averaged over the ground states. The time independence of $P_g(x)$ can be misinterpreted here as though the model is in the state of thermodynamic equiliubrium. In other words, the ergodicity can be mistaken for thermodynamic equilibrium. It is actually very common in the literature that the ergodicity and be confused with thermodynamic equilibrium. If the ergodicity and thermodynamic equilibrium were indeed equivalent, the concept of the ``ergodic theory of chaos'' \cite{RevModPhys.57.617} would not make sense (see also the last paragraph of Section \ref{ChaosTopBreaking}). In~other words, ergodicity is not equivalent to thermodynamic equilibrium. 

The point here is that the TPD in Equation (\ref{ExpectationFunction}) is not the wavefunctions themselves but rather the bra-ket combination. This situation is similar to that in quantum theory, where it is the bra-ket combinations of the eigenstates that are the TPDs. The fact that this combination (the diagonal element of the density matrix) is stationary in time by no means implies that the eigenstate itself has zero~eigenvalue.  

For models with spectra given in Figure \ref{Figure_3_2}c, the situation is more subtle because there is a~pair of Ruelle--Pollicott resonances with two different eigenvalues, \emph{i.e.}, ${\mathcal E}_g$ and ${\mathcal E}_g^*$, with the same ''attenuation rate'' $Re{\mathcal E}_g=Re{\mathcal E}^*_g=min_n Re{\mathcal E}_n$. These states are equally good candidates for the title of the ground state of the model. In the limit of the infinitely long temporal evolution, the DPF is $Z_{t^+-t^-} \approx 4 cos (t^+-t^-)Im{\mathcal E}_g e^{|Re{\mathcal E}_g|(t^+-t^-)}$. This invalidates Equations (\ref{Intermed2}) and (\ref{CorrelatorsHeisenberg}) unless some additional arguments can circumvent this problem.

\begin{figure*}[htb]
\centerline{\includegraphics[width=0.6\linewidth]{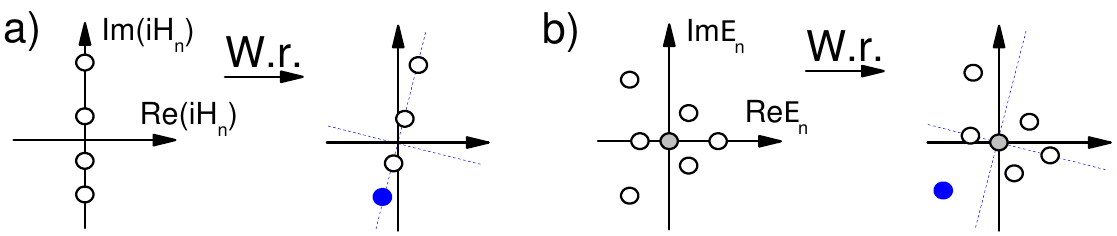}}
\caption{\label{Figure_4_2} (\textbf{a}) In quantum theory, the ground state is the one with the lowest energy. This can be justified by Wick rotating time a ``little'', \emph{i.e.}, $t \to t + i0 $, so that in the limit of infinitely long temporal evolution, the generating functional $Z_t = Tr e^{-i\hat H t}$ receives a contribution only from the ground state with the least eigenvalue: $\left.Z_t\right|_{t\to\infty} \to e^{-i H_g t}$; (\textbf{b}) A similar approach can be used in the supersymmetric theory of stochastics (STS). If a Ruelle--Pollicott resonance has the smallest attenuation rate $Re \mathcal{E}$, the ground state can be thought to be the one with the smallest $Im \mathcal{E}$, \emph{i.e.}, the parameter analogous to the quantum energy.}
\end{figure*}

The same problem concerning the identification of the ground state exists in quantum theory. There, the finite-time quantum evolution operator is $\hat U_{t^+ t^-} = e^{-(i\hat H_q)(t^+-t^-)}$, where $\hat H_q$ is some Hermitian Hamiltonian with real spectra. To ensure that the generating functional in the long time limit receives contribution only from the ground state(s) with the lowest possible eigenvalue of $\hat H_q$, one can Wick-rotate time ``a little'', \emph{i.e.}, $t\to t + i 0^+$, with $0^+$ being a vanishingly small positive constant. This approach can be borrowed for the STS, as illustrated in Figure \ref{Figure_4_2}. After Wick rotating time a little, only the ground states with the least ``energy'', \emph{i.e.}, $Im{\mathcal E}_n$, survive the limit of the infinitely long temporal evolution. Once this is done, the passage from Equations~(\ref{correlators}) to (\ref{CorrelatorsSchoedinger}) and all the later formulas become valid.

As already mentioned in Section \ref{Sec:TimeReversalSymmetry}, the SEO possesses the $\eta$T-symmetry, and each eigenstate with a complex eigenvalue must break this symmetry. By declaring one of the Ruelle--Pollicott resonances as the ground state of the model, one seemingly breaks the $\eta$T-symmetry spontaneously, as discussed in Section \ref{Sec:TimeReversalSymmetry}. The physical implications of this mechanism of spontaneous $\eta$T-symmetry breaking is not clear at this moment to the present author.

\subsection{One Way to a Unique Ground State}
\label{Sec:OneWayForUnique}

At this point of the discussion of the STS, the ground states are not unique for all three types of spectra given in Figure \ref{Figure_3_2}. Indeed, for models with unbroken topological supersymmetry, there may be many supersymmetric states, each of which may be viewed as a ground state of the model. For~models with spontaneously broken topological supersymmetry (Figure \ref{Figure_3_2}b,c), the non-$\hat d$-symmetric ground state is doubly degenerate because it is a boson-fermion pair of eigenstates. 

\begin{figure}[h]
\centerline{\includegraphics[height=4.1cm, width=3cm]{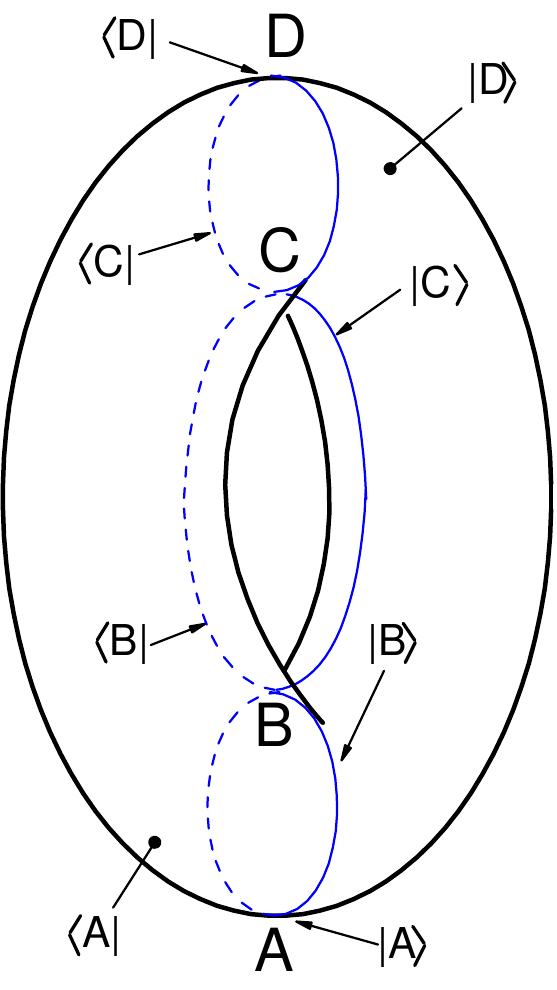}}
\caption{\label{Figure_5_3} The global ground states of the Langevin SDE on a torus in the deterministic limit and with the Langevin potential being the ``height''. There are four global ground states in each of the four cohomology classes. Each ground state is the Poincar\'e dual of a global unstable manifold of one of the four critical points denoted as $A, B, C$, and $D$. The bras of the ground state are the Poincar\'e duals of the corresponding global stable manifolds. The bra-ket combination for each ground state is a~delta-functional distribution on the corresponding critical point. As discussed in Section \ref{Sec:OneWayForUnique}, among these four ground states, the one corresponding to the critical point $A$ must be considered the true ground state of the model. This is the ground state of the thermodynamic equilibrium.}
\end{figure}

The ground state can be made unique using yet another additional reasoning. This reasoning follows from the analysis of the  supersymmetric states of the integrable models in the deterministic limit in Section \ref{DetIntModles}. There, it will be discussed that the supersymmetric states of integrable deterministic models are the so-called Poincar\'e duals of the global unstable manifolds of the flow. One example of this situation is given in {Figure} \ref{Figure_5_3} for the case of the Langevin SDE on a 2D torus. In~this model, there are four supersymmetric states, each being the Poincar\'e dual of the global unstable manifolds of the four critical points denoted as A, B, C and D. The corresponding bras of these supersymmetric states are the Poincar\'e duals of the global stable manifolds. 
The expectation value of a~function in Equation (\ref{ExpectationFunction}) reads
\begin{eqnarray}
\langle f(x(t)) \rangle = 4^{-1} (f(A)+f(B)+f(C)+f(D)),\label{ExpectationFunction1}
\end{eqnarray}
where the fact that the bra-ket combination of each of these supersymmetric states is a $\delta$-functional TPD on the corresponding critical points has been used. 

On the other hand, it is intuitively clear that this expectation value must equal $f(A)$. To bypass this controversy, one can propose to view the ground state with the maximal number of fermions as the true ground state of the model. This rule can be called the principle of ``minimal knowledge'' for the following reason. The presence of a fermion in a wavefunction means that the wavefunction is a distribution in the corresponding bosonic variable, whereas the absence of a fermion suggests that the corresponding bosonic variable is not ``thermalized'' so that something else (\emph{e.g.}, the bra of the wavefunction or an external observer) must know with certainty the value of this bosonic variable. In other words, the more fermions a wavefunction has, the less external knowledge one needs to view the wavefunction as a ``complete'' probability distribution.  

With this principle at hand, the ground state of the model is unique. When the topological supersymmetry is unbroken, the ground state is the TE state. For the broken supersymmetry case, the ground state is $\hat d$-exact, \emph{i.e.}, $|\vartheta'_G\rangle = \hat d| {\underline{\vartheta}}_G\rangle$, where the notations of Equation (\ref{SecondType}) have been used. Now~that the ground state is unique, the correlators take the familiar field-theoretic form of the ''vacuum'' correlators:
\begin{eqnarray}
&\langle O^{\alpha_k}(t_k)...O^{\alpha_1}(t_1) \rangle \nonumber = \\ & 
\langle \vartheta'_G| {\mathcal T} \left(\hat O_H^{\alpha_k}(t_k)...\hat O_H^{\alpha_1}(t_1)\right) |\vartheta'_G\rangle.\label{ExpectGround}
\end{eqnarray}

\subsection{Response and the Butterfly Effect}
\label{Butterfly}

Of special interest are the correlators that reveal the response of the model to external perturbations. To understand what these response correlators are, one notes that the only physical way to perturb a model is to perturb it on the level of the SDE. This can be done with the following modification of the flow vector field in Equation (\ref{SDE}):
\begin{eqnarray}
F^i(x(t)) \to F^i(x(t)) - J_c(t) f^{c,i}(x(t)),
\end{eqnarray}
where $J_c(t), c=1,2...$, is a set of probing fields and $f$ is a set of predetermined vector fields. The action of the model transforms accordingly:
\begin{eqnarray}
&\{{\mathcal Q},\Psi\} \to \{{\mathcal Q},\Psi\} + \nonumber \\&  + \int d\tau J_c(t) \left\{ {\mathcal Q}, i\bar\psi_i(\tau) f^{c,i}(x(\tau))\right\}.\label{ModifiedAction}
\end{eqnarray}

The methodology of Section \ref{GeneratFunctional} applies, with the perturbation operators or observables~being
\begin{eqnarray}
{\mathcal L}^c(t) = \{{\mathcal Q}, i\bar\psi_i(t) f^{c,i}(x(t)) \}.
\end{eqnarray}

Here, the notation is switched from $O$ to ${\mathcal L}$ to note that these perturbation operators are Lie derivatives in the operator representation:
\begin{eqnarray}
\hat {\mathcal L}^c = [\hat d, \hat \imath_{f^c}],
\end{eqnarray}
with $\hat \imath_{f^c} = f^{c,i}(x(t)) \partial/\partial \chi^i$. In the Heisenberg representation,  
\begin{eqnarray}
\hat {\mathcal L}_H^c(t) = [\hat d, \hat \imath_{f^c,H}(t)]\label{HeisenbergLie}
\end{eqnarray}
because $\hat d$ is commutative with $\hat H$ so that $\hat d_H(t) = e^{(t-t^*)\hat H} \hat d  e^{(t-t^*)\hat H} = \hat d$. The expression for the response correlators follows now from Equation (\ref{ExpectGround}):
\begin{eqnarray}
\langle {\mathcal L}^{c_k}(t_k)...{\mathcal L}^{c_1}(t_1) \rangle &=& \langle \vartheta'_G| {\mathcal T} \left(\hat {\mathcal L}_H^{c_k}(t_k)...\hat {\mathcal L}_H^{c_1}(t_1)\right) |\vartheta'_G\rangle \nonumber \\ &=& \langle \vartheta'_G| [\hat d, \hat R] |\vartheta'_G\rangle,
\label{Response}
\end{eqnarray}
with
\begin{eqnarray}
\hat R = {\mathcal T} \left(\hat \imath_{f^{c_k},H}(t_k) \hat {\mathcal L}_H^{c_{k-1}}(t_{k-1})...\hat {\mathcal L}_H^{\alpha_1}(t_1)\right).
\end{eqnarray}

Here, the fact that a product of the $\hat d$-exact Lie derivatives from Equation (\ref{HeisenbergLie}) is a $\hat d$-exact operator itself, as can be shown using the fact that $\hat d$ is a nilpotent (bi-graded) differentiation, has been used.

When the topological supersymmetry is unbroken and the ground state is $\hat d$-symmetric, the response correlators in Equation (\ref{Response}) vanish by the definition of the $\hat d$-symmetric states in Equation~(\ref{ZeroExactOperators}). In other words, in the infinitely long temporal evolution limit, the model does not respond to perturbations. It can be said that the model forgets perturbations. 

On the contrary, if the topological supersymmetry is spontaneously broken, some of the perturbation correlators do not vanish. This can be interpreted as though the model ``remembers'' perturbations even in the limit of the infinitely long temporal evolution. This is how the STS reveals the famous butterfly effect. It is worth noting that the butterfly effect was previously often viewed as an intrinsic part of the definition of (deterministic) chaos, whereas within the STS, it is a~derivable~consequence.

The butterfly effect derived above is a part of a more general statement known as the Goldstone theorem. This theorem states that a model must exhibit a long-range order under the conditions of the spontaneous breakdown of a continuous global symmetry. In spatially extended models, this tailors the existence of a gapless excitation called the Goldstone--Nambu boson for bosonic symmetries and the goldstino for supersymmetries. This long-range order associated with the $\hat d$-symmetry breaking is the DLRO discussed in the Introduction. 

%
%
%

\section{Classification of Ergodic Stochastic Dynamics}

\label{Chap:DynamicsTypes}
\vspace{-6pt}

\subsection{Transient {\em vs.} Ergodic Dynamics}

Before turning to the discussion of ergodic dynamics, it is worth addressing the following issue. One important type of dynamics is called transient dynamics. Roughly speaking, transient dynamics begins at one point of the phase space and ends at another point. Physical examples of transient dynamics include various quenches as well as processes that can be identified as ``slow'' quenches, e.g., the Barkhausen effect and crumpling paper. Another example is glasses: it is often said that (at non-zero temperature) a glass will eventually crystallize. This crystallization process, however, may take a very long time, and at the moment of observation, an external observer observes transient (noise-assisted) dynamics from some initial point in the phase space corresponding to the disordered lattice to the state of crystallization.


It is well known that quenches and other transient processes also exhibit the long-range dynamical behavior (LRBD). One example is the power-law statistics of the Barkhausen jumps in ferromagnets. The mathematical origin of this LRDB has never been explained in the general case. For quenches across phase transitions, this LRDB is often attributed to the ``criticality'' of the DS, \emph{i.e.}, to the proximity of the phase transition. This may be a misleading explanation because quenches that are not across a~phase transition also exhibit LRDB, and the criticality arguments are not valid for them. At the same time, it is natural to expect that the origin of LRDB  must be the same for all quenches.

Within the STS, this LRDB is the result of the intrinsic breakdown of the topological supersymmetry within instantons. More specifically, it has been shown \cite{Frenkel2007215} that a model must be log-conformal when instantons condense, \emph{i.e.}, when the dynamics is a composite instanton or rather a composition of fundamental instantons.

Transient dynamics is often referred to as out-of-equilibrium dynamics. The same term is often used for the characterization of chaotic behavior. In these two situations, the term ``out-of-equilibrium'' has two different meanings. In one case, it means non-ergodic dynamics out of the global ground state of the DS, whereas in the second case, it denotes dynamics out of the $\hat d$-symmetric state of the thermodynamic equilibrium but within the global non-$\hat d$-symmetric ground state. This second type of the ``out-of-equilibrium'' dynamics is often called ``self-sustained'' dynamics, \emph{i.e.}, happening forever.

In this section, only ergodic or self-sustained dynamics is addressed. Transient dynamics is beyond the scope of this paper. It is worth mentioning, however, that in some cases (glasses, for example) it must be possible to map transient dynamics in a model onto an ergodic dynamics in another model with the spontaneously broken supersymmetry.

\subsection{Unstable Manifolds and Ground States: Langevin SDEs}
\label{LangevinSDE}

In this subsection, the relation between $\hat d$-symmetric ground states in the weak noise limit and unstable manifolds of flow vector fields will be discussed. It is convenient to start the discussion with Langevin SDEs - the most studied class of SDEs closely related to $N=2$ supersymmetric quantum mechanics (see, e.g., \cite{MirrorSymmetry}). For simplicity, the noise-induced metric is assumed to be Euclidean: $e^i_a  =\delta^i_a, g^{ij} = \delta^{ij}$. The flow vector field  $F^i(x)= - \delta^{ij}U_{'j}(x)$ is defined via the Langevin potential $U(x)$, and $U_{'j} = \partial U/\partial x^j$. The SEO of this model is given by Equation (\ref{DExactFPOperator}), with
\begin{eqnarray}
\hat {\bar d} = \frac\partial{\partial\chi^i} \delta^{ij} \left(-U_{'j} - \Theta \frac\partial{\partial x^j}\right).
\end{eqnarray}

The similarity transformation $\hat A \to \hat A_U = e^{U/(2\Theta)}\hat A e^{-U/(2\Theta)}$ acts on the SEO as
\begin{eqnarray}
\hat H \to \hat H_U = \Theta [\hat d_U, \hat d_U^\dagger], \label{HU}
\end{eqnarray}
where
\begin{eqnarray}
\hat d_U &=& e^{U/(2\Theta)}\hat d e^{-U/(2\Theta)} = \chi^i\left(\frac\partial{\partial x^i} - U_{'i}/2\Theta\right),\\
\hat {\bar d}_U &=& \frac\partial{\partial\chi^i} \delta^{ij} \left(-U_{'j}/2 - \Theta \frac\partial{\partial x^j}\right) = \Theta\hat d_U^\dagger,
\end{eqnarray}
with $(\chi^i)^\dagger=\delta^{ij} \partial/\partial\chi^j$ and $(\partial/\partial x^i)^\dagger = - \partial/\partial x^i$.

Because $\hat H$ and $\hat H_U$ are related via a similarity transformation, their spectra are identical. As to the eigenstates, they are related as
\begin{eqnarray}
| \psi \rangle = e^{-U/2\Theta}|\psi_{U}\rangle, \text{ and }
\langle \psi | = \langle\psi_{U}| e^{U/2\Theta}.\label{relationWF}
\end{eqnarray}

Up to the factor $\Theta$, the operator $\hat H_U$ is the Hermitian Hamiltonian of $N=2$ supersymmetric quantum mechanics. Its spectrum is real and non-negative. This implies that the topological supersymmetry is never broken in this class of models as long as there exists at least one $\hat d$-symmetric ground state of the thermodynamic equilibrium (see Section \ref{SubSec:TermEquil}). 


In the single-variable case with the harmonic potential $U = \omega x^2$, the zero-eigenvalue ground state of $\hat H_U$ is (see, e.g., Section 10.2.4 in \cite{MirrorSymmetry})
\begin{eqnarray}
\psi_{g,U} = \star \bar\psi_{g,U}^* \propto
\left\{\begin{array}{cc}
\chi e^{-|\omega|x^2/2\Theta}, & \omega >0,\\
e^{-|\omega|x^2/2\Theta}, & \omega < 0.
\end{array}\right.\label{WaveFunction}
\end{eqnarray}

Here, the relation between bras and kets is trivial because $\hat H_U$ is Hermitian. In terms of the eigensystem of the original non-Hermitian $\hat H$, the bra and ket are different. Using Equation (\ref{relationWF}), one~has
\begin{eqnarray}
\psi_{g} \propto
\left\{\begin{array}{cc}
\chi e^{-|\omega|x^2/\Theta}, & \omega >0,\\
1, & \omega < 0,
\end{array}\right.
\end{eqnarray}
and 
\begin{eqnarray}
\bar\psi_{g}\propto
\left\{\begin{array}{cc}
1, & \omega >0,\\
\chi e^{-|\omega|x^2/\Theta}, & \omega < 0.
\end{array}\right.
\label{WaveFunction1}
\end{eqnarray}

These are the ground state wavefunctions of the two models in Figure \ref{Figure_2_3}. For the stable variable case ($\omega>0$), the ket of the ground state is the narrow distribution around the stationary position $x=0$, and the bra is not a distribution; rather, it is a constant function. In the unstable case ($\omega<0$), the bra and ket are switched.



\begin{figure}[h]
\centerline{\includegraphics[width=0.75\linewidth]{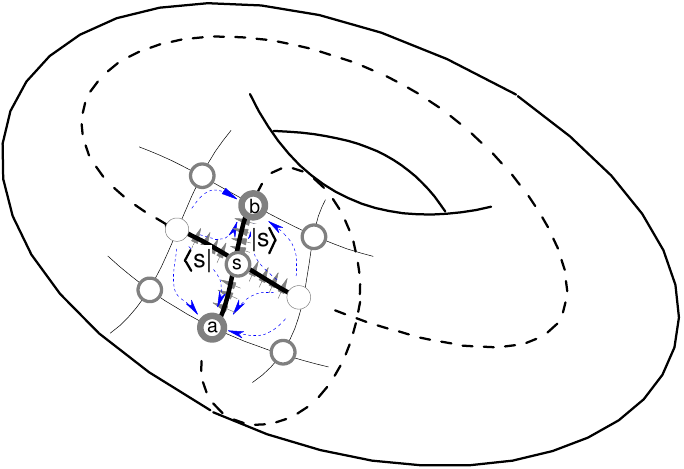}}
\caption{\label{Figure_5_1} The bra $\langle s|$ and ket $|s\rangle$ of the perturbative (or local supersymmetric) ground state on a saddle point $s$ are the Poincar\'e duals of the local stable and unstable manifolds, respectively. The small gray arrows in the transverse directions represent differentials/fermions. For integrable (non-chaotic) flow vector fields, local unstable/stable manifolds can be glued into the global unstable/stable manifolds indicated as closed dashed  curves from the two first homology classes of the phase space, which is assumed here to be a 2D torus. The exterior derivative annihilates the Poincar\'e duals of the closed global unstable manifolds (see Figure \ref{Figure_5_2}), which in this case is the wavefunction of (one of) the global $\hat d$-symmetric ground state(s).}
\end{figure}

This analysis can be extended now to multiple-variable Langevin SDEs. Consider a vicinity of a non-degenerate critical point where the Langevin potential can be approximated as a quadratic form. With the appropriate coordinate rotation, this quadratic form can be diagonalized as  \linebreak $U = \sum_i \omega_i(x^i)^2/2, \omega_i\ne0, i = 1...D$. The wavefunction of the (local) $\hat d$-symmetric ground state factorizes in all coordinates, and each coordinate provides a factor of the Form (\ref{WaveFunction1}). As a result, the wavefunction is a narrow distribution in stable variables and is a constant function in unstable variables of the unstable manifold of this critical point, as illustrated in Figure \ref{Figure_5_1}.

\vspace{-6pt}

The so-emerged wavefunctions are known as Poincar\'e duals. They appear in one version of Poincar\'e duality stating that, for each $k$-dimensional submanifold $c_k$, there exists a differential form $\underline{\psi}_{c_k}\in\Omega^{(D-k)}$ such that $\int_{c_k}\varphi^{(k)} = \int_X\varphi^{(k)}\wedge \underline{\psi}_{c_k}$ for all $\varphi^{(k)}\in\Omega^{(k)}$. Using this terminology, the bra and ket of the local $\hat d$-symmetric ground state on a non-degenerate critical point of a Langevin SDE and in the weak noise limit are the Poincar\'e duals of the local stable and unstable manifolds, respectively.

The local unstable manifolds have boundaries on the lower dimensional local unstable manifolds of more stable critical points. For example, in Figure \ref{Figure_5_1}, the local unstable manifold of the unstable critical point ($s$) terminates at the stable critical points ($a$) and ($b$).

\begin{figure}[t]
\centerline{\includegraphics[width=0.8\linewidth]{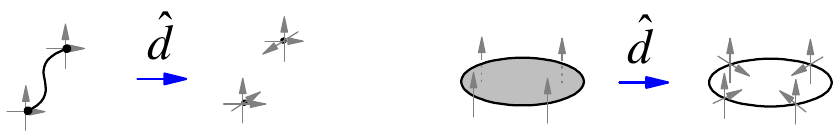}}
\caption{\label{Figure_5_2} The operator of the exterior derivative acts on Poincar\'e duals of submanifolds (a curve and a disk given as examples) as the boundary operator would have acted on the submanifolds themselves.}
\end{figure}

The collection of the local unstable manifolds of different dimensionality is known as the Morse complex, whereas the collection of the corresponding perturbative (or local) $\hat d$-symmetric states is known as the Morse--Witten complex. The operator $\hat d$ acts on the perturbative $\hat d$-symmetric states as the boundary operator would have acted on the local unstable manifolds themselves (see Figure \ref{Figure_5_2}). The~local unstable manifolds have boundaries and the corresponding perturbative $\hat d$-symmetric ground states are non-$\hat d$-symmetric in the global sense because $\hat d$ does not annihilate them. For example, the ket of the perturbative ground state of the critical point $(s)$ in Figure \ref{Figure_5_1} satisfies
\begin{eqnarray}
\hat d |\psi_s\rangle = |\psi_a\rangle-|\psi_b\rangle.
\end{eqnarray}

To obtain the global $\hat d$-symmetric ground states, one must glue local unstable manifolds into the global unstable manifolds with no boundaries. The Poincar\'e duals of the global unstable manifolds are the global $\hat d$-symmetric ground states of the model. The discussion can be generalized to the Morse--Bott situation, in which critical points of the gradient flow vector field are not isolated but form closed submanifolds of $X$. In this case, the local $\hat d$-symmetric states must be complemented by the factors from the de Rham cohomology of these critical submanifolds.

\subsection{Deterministic Models}
\vspace{-6pt}
\subsubsection{Integrable Models}

\label{DetIntModles}

The existence of the well-defined global (un)stable manifolds that are said to provide foliations of the phase space is essentially the definition of the integrability of a flow vector field in the sense of DS theory. From the point of view of the STS, the Poincar\'e duals of these unstable manifolds are the kets of the global $\hat d$-symmetric ground states. These global $\hat d$-symmetric ground states are invariant with respect to the (deterministic) flow. Indeed, by definition, an (un)stable manifold consists of points that remain on it at all times of the flow. Therefore, a Poincar\'e dual being a constant function on an (un)stable manifold is unchanged by the flow. The squeezing in the transverse directions will provide a corresponding Jacobian from the $\delta$-functional dependence on the transverse coordinates. This Jacobian will be compensated in the supersymmetric manner by the Jacobian provided by the corresponding transformation of the differentials/fermions. As a result, the Poincar\'e duals of the global (un)stable manifolds are invariant under the flow, \emph{i.e.}, they have zero eigenvalues. That these states are $d$-symmetric follows trivially from the fact that the global (un)stable manifolds have no boundaries. This picture suggests that the integrability of the flow vector field must be equivalent to the unbroken topological supersymmetry in the corresponding STS.

Each de Rham cohomology class may contain more than one global $\hat d$-symmetric ground state. This can be the case only in the strict deterministic limit for the following reason. Each of such $\hat d$-symmetric ground states is a superposition of one $\hat d$-symmetric ground state and a $\hat d$-exact piece. This means that pairs of non-$\hat d$-symmetric states accidentally have zero eigenvalues. Any noise will introduce exponentially weak tunneling effects that must lift this accidental degeneracy, leaving only one $\hat d$-symmetric ground state in each de Rham cohomology class.

The example of the global supersymmetric eigenstates for the Langevin SDE in the deterministic limit on a 2D torus is given in Figure \ref{Figure_5_3}. In Section \ref{GeneratFunctional}, this example was used to argue that, among all the supersymmetric states of a model with unbroken supersymmetry, one should choose the state of the thermodynamic equilibrium as the true ground state, within which various correlators and observables should be calculated.

\subsubsection{Chaotic Models}
\label{DeterministicChaos}

\begin{figure}[h]
\centerline{\includegraphics[width=0.9\linewidth]{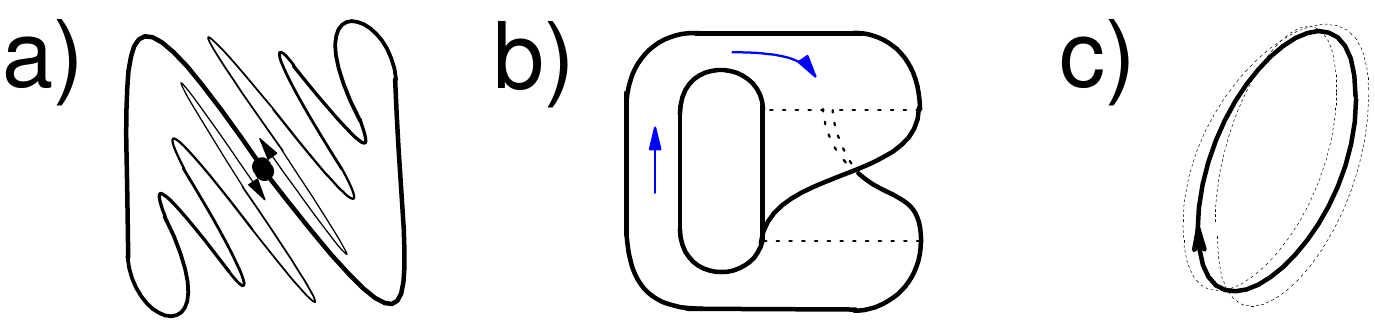}}
\caption{
\label{Figure_5_4} (\textbf{a}) Poincar\'e section of the unstable manifold of the deterministic chaotic behavior known as homoclinic tangle. The unstable manifold recursively folds on itself and accumulates at the origin, where its orientation is ambiguous unless the wavefunction vanishes, as indicated by the fading width of the curve. This coordinate dependence on the position on the unstable manifold suggests that $\hat d$ does not annihilate the would-be Poincar\'e dual, and thus, the topological supersymmetry is broken;  (\textbf{b}) Schematic representation of the unstable manifold in the R\"ossler model in the topological theory of chaos (see, e.g., \cite{Gil98}). The manifold is a branching manifold with self-intersection. The~ground state's wavefunction must be its Poincar\'e dual modified by a continuous function that vanishes at the self-intersection. Such a wavefunction is non-$\hat d$-symmetric, and the topological supersymmetry must be broken; (\textbf{c}) Strange attractors consist of an infinite number of unstable periodic orbits, some of which have non-orientable local unstable manifolds. The would-be Poincar\'e dual must have such a coordinate dependence that circling around the orbit produces a sign change. This functional dependence on the position of the orbit also suggests that the topological supersymmetry is broken. }
\end{figure}

The next goal is to analyze qualitatively the structure of the ground states in chaotic or non-integrable deterministic models. These ground state(s) must represent the dynamics on fractal or strange attractors. Just like in the integrable models above, strange attractors are formed by the intersection of the stable and unstable manifolds. The bra/ket of the ground state must represent (or~rather be) the Poincar\'e duals of these manifolds. The (un)stable manifolds in chaotic deterministic models are not well-defined topological manifolds however. They can fold on themselves in a~recursive manner, as illustrated for the class of models known as ``homoclinic tangle'' in Figure \ref{Figure_5_4}a. The~straightforward attempt to construct a Poincar\'e dual for such an unstable manifold leads to the ambiguity in the orientation of the manifold at the point of the accumulation of self-folding.

This ambiguity has its analogues in quantum theory. For example, a non-rotationally symmetric electron wavefunction on a rotationally symmetric atom ({p, d, f}, ... orbitals) 
would be ambiguous at the origin if it did not vanish there, which is always the case. For the same reason, in the theory of superfluids, the superfluidic order parameter of the Bose condensate at the core of a vortex must~vanish.

In the case of the homoclinic tangle in Figure \ref{Figure_5_4}a, the ambiguity of the Poincar\'e dual of the unstable manifold can be remedied by modifying it with a continuous function that vanishes at the origin. This will introduce the coordinate dependence along the unstable manifold, and this coordinate dependence automatically suggests that the wavefunction is not annihilated by $\hat d$; thus, the ground state wavefunction representing the unstable manifold is non-$\hat d$-symmetric.

Another way to see that the ground state in a chaotic model is non-$\hat d$-symmetric can be borrowed from the topological theory of chaos \cite{Gil98}. There, the global unstable manifold is qualitatively represented by a branched manifold that has self-intersections (see Figure \ref{Figure_5_4}b). The action by $\hat d$ on the Poincar\'e dual of this branched manifold is the Poinar\'e dual of its self-intersection. Thus, such a~wavefunction is non-$\hat d$-symmetric.

Yet another way to convince oneself that chaotic deterministic models have non-$\hat d$-symmetric ground states is to recall that strange attractors contain an infinite number of unstable periodic orbits with arbitrary large periods. Some of these orbits have non-orientable local unstable manifolds, as illustrated in Figure \ref{Figure_5_4}c. The Poincar\'e duals of these local unstable manifolds must be modified by such coordinate dependence that going along the periodic orbit changes the sign of the wavefunction. Again, such a wavefunction is non-$\hat d$-symmetric.    

The above qualitative analysis of the ground states of the deterministic chaotic models is only an indication that the topological supersymmetry breaking must be the field-theoretic essence of deterministic chaos. The rigorous proof of this statement is given by Equation (\ref{ExponentialGrowth}), which establishes the exponential growth of periodic solutions, being definitive for chaos, as well as by the emergence of the butterfly effect discussed in Section \ref{Butterfly}.

\subsection{Stochastic Models: Two Types of ``Border of Chaos''}
\label{SecPhaseDiagram}

In deterministic models, the $\hat d$-symmetry is spontaneously broken or not depending on whether its flow vector field is non-integrable (chaotic) or integrable in the sense of DS theory. The stochastic generalization of this picture is the subject of interest in this subsection.

One important thing to note is that in the high-temperature limit, the SEO (\ref{FPOp}) is dominated by the diffusion Laplacian. In a wide class of models (e.g., torsion-free vielbeins \cite{Ovc14}) the diffusion Laplacian equals the Hodge Laplacian (\ref{HodgeLaplacian}). The latter has real and non-negative spectra, which correspond to the unbroken $\hat d$-symmetry, so that the $\hat d$-symmetry must always be unbroken at sufficiently large temperatures. Only models of this type are of interest here. It can be said that the noise destroys the DLRO at sufficiently high temperatures in this class of models.

\begin{figure}[h]
\centerline{\includegraphics[width=0.8\linewidth]{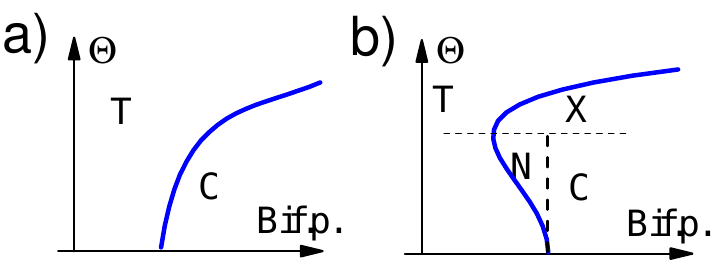}}
\caption{\label{Figure_5_5} Two types of the ``border of chaos''. 
 (\textbf{a}) In type I, there are only two phases: the chaotic phase (C) and the phase of the thermodynamics equilibrium (T). In the low-temperature limit, the topological supersymmetry of the C-phase is broken by the non-integrable flow vector field. As the temperature increases, the border moves to the ``right'' because the noise has the tendency to destroy the DLRO; (\textbf{b})~In type-II phase diagrams, there is an additional phase of the noise-induced chaos (N) where the flow vector field is integrable but where the topological supersymmetry is spontaneously broken by the condensation of (anti-)instantons, \emph{i.e.}, the noise-induced tunneling processes between, e.g., different attractors. One type of  dynamical behavior in the N-phase is such that an external observer sees a~sequence of unpredictable jumps between patterns of  ``regular'' behavior and/or attractors. This type of dynamics can be recognized as that of self-organized criticality. At higher temperatures, the sharp boundary between the N- and C-phases must smear out into a crossover because the perturbative supersymmetric ground states overlap significantly, and it is not possible for an external observer to tell one (anti-)instantonic process from another. The N- and C-phases must merge into a complicated phase (X) with the spontaneously broken topological supersymmetry.}
\end{figure}

Two qualitatively different types of the ``border of chaos'' exist for this class of models (see~Figure~(\ref{Figure_5_5})). For the first type, the $\hat d$-broken phase gradually narrows with increasing temperature, which corresponds to the situation discussed, e.g., in \cite{Kap90}. The second type (Figure \ref{Figure_5_5}b) is more involved. There, the $\hat d$-broken phase first widens with increasing temperature before shrinking. In~other words, there exists a phase with an integrable flow vector field on one hand and with $\hat d$-symmetry spontaneously broken on the other. This peculiar phase can be called noise-induced chaos (N-phase) because the supersymmetry can be restored by decreasing the temperature. In the deterministic limit, the N-phase collapses into the boundary of the deterministic chaos.

\subsubsection{Low-Temperature Regime and Self-Organized Criticality}
\label{LowTemp}

In the low-temperature regime of the type-II phase diagram, the $\hat d$-broken phase consists of two major subphases: the ordinary chaotic phase (C-phase), where the $\hat d$-symmetry is broken by the non-integrable flow vector field, and the N-phase, where the $\hat d$-symmetry is broken by some other mechanism. There are two other known mechanisms for the spontaneous breakdown of a~symmetry. The first one is an anomaly, \emph{i.e.}, the possibility that a symmetry is broken by perturbative or fluctuational corrections. Supersymmetries, however, are difficult to break via anomaly. This fact is related to the so-called  supersymmetry non-renormalization theorems  \cite{Seinberg,Weinberg}. This suggests that the $\hat d$-symmetry breaking in the N-phase must be due to the other mechanism. This other mechanism of supersymmetry breaking is known as the condensation of (anti-)instantonic configurations \cite{DynSusyBrWitten}. Due~to the renormalization theorems, the ``dynamical'' supersymmetry breaking by (anti-)instantons is considered as one of the most reliable mechanisms of supersymmetry breaking in high-energy physics~models. 

In case of stochastic dynamics, these (anti-)instantonic configurations are the tunneling processes that appear due to the noise-induced and exponentially weak overlap between perturbative ground states on unstable manifolds. One of the effects that the noise-induced tunneling processes will provide is the removal of the degeneracy of the deterministic zero-eigenvalue eigenstates representing Poincar\'e duals of ``parallel'' global unstable manifolds within the same de Rham cohomology class discussed in the last paragraph of Section \ref{DetIntModles}. As a result, each de Rham cohomology class will have only one $\hat d$-symmetric eigenstate, whereas other eigenstates will acquire (exponentially small) non-zero eigenvalues. This removal of degeneracy does not necessary suggest that $\hat d$-symmetry is spontaneously broken. Clearly, the very existence of the noise-induced tunneling processes is insufficient. Indeed, for a Langevin SDE (see Section \ref{LangevinSDE}) with a Langevin potential with multiple local minima, the tunneling processes between these local minima certainly exist at non-zero temperatures. Nevertheless, the $\hat d$-symmetry is never broken for this class of models. In other words, the weak-noise tunneling processes can only help the spontaneous $\hat d$-symmetry breaking in models with flow vector fields that are close to being chaotic on their own. This is why the N-phase resides on the ``border of chaos''.

The physical picture of one type of dynamics in the N-phase is as follows. The fluctuating dynamics is mostly around unstable manifolds such as point attractors or limit cycles. The dynamics is sporadically interrupted by noise-induced tunneling processes or jumps between different attractors. Because it is the noise-induced tunneling processes that break the $\hat d$-symmetry, the jumps must exhibit signatures of long-range dynamical behavior such as the power-law statistics. This power-law statistics of jumps, or avalanches as they also called in the literature, is a well-established phenomenon with the Richter scale for earthquakes being perhaps the best known example. 

The ubiquitous power-law statistics of avalanches in nature was previously proposed to explain via the concept of self-organized criticality \cite{Bak87}. There, the power-law statistics is believed to be the signature of a gapless \emph{soft mode} (see discussion in Section \ref{Sec:Emergence}) associated with the ongoing phase transition into chaos, whereas the conspicuous contradiction with the fact that, unlike phase transitions, the N-phase has a finite width is circumvented by postulating of the existence of a \emph{mysterious} force that fine-tunes the parameters of the \emph{stochastic} model into the phase transition into chaos. This~understanding of the essence of stochastic dynamics in the N-phase is all but scientific. The~\emph{Goldstone mode} explanation by the STS discussed above resolves this issue. 

\subsubsection{High-Temperature Regime}

In the previous discussion of the weak-noise regime, the concept of noise-induced tunneling processes is well defined because the overlap between the perturbative ground states is exponentially weak. As a result, an external observer will be able to differentiate between tunneling process. At~higher temperatures, the overlap is no longer weak, and it may become difficult for an external observer to differentiate between tunneling events. This suggests that the sharp boundary between the C- and N-{phases} must smear out into a crossover. Note that the boundary between the N- and C-phases is not a $\hat d$-symmetry-breaking phase transition; thus, its disappearance does not contradict any symmetry-based argument.

It can be said that, above a certain temperature, the C- and N-phases must merge into a~complicated phase with spontaneously broken $\hat d$-symmetry. In Figure \ref{Figure_5_5}b, this phase is indicated as an X-{phase}. Borrowing from the terminology of high-energy physics, one way to identify this phase is as  stochastic chaos in the ``strongly coupled'' regime, where strong coupling would mean the strong overlap between the perturbative ground states.

\section{Conclusions and Outlook}
\label{Chap:SecConclusion}


This paper offers a brief introduction to the current state of the recently proposed approximation-free supersymmetric theory of stochastic differential equations (STS). This theory provides several novel theoretical insights into stochastic dynamics. It establishes a rigorous stochastic generalization of the concept of dynamical chaos, which is found to be the phenomenon of the spontaneous breakdown of topological or de Rham supersymmetry that all SDEs possess. This paper also reveals that stochastic chaos is complementary to thermodynamic equilibrium, corresponding in  turn to the unbroken topological supersymmetry. Being the low-symmetry or the ordered phase, the chaotic phase has what can be called a dynamical long-range order, whereas the phase of the thermodynamic equilibrium does not. The presence of this order is the reason why many natural, engineered, and social DSs exhibit emergent long-range dynamical behavior such as $1/f$ noise, \emph{i.e.}, long-term memory effects; the butterfly effect, \emph{i.e.}, sensitivity to the initial conditions; and the algebraic, \emph{i.e.}, scale-free, statistics of sudden or instantonic processes. These and a few other qualitative findings, such as the clarification of the concept of ergodicity, are the main outcomes of the STS so far. Further~work on the STS may lead to more specific and valuable results. As is discussed next, one of the most fruitful directions of further investigation is the work on the methodology of the identification of the dynamical long-range order parameter and construction of the low-energy effective theory (LEET) for it in spatially extended models such as hydrodynamical models. 

The most important qualitative aspect of dynamics under the conditions of the spontaneous breakdown of a global symmetry occupies a reduced phase space. In other words, this aspect occurs in a reduced number of ``low-energy'' variables called the order parameter, and the LEETs are the theories describing this dynamics. In ferromagnets, for instance, the order parameter is the (local) magnetization of the electron liquid, and the LEET (or rather the equations of motion of the LEET) is the Landau--Lifshitz--Gilbert equation. In superconductors, the order parameter is the wavefunction of the Bose--Einstein condensate of the Cooper pairs, and the LEET is the corresponding Ginzburg--Landau theory. In solids, in which the global translational symmetry is broken by the lattice structure, the order parameter is the local displacement of atoms from their average positions in the lattice, and the LEET is the low-energy theory describing, say, the propagation of transverse sound, which is the Goldstone--Nambu particle in this case.

Concerning chaotic DSs, the most important variables are the unstable and/or unthermalized variables of the wavefunction of the non-$d$-symmetric ground state. It is in these variables that a chaotic DS exhibits the infinite memory of perturbations. The local order parameter must be the gapless fermions or goldstinos that are the supersymmetric partners of the unstable bosonic variables. In spatially extended nonlinear models, the unstable variables must be the moduli of the solitonic configurations consisting of fundamental solitons such as kinks, domain walls, and vortices. The processes of the creation/annihilation of (pairs of) the fundamental solitons are the (anti-)instatonic processes, the condensation of which is the essence of the noise-induced chaotic phase discussed in Section \ref{LowTemp}. One of the candidates for such models is a two-dimensional vortex-mediated turbulence, wherein the goldstinos must be the supersymmetric partners of the spatial positions of the~(anti-)vortices.

It is intuitively appealing to believe that the dynamical long-range order of the spontaneously broken topological supersymmetry must at least partially possess some global or topological features. These features may appear on the level of the inter-goldstino interactions. In the above example of the vortex-mediated 2D turbulence, the interaction that remembers the braining between the vortices may as well be approximated as a (chiral) gauge field. Such an LEET would appear somewhat reminiscent of the Schwartz-type topological field theories used in models related to the concept of topological quantum computing. Somewhere down this line of thinking it may turn out that some complex DSs may be useful for the purposes of natural computing. Thus, the development of the methodology of the LEET for the STS may prove fruitful from the point of view of the recent search for new computational~paradigms.

Apart from the methodology of the LEET, there are many other open questions in the STS, even in the interpretational side of the theory. For example, what happens to the wavefunction when one~observes/measures variables? It was argued in Section \ref{SecHilertSpace} that one of the possible (local) interpretations of the wavefunction is that of a generalized probability distribution. Therefore, one~may as well expect that the wavefunction may change suddenly upon observation in a Bayesian update manner. If this is indeed true, yet another question arises of how this change is related to wavefunction collapse in quantum theory. Hopefully, future work will reveal answers to these and other open questions in the STS. It would also be interesting to see how the STS relates to other modern approaches to stochastics such as Stochastic Thermodynamics \cite{Therm1,Therm2,Therm3,Therm3,Sekimoto}.

\vspace{6pt}
\acknowledgments{The author's sincerest gratitude is offered to Kang L. Wang for valuable discussions, support, and encouragement.}


\section*{\noindent Abbreviations}\vspace{6pt}\noindent

The following abbreviations are used in this manuscript:\\

\noindent 
DLRO ---dynamical long-range order \\
DPF---dynamical partition function\\
DS---dynamical system\\
FP operator---Fokker--Planck operator\\
KD---kinematic Dynamo\\
LRDB---long-range dynamical behavior\\
ODE---ordinary differential equation\\
SFE---stochastic flow equation\\
SDE---stochastic \emph{differential} equation\\
SdE---stochastic \emph{difference} equation\\
SEO---stochastic evolution operator\\
STS---supersymmetric theory of stochastics\\
TPD---total probability distribution\\

\appendix
\section{} 
\vspace{-6pt}

\subsection{Differential {\em vs.} Difference Equations: Ito--Stratonovich Dilemma}
\label{ItoStratonovichDilemma}

The goal of this Appendix is to derive the FP equation for the SdE (\ref{WienerSdE}). The latter can be given a~more compact form,
\begin{eqnarray}
\frac{\Delta x}{\Delta t} = {\mathcal F}_n(x + \alpha \Delta x), \label{ItoStratonivichAmb}
\end{eqnarray}
where the subscripts are dropped in $x_{n-1}$ and $\Delta x_{n}$, which herein are simply $x$ and $\Delta x$, and 
\begin{eqnarray}
{\mathcal F}_n(x) = F(x) + (2\Theta)^{1/2} e_a(x)\xi^a_n.\label{mathfn}
\end{eqnarray}

Now, it is assumed that, at time moment $t_{n-1}$, the model is described by the total probability function $P_{n-1}(x)$. The expectation value of some function $f:X\to\mathbb{R}$ is given at this time moment as
\begin{eqnarray}
\overline{f} (t_{n-1}) = \int f(x) P_{n-1}(x)d^Dx.
\end{eqnarray}

At $t_n=t_{n-1}+\Delta t$, this expectation value becomes
\begin{eqnarray}
\overline{f} (t_{n}) &=& \left\langle \int f(x+ \Delta x) P_{n-1}(x)d^Dx \right\rangle_\text{Ns},\label{ftn1}
\end{eqnarray}
where the stochastic averaging is over $\xi_n$. After this stochastic averaging is performed, Equation (\ref{ftn1}) takes the following form:
\begin{eqnarray}
\overline{f} (t_{n}) =\int f(x) P_{n}(x)d^Dx,\label{ftn}
\end{eqnarray}
where
\begin{eqnarray}
P_{n}(x)d^D x = (\hat 1 - \Delta t \hat H^{(D)} +...) P_{n-1}(x)d^Dx,\label{pnpn1}
\end{eqnarray}
with dots denoting terms of higher order in $\Delta t$ and $\hat H^{(D)}$ being the sought after FP operator. In the continuous-time limit, the above expression can be given the familiar form of the FP equation:
\begin{eqnarray}
\partial_t P(t)d^Dx = - \hat H^{(D)} P(t)d^Dx.\label{FPEqD}
\end{eqnarray}

The task now it to establish the explicit expression for $\hat H^{(D)}$.

It is understood that, for small $\Delta t$, $\Delta x$ is also small. In other words, the Taylor expansion of~$\Delta x$ in $\Delta t$ begins with the first-order term. Equation (\ref{pnpn1}) also seemingly implies that it suffices to retain only terms of first order in $\Delta t$. This would indeed be true if it was not for the stochastic averaging over~$\xi_n$. This averaging will transform the terms that are second order in $\Delta t$ and contain two $\xi$ into the first-order terms in $\Delta t$, as is clear from Equation (\ref{AverageNoise}). Therefore, the Taylor expansion up to  second order in $\Delta x$ must suffice for the derivation of the FP operator.

The Taylor expansion of $f$ in Equation (\ref{ftn1}) up to second order in $\Delta x$ gives
\begin{widetext}
\begin{eqnarray}
\overline{f}(t_n) &=& \left\langle \int f(x + \Delta x) P_{n-1}(x)d^D x \right\rangle_\text{Ns}
=\left\langle \int \left(f(x) + f_{'i}(x) \Delta x^i + (1/2) f_{'ij}(x) \Delta x^i\Delta x^j +...\right) P_{n-1}(x)d^D x \right\rangle_\text{Ns}
\nonumber\\
&=&\int f(x)\left\langle \hat 1 - \frac\partial{\partial x^i} \Delta x^i + (1/2) \frac{\partial^2}{\partial x^i\partial x^j} \Delta x^i\Delta x^j +...\right\rangle_\text{Ns} P_{n-1}(x)d^D x. \label{tn}
\end{eqnarray}
\end{widetext}
Here, $f_{'j}\equiv \partial f/\partial x^i$ and similar for $f_{'ij}=\partial^2 f/\partial x^i\partial x^j$, and the partial integration has been used. 

The next step is to Taylor expand $\Delta x$ up to  second order in $\Delta t$ and substitute this expansion into the above expression. Using Equation (\ref{ItoStratonivichAmb}), one has
\begin{eqnarray}
\Delta x^i = {\mathcal F}_n^i\Delta t + \alpha ({\mathcal F}^i_n)_{'j} {\mathcal F}_n^j\Delta t^2 + ...\label{DeltaX1}
\end{eqnarray}

Substituting this expression into Equation (\ref{tn}), using Equation (\ref{mathfn}), and performing the stochastic averaging over $\xi_n$ with the help of Equation (\ref{AverageNoise}), one arrives at
\begin{eqnarray}
\overline{f}(t_n) = \int f(x) (\hat 1 - \Delta t \hat H^{(D)}_\alpha + ...)P_{n-1}(x)d^D x,
\end{eqnarray}
with the FP operator being
\begin{eqnarray}
\hat H^{(D)}_\alpha = -  \frac\partial{\partial x^i} F^i_\alpha(x) - \Theta \frac\partial{\partial x^i} e_a^i(x) \frac\partial{\partial x^j} e^j_a(x)\label{FPOpDStrat}
\end{eqnarray}
and with the $\alpha$-dependent flow vector field from Equation (\ref{Falpha}).

In the above derivation of the FP operator, SdE (\ref{DeltaX1}) was used as a formal equation defining~$\Delta x$. One can take an alternative view on stochastic dynamics in which the dynamics is continuous in time and the noise is piece-wise constant, as  given in Figure \ref{Figure_2_1}. For a fixed noise configuration, one~has a continuous trajectory $x(t)$, defined by $\dot x = {\mathcal F}_n(x(t))$ with the initial condition $x(t_{n-1})=x_{n_1}$. Now,~there is no freedom in choosing $\alpha$ because $\Delta x$ is uniquely defined by the evolution according to the Picard-Lindel\"of theorem. In particular, $\Delta x$ has a unique Taylor expansion in $\Delta t$:
\begin{eqnarray}
\Delta x^i = \left.\frac{\partial x^i}{\partial t}\right|_{\Delta t = 0}\Delta t + \frac12 \left.\frac{\partial^2 x^i}{\partial^2 t}\right|_{\Delta t = 0}\Delta t^2 + ... \label{Taylorx} 
\end{eqnarray}

The first coefficient here is determined from the SDE itself,
\begin{eqnarray}
\left.\frac{\partial x^i}{\partial t}\right|_{\Delta t = 0} = {\mathcal F}_n^i(x),
\end{eqnarray}
whereas the second coefficient is obtained via one differentiation of the SDE over time
\begin{eqnarray}
\left.\frac{\partial^2 x^i}{\partial^2 t}\right|_{\Delta t = 0} = \left.\frac{\partial {\mathcal F}_n^i(x)}{\partial t}\right|_{\Delta t = 0} \nonumber = \\ = {\mathcal F}^i_{n'j}(x) \left.\frac{\partial x^i}{\partial t}\right|_{\Delta t = 0} = {\mathcal F}^i_{n'j}(x) {\mathcal F}_n^j(x);
\end{eqnarray}
thus, the quantity in Equation (\ref{Taylorx}) becomes
\begin{eqnarray}
\Delta x^i = {\mathcal F}_n^i(x)\Delta t + \frac12 {\mathcal F}^i_{n'j}(x) {\mathcal F}_n^j(x)\Delta t^2 + ... \label{Taylorx1} 
\end{eqnarray}

Comparing this equation with Equation (\ref{DeltaX1}), one concludes that  the Stratonovich choice of $\alpha=1/2$ must always be used for the continuous-time picture of temporal evolution.

Concerning the Ito interpretation of SDEs, it is often said that, unlike all other interpretations, the Ito approach respects the Markovian property in the sense that the increment $\Delta x_n$ or, equivalently, the final point $x_n=x_{n-1}+\Delta x_n$ is a function of only $x_{n-1}$ and not of $x_{n}$. This advantage of Ito SDEs, however, is a misinterpretation. Indeed, the very statement that $x_n$ is a function of itself for $\alpha>0$ does not make sense from the point of view of functional dependence. This sentence only tells us that $x_n$ as a function of $x_{n-1}$ is given only implicitly by Equation (\ref{WienerSdE}). For a fixed noise variable $\xi_n$, the final point $x_n$ together with the increment $\Delta x_n$ is always a function of $x_{n-1}$ only. Its explicit expression is given by Equation (\ref{DeltaX1}) up to second order in $\Delta t$, the only accuracy relevant in the continuous-time limit. 

Furthermore, the Markovian property of stochastic processes is concerned not with the trajectories (the variables $x_n$ and $x_{n-1}$) but rather with the temporal evolution of TPDs. In application to the SdEs (\ref{WienerSdE}), the Markovian property means that the TPD at time moment $t_n$ depends on the TPD at the previous time moment $t_{n-1}$ only and not on the TPD at earlier time moments. As clearly observed from Equation (\ref{pnpn1}), which is correct for all $\alpha$, all the interpretations of SDEs satisfy this requirement of Markovianity. In other words, Ito SDEs are just as Markovian as SDEs in all the other interpretations.

In other words, the only advantage of the Ito interpretation is the relative ease of its numerical implementation because the increment as a function of $x_{n-1}$ is given explicitly by the Ito SdE. This~convenience for numerical implementations, however, does not have any significance from the mathematical point of view.

\subsection{Perturbative Supersymmetric Eigenstates}
\label{PertubativeStates}

The correspondence between supersymmetric states and de Rham cohomology classes can be established using standard perturbation theory. The first step is to recall that the Hodge Laplacian from Equation (\ref{HodgeLaplacian}) has a real and non-negative spectrum. Each de Rham cohomology class provides one $\hat d$-symmetric harmonic eigenstate from the kernel of the Hodge Laplacian used previously in Equation~(\ref{Harmonic}):
\begin{eqnarray}
\hat \triangle_{H} |h_k\rangle = 0,
\langle h_k|\hat \triangle_{H} =0.
\end{eqnarray}

All the other eigenstates of the Hodge Laplacian are non-$\hat d$-symmetric and have real and positive eigenvalues. By analogy with Equations (\ref{FirstType}) and (\ref{SecondType}), these non-$\hat d$-symmetric pairs of eigenstates of $\triangle_{H}$ can be denoted as
\begin{subequations}
\label{HodgeEigenvec}
\begin{eqnarray}
&|\zeta_n\rangle=|\underline{\zeta}_n\rangle,
\langle\zeta_n|=\langle\underline{\zeta}_n|\hat d,\\
&\text{and}\nonumber\\
&|\zeta'_n\rangle=\hat d|\underline{\zeta}_n\rangle, \langle \zeta'_n|=\langle\underline{\zeta}_n|,
\end{eqnarray}
\end{subequations}
and their eigenvalues $\Delta_n>0$.

One can now split the SEO into two parts as
\begin{eqnarray}
\hat H = \hat H_0 + \hat V,  \text{ } \hat H_0 = \Theta \hat \triangle_{H}, \nonumber \\ \hat V  = [\hat d, \hat v], \hat v = \hat {\bar d} - \Theta \hat d^\dagger,
\end{eqnarray}
and view $\hat V$ as a perturbation. The ``zeroth-order'' SEO, \emph{i.e.}, $\hat H_0$, is elliptic, whereas the perturbation operator is only linear in spatial derivatives: $\hat V = \hat f^i \frac\partial{\partial x^i} + \hat g$. This implies that the perturbation series must be well defined, e.g., convergent, at least for some class of models and for sufficiently large $\Theta$.

Because $\hat V$ is $\hat d$-exact, the following is true:
\begin{eqnarray*}
\langle \zeta_n| \hat V | h_{k} \rangle &=& \langle \underline{\zeta}_n| \hat d [\hat d, \hat v] | h_{k} \rangle = 0, \\
\langle \zeta_n| \hat V | \zeta'_{m} \rangle &=& \langle \underline{\zeta}_n| \hat d[\hat d,\hat v] \hat d | \underline{\zeta}_{m} \rangle= 0, \\
\langle h_k|\hat V|h_i\rangle &=& \langle h_k|[\hat d, \hat v]|h_i\rangle = 0, \\
\langle h_k|\hat V|\zeta'_i\rangle &=& \langle h_k|[\hat d, \hat v]\hat d|\underline{\zeta}_i\rangle = 0,
\end{eqnarray*}
where Equation (\ref{HodgeEigenvec}) have been used. Using these equalities, it is now clear that, to all orders of the perturbation series, each harmonic form remains a $\hat d$-symmetric eigenstate
\begin{eqnarray}
|\theta_n\rangle = |h_{n}\rangle + \hat d |\tilde\theta_n\rangle,
\end{eqnarray}
where
\begin{eqnarray}
|\tilde \theta_n\rangle &=&
\sum\nolimits_{n_1} | \tilde \zeta_{n_1} \rangle \frac1{-\Theta\Delta_{n_1}}
\bigg(
\langle \tilde \zeta_{n_1} | \hat V | h_\text{n} \rangle \nonumber\\
&&+
\sum \nolimits_{n_2} \frac 1 {-\Theta \Delta_{n_2}}
\langle \tilde \zeta_{n_1} | \hat V \hat d | \tilde \zeta_{n_2} \rangle \langle \tilde \zeta_{m_2} | \hat V| h_{n} \rangle + ...
\bigg).
\end{eqnarray}

Similarly, the bra of this $\hat d$-symmetric state is
\begin{eqnarray}
\langle\theta_n| = \langle h_{n}| + \langle \tilde\theta_n|\hat d.
\end{eqnarray}

Thus, within the domain of the applicability of the perturbation theory, each de Rham cohomology class provides one supersymmetric eigenstate.

\subsection{Kinematic Dynamo as an Example of Both Types of Supersymmetry-Breaking Spectra}
\label{sec:Appendix_KD}

In this Appendix, it is discussed how the theory of KD is related to STS and how this relation provides examples of the two supersymmetry-breaking spectra in Figure \ref{Figure_3_2}b,c.

The KD is a part of the more general hydromagnetodynamical phenomenon of the magnetic dynamo. The latter is the ability of a moving conducting medium to generate and/or sustain a magnetic field \cite{1980opp..bookR....K}. Many astrophysical objects exhibit magnetic dynamos, including galaxies \cite{2007A&A...470..539B,2006A&A...453..447E}, galaxy clusters \cite{2014MNRAS.445.3706V}, stars \cite{2008ApJ...676.1262B}, and planets, including the Earth \cite{1997Natur.389..371K}. In turn, the KD is the linear regime of a magnetic dynamo when a relatively weak magnetic field is generated by a stationary flow of the conducting medium. The KD is realized, e.g., in the early stages of the formation of galaxies. 

The temporal evolution of the magnetic field within the KD effect is governed by the induction~equation
\begin{eqnarray}
\partial_{t}B & = & \hat{\partial}\times v\times B+\eta\hat{\triangle}B.\label{StochInductionEq}
\end{eqnarray}

Here, $\hat{\partial}$ is the gradient operator of the Euclidean space $X=\mathbb{R}^3$; $\hat{\triangle}=\hat \partial_i\hat \partial_i$ is the standard Laplace operator; $\times$ denotes the vector product; $\hat\partial\times$~is the curl of a vector; $B$ is the magnetic field vector; $v$~is the vector field of the underlying flow of the conducting medium; and $\eta=1/\sigma\mu$ is the magnetic diffusivity, with $\sigma$ and $\mu$ being the electrical conductivity and permeability, respectively. The first term represents the well-known magnetohydrodynamical phenomenon of the ``freezing'' of the magnetic field into the conducting medium, whereas the second term is the magnetic field diffusion.

In the theory of KD, the ``phase space'' is non-compact $\mathbb{R}^3$. On the other hand, in this paper, compact (and closed) phase spaces are under consideration. This problem can be circumvented. The~point is that the spatial structures of the KD magnetic fields have local support, and one can always compactify the phase space into the 3D sphere at spatial infinity without affecting the structure of the KD magnetic fields.

Equation (\ref{StochInductionEq}) can be presented in a coordinate-free form. Instead of the vector $B$, one can equivalently use the 2-form representing the magnetic field
\begin{eqnarray}
\mathcal{B}& = & \frac{1}{2!}\mathcal{B}_{ij}dx^{i}\wedge dx^{j}=\hat{d}A,\label{Vector Potential}
\end{eqnarray}
where $A=A_{i}dx^{i}$ is the 1-form of the vector potential. In components, $\mathcal{B}_{ij}$, called the magnetic field tensor, is
\begin{eqnarray*}
\mathcal{B}_{ij} & = & \epsilon_{ijk}B^{k}=\partial_{i}A_{j}-\partial_{j}A_{i}=\left(\begin{array}{ccc}
0 & B^{z} & -B^{y}\\
-B^{z} & 0 & B^{x}\\
B^{y} & -B^{x} & 0
\end{array}\right),
\end{eqnarray*}
where $\epsilon_{ijk}$ is the antisymmetric Levi-Civita tensor. Equation (\ref{StochInductionEq}) can now be expressed as
\begin{eqnarray*}
\partial_{t}B^{i}=e^{ipq}\hat{\partial}_{p}e_{qkl}v^{k}B^{l}+\eta\hat{\triangle}B^{i}.
\end{eqnarray*}

Lowering and raising the indexes in the Euclidean space has no effect on the values of the components of the antisymmetric tensor, e.g., $e_{ijk}=e^{ijk}$. Using the identity
\begin{equation}
e_{qkl}e^{ipq}=det\left(\begin{array}{cc}
\delta_{k}^{i} & \delta_{k}^{p}\\
\delta_{l}^{i} & \delta_{l}^{p}
\end{array}\right)\label{eq:determ_2}
\end{equation}
and $\partial_{i}B^{i}=0$, Equation (\ref{StochInductionEq}) can be rewritten as
\begin{eqnarray*}
\partial_{t}B^{i}=-\hat{\partial}_{j}v^{j}B^{i}+B^{j}v_{'j}^{i}+\eta\hat{\triangle}B^{i},
\end{eqnarray*}
where $v_{'j}^{i}=\partial_{j}v^{i}$. Now, using
\begin{eqnarray*}
B^{i}=\frac{1}{2}e^{ikl}\mathcal{B}_{kl},
\end{eqnarray*}
the induction equation can be further transformed as
\begin{eqnarray*}
\partial_{t}\frac{1}{2}e^{ikl}\mathcal{B}_{kl}=-\hat{\partial}_{j}v^{j}\frac{1}{2}e^{ikl}\mathcal{B}_{kl}+\frac{1}{2}e^{jkl}\mathcal{B}_{kl}v_{'j}^{i}+\eta\hat{\triangle}\frac{1}{2}e^{ikl}\mathcal{B}_{kl}.
\end{eqnarray*}

Multiplying both sides of this equation by $e_{iab}$ and summing over index $i$, one arrives at
\begin{eqnarray*}
\partial_{t}\mathcal{B}_{ab}=-\hat{\partial}_{j}v^{j}\mathcal{B}_{ab}+\frac{1}{2}e_{iab}e^{jkl}\mathcal{B}_{kl}v_{'j}^{i}+\eta\hat{\triangle}\mathcal{B}_{ab}.
\end{eqnarray*}

Using the identity
\begin{eqnarray*}
e_{iab}e^{jkl}=det\left(\begin{array}{ccc}
\delta_{i}^{j} & \delta_{a}^{j} & \delta_{b}^{j}\\
\delta_{i}^{k} & \delta_{a}^{k} & \delta_{b}^{k}\\
\delta_{i}^{l} & \delta_{a}^{l} & \delta_{b}^{l}
\end{array}\right),
\end{eqnarray*}
one has
\begin{eqnarray*}
\partial_{t}\mathcal{B}_{ab}=-\hat{\partial}_{j}v^{j}\mathcal{B}_{ab}+\frac{1}{2}\left(2\mathcal{B}_{ab}v_{'j}^{j}-2\mathcal{B}_{jb}v_{'a}^{j}-2\mathcal{B}_{aj}v_{'b}^{j}\right)+\eta\hat{\triangle}\mathcal{B}_{ab},
\end{eqnarray*}
or
\begin{eqnarray*}
\partial_{t}\mathcal{B}_{ab}=-\left(v^{j}\hat{\partial}_{j}\mathcal{B}_{ab}+v_{'a}^{j}\mathcal{B}_{jb}+\mathcal{B}_{aj}v_{'b}^{j}\right)+\eta\hat{\triangle}\mathcal{B}_{ab}.
\end{eqnarray*}

The first term here is the Lie derivative applied to the 2-form (\ref{Vector Potential}). Therefore, Equation (\ref{StochInductionEq}) can also be given as 
\begin{eqnarray}
\partial_{t}\mathcal{B} & = & -\hat{H}_{KD}\mathcal{B},\:\;\hat{H}_{KD}=\hat{\mathcal{L}}_{v}-\eta\hat{\triangle}.\label{EqNew1}
\end{eqnarray}

This result is rather natural. As previously mentioned, the first term in the R.H.S. of Equation~(\ref{StochInductionEq}) is the infinitesimal temporal evolution of the magnetic field ``frozen'' into the conducting medium. This freezing is the evolution solely due to the flow along $v$, and such an evolution is given by the Lie derivative. This also explains why the Lie derivative is also known as the physical derivative.

The Laplacian in the Euclidean space is given as \begin{eqnarray}
\hat{\triangle} & = & -[\hat{d},\hat{d}^{\dagger}],\label{Lapl}
\end{eqnarray}
where $\hat{d}^{\dagger}=-\imath_{i}\delta^{ij}\partial_{j}$ is the codifferential operator defined in Equation (\ref{dConjugate}). Thus, the KD evolution operator~is
\begin{eqnarray}
\hat{H}_{KD} & = & [\hat{d},\hat{\bar{d}}],\label{SUSYHOp}
\end{eqnarray}
where
\begin{eqnarray*}
\hat{\bar{d}} & = & \hat{\imath}_{v}-\eta\hat \imath_{i}\delta^{ij}\hat \partial_{j}
\end{eqnarray*}
and where the identity $\delta^{ij} = \delta^{i}_a\delta^{j}_a$ has been used. It is now clear that Equation (\ref{SUSYHOp}) is the SEO of the following SDE:
\begin{eqnarray}
\dot{r}^{i} & = & v^{i}+(2\eta)^{1/2}\delta_a^i\xi^{a}(t),
\end{eqnarray}
with $\xi(t)\in\mathbb{R}^3$ being Gaussian white noise.

This is the result needed to establish that the supersymmetry-breaking spectra of both types in Figure \ref{Figure_3_2}b,c are realizable. Indeed, it is well established that the eigenvalues of the KD operator with the lowest real part can be not only negative but also complex (see, e.g., \cite{ComplexDyn} and the references therein). The complex eigenvalues indicate that the spatial structure of the growing magnetic field is also rotating. 






\begin{thebibliography}{999}

\bibitem[Aschwanden(2011)]{Asc11}
Aschwanden, M.
\newblock {\em Self-Organized Criticallity in Astrophysics: Statistics of
  Nonlinear Processes in the Universe}; Springer: Berlin/Heidelberg, Germany,
  2011.

\bibitem[Gutenberg and Richter(1955)]{Gut55}
Gutenberg, B.; Richter, C.F.
\newblock Magnitude and energy of earthquakes.
\newblock {\em Nature} {\bf 1955}, {\em 176},~795, doi:10.1038/\linebreak 176795a0.

\bibitem[Beggs and Plenz(2004)]{Beg04}
Beggs, J.M.; Plenz, D.
\newblock Neuronal avalanches are diverse and precise activity patterns that are stable for many hours in cortical slice cultures.
\newblock {\em J. Neurosci.} {\bf 2004}, {\em 24},~5216--5229.

\bibitem[Chialvo(2010)]{Chialvo10}
Chialvo, D.R.
\newblock Emergent complex neural dynamics.
\newblock {\em Nat. Phys.} {\bf 2010}, {\em 6},~744--750.

\bibitem[Preis \em{et~al.}(2011)Preis, Schneider, and Stanley]{Pre11}
Preis, T.; Schneider, J.J.; Stanley, H.E.
\newblock Switching processes in financial markets.
\newblock {\em Proc. Natl. Acad. Sci. USA} {\bf 2011},
  {\em 108},~7674--7678.

\bibitem[Lorenz(1963)]{ButterFly}
Lorenz, E.N.
\newblock Deterministic nonperiodic flow.
\newblock {\em J. Atmos. Sci.} {\bf 1963}, {\em  20},~130???141.

\bibitem[Kogan(1996)]{Kog96}
Kogan, S.
\newblock {\em Electronic Noise and Fluctuations in Solids}; Cambridge University Press: Cambridge, UK, 1996.

\bibitem[Dana \em{et~al.}(2009)Dana, Roy, and Kurths]{BookHeartBrainNoise}
Dana, S.K.; Roy, P.K.; Kurths, J. (Eds.)
\newblock {\em Complex Dynamics in Physiological Systems: From Heart to Brain}; Springer: Berlin/Heidelberg, Germany, 2009.

\bibitem[Musha and Mitsuaki(1997)]{Biology1fNoise}
Musha, T.; Mitsuaki, Y.
\newblock $1/f$ Fluctuations in Biological Systems.
\newblock  In Proceedings of the 19th Annual International Conference of the IEEE on Engineering in Medicine and Biology Society, Chicago, IL, USA, 30~October--2~November~1997; Volume~6, pp. 2692--2697.

\bibitem[Ruelle(2014)]{Rue14}
Ruelle, D.
\newblock Early chaos theory.
\newblock {\em Phys. Today} {\bf 2014}, {\em 67},~9--10, doi:10.1063/PT.3.2291.

\bibitem[Motter and Campbell(2014)]{Mot14}
Motter, A.E.; Campbell, D.K.
\newblock Chaos at fifty.
\newblock {\em Phys. Today} {\bf 2013}, {\em 66},~27--33.

\bibitem[Shepelyansky(2014)]{Shep14}
Shepelyansky, D.
\newblock Early chaos theory.
\newblock {\em Phys. Today} {\bf 2014}, {\em 67},~10, {doi:10.1063/PT.3.2292} .

\bibitem[Ruelle(1995)]{RuelleTurb}
Ruelle, D.
\newblock {\em Turbulence, Strange Attractors, and Chaos}; World Scientific: Singapore, Singapore, 1995.

\bibitem[Davidson(2004)]{Turbulence}
Davidson, P.
\newblock {\em Turbulence: An Introduction for Scientists and Engineers}; Oxford University Press: New York, NY, USA,~2004.

\bibitem[Lewin(1999)]{DynamicalComplexity}
Lewin, R.
\newblock {\em Complexity: Living on the Edge of Chaos}; University of Chicago Press: Chicago, IL, USA, 1999.

\bibitem[Kauffman(1993)]{SelfOrganization}
Kauffman, S.A.
\newblock {\em The Origins of Order: Self-Organization and Selection in Evolution}; Oxford University Press: Oxford, UK, 1993.

\bibitem[Hoyle(2006)]{patternFomration}
Hoyle, R.
\newblock {\em Pattern Formation: An Introduction to Methods}; Cambridge University Press: Cambridge, UK, 2006.

\bibitem[Bak \em{et~al.}(1987)Bak, Tang, and Wiesenfeld]{Bak87}
Bak, P.; Tang, C.; Wiesenfeld, K.
\newblock Self-organized criticality: An explanation of the $1/f$ noise.
\newblock {\em Phys. Rev. Lett.} {\bf 1987}, {\em 59},~381--384.

\bibitem[Breuer and Petruccione(2007)]{QuantumOpt}
Breuer, H.; Petruccione, F.
\newblock {\em The Theory of Open Quantum Systems}; Oxford University Press: Oxford, UK, 2007.

\bibitem[Mandt \em{et~al.}(2015)Mandt, Sadri, Houck, and T??reci]{SDEQuuantum}
Mandt, S.; Sadri, D.; Houck, A.A.; T\"ureci, H.E.
\newblock Stochastic differential equations for quantum dynamics of spin-boson networks.
\newblock {\em New J. Phys.} {\bf 2015}, {\em 17},~053018.

\bibitem[Tien(2013)]{GinzburgLandauSDE}
Tien, D.N.
\newblock A stochastic Ginzburg-Landau equation with impulsive effects.
\newblock {\em Physica A} {\bf 2013}, {\em 392},~1962--1971.

\bibitem[Ringel and Gritsev(2013)]{Ringel}
Ringel, M.; Gritsev, V.
\newblock Dynamical symmetry approach to path integrals of quantum spin systems.
\newblock {\em Phys.~Rev.~A} {\bf 2013}, {\em 88},~062105.

\bibitem[{\O}ksendal(2010)]{Oks10}
{\O}ksendal, B.
\newblock {\em Stochastic Differential Equations: An Introduction with Applications}; Springer: Berlin/Heidelberg, Germany, 2010.

\bibitem[Kunita(1997)]{Kunita1}
Kunita, H.
\newblock {\em Stochastic Flows and Stochastic Differential Equations}; Cambridge University Press:  Cambridge, UK, 1997.

\bibitem[Baxendale and Lototsky(2007)]{Baxendale1}
Baxendale, P.H.; Lototsky, S.V.
\newblock {\em Stochastic Differential Equations: Theory and Applications}; World Scientific: Singapore, Singapore,~2007.

\bibitem[Arnold(2003)]{Arn03}
Arnold, L.
\newblock {\em Random Dynamical Systems}; Springer: Berlin/Heidelberg, Germany, 2003.

\bibitem[Nobuyuki and Watanabe(1989)]{Watanabe1}
Ikeda, N.; Watanabe, S.
\newblock {\em Stochastic Differential Equations and Diffusion Processes}; North-Holland: Amsterdam, The Netherlands, 1989.

\bibitem[Crauel and Gundlach(1999)]{Crauel1}
Crauel, H.; Gundlach, M.
\newblock {\em Stochastic Dynamics}; Springer: New York, NY, USA, 1999.

\bibitem[Kapitaniak(1990)]{Kap90}
Kapitaniak, T.
\newblock {\em Chaos in Systems with Noise}; World Scientific: Singapore, Singapore, 1990.

\bibitem[{Le Jan}(1984)]{LaJen1}
{Le Jan}, Y.; Watanabe, S.
\newblock Stochastic Flows of Diffeomorphisms. In {\em Stochastic Analysis}, Proceedings of the Taniguchi International Symposium on Stochastic Analysis, Kyoto, Japan, day-month 1982; North-Holland: Amsterdam, The Netherlands, 1984; Volumne~32, pp. 307--332.

\bibitem[Parisi and Sourlas(1979)]{ParSour}
Parisi, G.; Sourlas, N.
\newblock Random magnetic fields, supersymmetry, and negative dimensions.
\newblock {\em Phys. Rev. Lett.} {\bf 1979}, {\em 43},~744--745.

\bibitem[Parisi and Sourlas(1982)]{ParSour1}
Parisi, G.; Sourlas, N.
\newblock Supersymmetric field theories and stochastic differential equations
\newblock {\em Nucl. Phys. B} {\bf 1982}, {\em 206},~321--332.

\bibitem[Cecotti and Girardello(1982)]{CG}
Cecotti, S.; Girardello, L. 
\newblock Stochastic and parastochastic aspects of supersymmetric functional measures: A~new non-perturbative approach to supersymmetry.
\newblock {\em Ann. Phys.} {\bf 1983}, {\em 145}, 81--99.

\bibitem[Cecotti and Girardello(1982)]{CG1}
Cecotti, S.; Girardello, L. 
\newblock A supersymmetry anomaly and the fermionic string.
\newblock {\em Nucl. Phys. B} {\bf 1984}, {\em  239}, 573--582.

\bibitem[Drummond and Horgan(2012)]{DH}
Drummond, I.T.; Horgan, R.R.
\newblock Stochastic processes, slaves and supersymmetry.
\newblock {\emph{J. Phys. A}} {\bf 2012}, {\em 45}, 095005.

\bibitem[Kleinert and Shabanov(1997)]{KS}
Kleinert, H.; Shabanov, S.V. 
\newblock Supersymmetry in stochastic processes with higher-order time derivatives.
\newblock {\emph{Phys.~Lett.~A}} {\bf 1997}, {\em 235},~105--112. 

\bibitem[Olemskoi, Khomenko, Olemskoi (2006)]{Olenskoi1}
Olemskoi, A.I.; Khomenko, A.V.; Olemskoi, D.A.
\newblock Field theory of self-organization.
\newblock {\em Physica A} {\bf 2004}, {\em 332},~185--206.

\bibitem[Kurchan (1992)]{KurchanSpin}
Kurchan, J.
\newblock Supersymmetry in spin glass dynamics.
\newblock {\em Journal de Physique I} {\bf 1992}, {\em 2}, 1333--1352.

\bibitem[Dijkgraaf, Orlando, Reffert(2010)]{Dijkgraaf}
Dijkgraaf, R.; Orlando, D.; Reffert, S.
\newblock Relating field theories via stochastic quantization.
\newblock {\em Nucl. Phys. B} {\bf 2010}, {\em 824},~365--386. 

\bibitem[Gozzi(1984)]{Gozzi0}
Gozzi, E.
\newblock Onsager principle of microscopic reversibility and supersymmetry.
\newblock {\em Phys. Rev. D} {\bf 1984}, {\em 30},~1218, doi:10.1103/PhysRevD.30.1218.

\bibitem[Zinn-Justin(1986)]{ZinnJustin}
Zinn-Justin, J.
\newblock Renormalization and stochastic quantization.
\newblock {\em Nucl. Phys. B {\bf 1986}, 275},~135--159.

\bibitem[Baulieu(1988)]{Bau88}
Baulieu, L.; Grossman, B. 
\newblock A topological interpretation of stochastic quantization.
\newblock {\em Physics Letters B} {\bf 1988}, {\em 212},~351--356.

\bibitem[Baulieu(1989-1)]{Bau89-1}
Baulieu, L.
\newblock Stochastic and topological gauge theories
\newblock {\em Physics Letters B {\bf 1989}, 232},~479--485

\bibitem[Baulieu(1993)]{Bau93}
Baulieu, L. 
\newblock Extended Supersymmetry For Path Integral Representations of Langevin Type Equations 
\newblock {\em Prog. Theor. Phys. Suppl. {\bf 1993}, 111 },~151--162.


\bibitem[Nicolai (1980a)]{Nicolai1}
Nicolai, H.
\newblock Supersymmetry and functional integration measures. 
\newblock {\em Nucl. Phys. B} {\bf 1980}, {\em 176},~419--428.

\bibitem[Nicolai (1980b)]{Nicolai2}
Nicolai, H.
\newblock On a new characterization of scalar supersymmetric theories. 
\newblock {\em Phys. Lett. B}  {\bf 1980}, {\em 89},~341--346.

\bibitem[Frenkel, Losev, Nekrasov(2007)]{Frenkel2007215}
Frenkel, E.; Losev, A.; Nekrasov, N.
\newblock Notes on instantons in topological field theory and beyond.
\newblock {\em Nucl.~Phys.~B} {\bf 2007}, {\em
  171},~215--230.

\bibitem[Birmingham \em{et~al.}(1991)]{Birmingham1991129}
Birmingham, D.; Blau, M.; Rakowski, M.; Thompson, G.
\newblock Topological field theory.
\newblock {\em Phys. Rep.} {\bf 1991}, {\em 209}, 129--340.

\bibitem[Labastida(1989)]{labastida1989}
Labastida, J.M.F.
\newblock Morse theory interpretation of topological quantum field theories.
\newblock {\em Commun. Math. Phys.} {\bf 1989}, {\em 123},~641--658.

\bibitem[Witten(1988{\natexlab{a}})]{Witten98}
Witten, E.
\newblock Topological quantum field theory.
\newblock {\em Commun. Math. Phys.} {\bf 1988}, {\em 117},~353--386.

\bibitem[Witten(1988{\natexlab{b}})]{Witten981}
Witten, E.
\newblock Topological sigma models.
\newblock {\em Commun. Math. Phys.} {\bf 1988}, {\em
  118},~411--449.

\bibitem[Witten(1982)]{Wit82}
Witten, E.
\newblock Supersymmetry and Morse theory.
\newblock {\em J. Differ. Geom.} {\bf 1982}, {\em
  17},~661--692.

\bibitem[Witten(1981)]{DynSusyBrWitten}
Witten, E.
\newblock Dynamical breaking of supersymmetry.
\newblock {\em Nucl. Phys. B} {\bf 1981}, {\em 188},~513--554.

\bibitem[Baulieu(1989)]{Bau89}
Baulieu, L.; Singer, I.M.
\newblock The topological sigma model.
\newblock {\em Commun. Math. Phys.} {\bf 1989}, {\em 125},~227--237.

\bibitem[Gozzi (1998)]{Gozzi1}
Gozzi, E. 
\newblock Universal Hidden Supersymmetry in Classical Mechanics and Its Local Extension.
\newblock In {\em Supersymmetry and Quantum Field Theory};
\newblock Akulov, V.P., Wess, J., Eds.; Springer: Berlin/Heidelberg, Germany, 1998; pp.~166--172.

\bibitem[Gozzi (1989)]{Gozzi2}
Gozzi, E.; Reuter, M.
\newblock Algebraic characterization of ergodicity.
\newblock {\em Phys. Lett. B} {\bf 1989}, {\em 233},~383--392. 

\bibitem[Gozzi (1994)]{Gozzi_New} 
Gozzi, E.; Reuter, M.
\newblock
Lyapunov exponents, path-integrals and forms
\newblock 
{\em Chaos Solitons Fractals} {\bf 1994}, {\em 4}, 1117--1139.

\bibitem[Deotto et.al. (2003)]{Deotto_1} 
Deotto, E.; Gozzi, E.; Mauro, D.
\newblock
Hilbert space structure in classical mechanics. I.
\newblock
{\em J. Math. Phys. {\bf 2003}, 44}, 5902--5936.

\bibitem[Gozzi (1990)]{Gozzi3}
Gozzi, E.; Reuter, M.
\newblock Classical mechanics as a topological field theory.
\newblock {\em Phys. Lett. B} {\bf 1990}, {\em 240},~137--144. 

\bibitem[Deotto and Gozzi(2001)]{Gozzi4}
Deotto, E.; Gozzi, E.
\newblock On the ``Universal'' $N=2$ Supersymmetry of Classical Mechanics.
\newblock {\em Int. J. Mod. Phys. A} {\bf 2001}, {\em 16},~2709--2746.

\bibitem[Niemi and Pasanen(1996)]{Niemi1}
Niemi, A.J.; Pasanen, P.
\newblock Topological $\sigma$-model, Hamiltonian dynamics and loop space Lefschetz number. 
\newblock {\em Phys. Letts. B} {\bf 1996}, {\em 386},~123--130.

\bibitem[Niemi (1996)]{Niemi2}
Niemi, A.J.
\newblock A lower bound for the number of periodic classical trajectories.
\newblock {\em Phys. Letts. B} {\bf 1996}, {\em 386},~123--130. 

\bibitem[Tailleur, {T\"anase-Nicola}, and Kurchan (2006)]{Kurchan}
Tailleur, J.; {T\"anase-Nicola}, S.; Kurchan, J.
\newblock Kramers equation and supersymmetry.
\newblock {\em J. Stat. Phys.} {\bf 2006}, {\em 122}, 557--595.




\bibitem[Mostafazadeh(2002{\natexlab{a}})]{Mos02}
Mostafazadeh, A.
\newblock Pseudo-supersymmetric quantum mechanics and isospectral pseudo-Hermitian Hamiltonians.
\newblock {\em Nucl. Phys. B} {\bf 2002}, {\em 640},~419--434.

\bibitem[Mostafazadeh(2002{\natexlab{c}})]{Mos022}
Mostafazadeh, A.
\newblock Pseudo-Hermiticity {\em versus} PT symmetry: The necessary condition for the reality of the spectrum of a non-Hermitian Hamiltonian.
\newblock {\em J. Math. Phys.} {\bf 2002}, {\em 43},~205--214.

\bibitem[Mostafazadeh(2002{\natexlab{b}})]{Mos021}
Mostafazadeh, A.
\newblock Pseudo-Hermiticity {\em versus} PT-symmetry II: A complete characterization of non-Hermitian Hamiltonians with a real spectrum.
\newblock {\em J. Math. Phys.} {\bf 2002}, {\em 43},~2814--2816.

\bibitem[Mostafazadeh(2002{\natexlab{d}})]{Mos023}
Mostafazadeh, A.
\newblock Pseudo-Hermiticity {\em versus} PT-symmetry III: Equivalence of pseudo-Hermiticity and the presence of antilinear symmetries.
\newblock {\em J. Math. Phys.} {\bf 2002}, {\em 43},~3944--3951.

\bibitem[Mostafazadeh(2013)]{Mos13}
Mostafazadeh, A.
\newblock Pseudo-Hermitian quantum mechanics with unbounded metric operators.
\newblock {\em Philos. Trans. R. Soc. A} {\bf 2013}, {\em 371},~20120050.

\bibitem[Bender \em{et~al.}(1998)Bender, Boettcher, and Meisinger]{Bend98}
Bender, C.; Boettcher, S.; Meisinger, P.
\newblock PT-symmetric quantum mechanics.
\newblock {\em J. Math. Phys.} {\bf 1998}, {\em 40},~2201--2229.

\bibitem[Bender and Boettcher(1998)]{Bend981}
Bender, C.; Boettcher, S.
\newblock Real spectra in non-Hermitian Hamiltonians having PT symmetry.
\newblock {\em Phys. Rev. Lett.} {\bf 1998}, {\em 80},~5243--5246.

\bibitem[Fernandez \em{et~al.}(1998)Fernandez, Guardiola, Ros, and
  Znojil]{Fernandez98}
Fernandez, F.; Guardiola, R.; Ros, J.; Znojil, M.
\newblock Strong-coupling expansions for the PT-symmetric oscillators V(x)=a(ix)+b(ix)(2)+c(ix)(3).
\newblock {\em J. Phys. A} {\bf 1998}, {\em 31},~10105--10112.

\bibitem[Bender \em{et~al.}(1999)Bender, Dunne, and Meisinger]{Bend99}
Bender, C.; Dunne, G.; Meisinger, P.
\newblock Complex periodic potentials with real band spectra.
\newblock {\em Phys. Lett. A} {\bf 1999}, {\em 252},~272--276.

\bibitem[Mezincescu(2000)]{Mezincescu2000}
Mezincescu, G.
\newblock Some properties of eigenvalues and eigenfunctions of the cubic
  oscillator with imaginary coupling constant.
\newblock {\em J. Phys. A} {\bf 2000},
  {\em 33},~4911--4916.

\bibitem[Gawedzki and Kupiainen(1986)]{Gaw86}
Gawedzki, K.; Kupiainen, A.
\newblock Critical behaviour in a model of stationary flow and supersymmetry breaking.
\newblock {\em Nucl. Phys. B} {\bf 1986}, {\em 269},~45--53.

\bibitem[Ovchinnikov(2011)]{Ovc11}
Ovchinnikov, I.V.
\newblock Self-organized criticality as Witten-type topological field theory
  with spontaneously broken Becchi--Rouet--Stora--Tyutin symmetry.
\newblock {\em Phys. Rev. E} {\bf 2011}, {\em 83},~051129.

\bibitem[Ovchinnikov(2012)]{Ovc12}
Ovchinnikov, I.V.
\newblock Topological field theory of dynamical systems.
\newblock {\em Chaos Interdiscip. J. Nonlinear Sci.} {\bf 2012}, {\em 22},~033134.

\bibitem[Ovchinnikov(2013)]{Ovc13}
Ovchinnikov, I.V.
\newblock Topological field theory of dynamical systems. II.
\newblock {\em Chaos Interdiscip. J. Nonlinear Sci.} {\bf 2013}, {\em 23},~013108.

\bibitem[Ovchinnikov(2013)]{Ovc14}
Ovchinnikov, I.V.
\newblock Transfer operators and topological field theory.
\newblock  {\bf 2013}, arXiv:1308.4222.

\bibitem[Ovchinnikov(2016)]{Ovc16}
Ovchinnikov, I.V.
\newblock Supersymmetric Theory of Stochastics: Demystification of Self-Organized Criticality. In~{\em Handbook of Applications of Chaos Theory}; Skiadas, C.H., Skiadas, C., Eds.; Chapman and Hall/CRC, 2016.

\bibitem[Gilmore(1998)]{Gil98}
Gilmore, R.
\newblock Topological analysis of chaotic dynamical systems.
\newblock {\em Rev. Mod. Phys.} {\bf 1998}, {\em 70},~1455--1529.

\bibitem[Hilborn(2000)]{UniversalityInChaos}
Hilborn, R.C.
\newblock {\em Chaos and Nonlinear Dynamics: An Introduction for Scientists and
  Engineers}; Oxford University Press: New York, NY, USA, 2000.

\bibitem[Ruelle(2002)]{Rue02}
Ruelle, D.
\newblock Dynamical zeta functions and transfer operators.
\newblock {\em Not. AMS} {\bf 2002}, {\em 49},~887--895.

\bibitem[Gozzi(1993)]{Gozzi5}
Gozzi, E.
\newblock Stochastic and Non-Stochastic Supersymmetry.
\newblock {\em Prog. Theor. Phys. Suppl.} {\bf 1993}, {\em 111},~115--150.

\bibitem[Intriligator and Seiberg(2007)]{AFewMechanismsForSusybreaking}
Intriligator, K.; Seiberg, N.
\newblock Lectures on Supersymmetry Breaking.
\newblock {\em Class. Quantum Gravity} {\bf 2007}, {\em 24},~S741--S772.

\bibitem[Chung \em{et~al.}(2005)]{SoftSusyBreaking}
Chung, D.J.H.; Everett, L.L.; Kane, G.L.; King, S.F.; Lykken, J.; Wang, L.T.
\newblock The soft supersymmetry-breaking Lagrangian: Theory and applications. 
\newblock {\em Phys. Rep.} {\bf 2005}, {\em 407},~1--203, doi:10.1016/j.physrep.2004.08.032.

\bibitem[Polettini(2013)]{ManifoldsSDE}
Polettini, M.
\newblock Generally covariant state-dependent
diffusion.
\newblock {\em J. Stat. Mech.} {\bf 2013}, {\em 2013}, P07005.

\bibitem[Nakahara(1990)]{Nakahara}
Nakahara, M.
\newblock {\em Geometry, Topology, and Physics}; IOP Publishing: Bristol, UK, 1990.

\bibitem[Coddington and Levinson(1955)]{Theory_Of_ODE}
Coddington, E.A.; Levinson, N.
\newblock {\em Theory of Ordinary Differential Equations}; McGraw-Hill: New York, NY, USA,~1955.

\bibitem[Eckmann and Ruelle(1985)]{RevModPhys.57.617}
Eckmann, J.P.; Ruelle, D.
\newblock Ergodic theory of chaos and strange attractors.
\newblock {\em Rev. Mod. Phys.} {\bf 1985}, {\em 57},~617--656.

\bibitem[Combescure and Robert(2012)]{GrassmannNumbers}
Combescure, M.; Robert, D.
\newblock Fermionic coherent states.
\newblock {\em J. Phys. A} {\bf 2012}, {\em 45},~244005.

\bibitem[Lau and Lubensky(2007)]{Isothermal}
Lau, A.W.C.; Lubensky, T.C.
\newblock State-dependent diffusion: Thermodynamic consistency and its path integral formulation.
\newblock {\em Phys. Rev. E} {\bf 2007}, {\em 76},~011123.

\bibitem[It\'o(1944)]{Ito}
It\^o, K.
\newblock Stochastic integral.
\newblock {\em Proc. Imp. Acad.} {\bf 1944}, {\em 20},~519--524.

\bibitem[Stratonovich(1966)]{Stratonovich}
Stratonovich, R.
\newblock A new representation for stochastic integrals and equations.
\newblock {\em SIAM J. Contr.} {\bf 1966}, {\em 4},~362--371.

\bibitem[Kampen(1981)]{Kampen}
Kampen, N.
\newblock It\'o {\em versus} Stratonovich.
\newblock {\em J. Stat. Phys.} {\bf 1981}, {\em 24},~175--187.

\bibitem[Wong and Zakai(1965)]{Wong}
Wong, E.; Zakai, M.
\newblock On the convergence of ordinary integrals to stochastic integrals.
\newblock {\em Ann. Math. Stat.} {\bf 1965}, {\em
  36},~1560--1564.

\bibitem[Shreve \em{et~al.}(2004)Shreve, Chalasani, and Jha]{Shreve}
Shreve, S.; Chalasani, P.; Jha, S.
\newblock {\em Stochastic Calculus for Finance}; Springer: New York, NY, USA, 2004; Volume 1.

\bibitem[Moon and Wettlaufer(2014)]{Moon14}
Moon, W.; Wettlaufer, J.S.
\newblock On the interpretation of Stratonovich calculus.
\newblock {\em New J. Phys.} {\bf 2014}, {\em 16},~055017.

\bibitem[West \em{et~al.}(1979)]{West}
West, B.J.; Bulsara, A.R.; Lindenberg, K.; Seshadri, V.; Shuler, K.E.
\newblock Stochastic processes with non-additive fluctuations: I. It\^o and Stratonovich calculus and the effects of correlations.
\newblock {\em Physica A} {\bf 1979}, {\em 97},~211--233.

\bibitem[Teschl(2012)]{Teschl}
Teschl, G.
\newblock {\em Ordinary Differential Equations and Dynamical Systems}; American Mathematical Society: Providence, RI, USA, 2012; Volume 140.

\bibitem[Losev(2005)]{TQM} 
Losev, A. \newblock Topological quantum mechanics for physicists. \newblock 
{\em JETP. Lett.} {\bf 2005} {\em 82},~335--342.

\bibitem[Borisov and Ilinski(1994)]{SQM} 
Borisov, N.V.; Ilinski, K.N. 
\newblock $N=2$ supersymmetric quantum mechanics on Riemann surfaces with meromorphic superpotentials. \newblock  
{\em Commun. Math. Phys. } {\bf 1994}, {\em 161},~177--194.

\bibitem[Ovchinnikov and Ensslin(2015)]{Torsten}
Ovchinnikov, I.V.; Ensslin, T.A.
\newblock Kinematic dynamo, supersymmetry breaking, and chaos.
\newblock {\bf 2015}, arXiv:1512.01651.

\bibitem[Hori \em{et~al.}(2003)]{MirrorSymmetry}
Hori, K.; Katz, S.; Klemm, A.; Pandharipande, R.; Thomas, R.; Vafa, C.; Vakil, R.; Zaslow, E.
\newblock {\em Mirror Symmetry}; American Mathematical Society and Clay Mathematics Institute: Cambridge, MA, USA, 2003.

\bibitem[Seinberg(1993)]{Seinberg}
Seinberg, N.
\newblock Naturalness {\em versus} supersymmetric non-renormalization theorems.
\newblock {\em Phys. Lett. B} {\bf 1993}, {\em 318}, 469--475.

\bibitem[Weinberg(1998)]{Weinberg}
Weinberg, S.
\newblock Nonrenormalization theorems in nonrenormalizable theories.
\newblock {\em Phys. Rev. Lett.} {\bf 1998}, {\em 80}, 3702--3705.

\bibitem[{Krause} and {Raedler}(1980)]{1980opp..bookR....K}
{Krause}, F.; {Raedler}, K.H.
\newblock {\em Mean-Field Magnetohydrodynamics and Dynamo Theory}; Elsevier: Amsterdam, The~Netherlands, 1980.

\bibitem[{Beck}(2007)]{2007A&A...470..539B}
{Beck}, R.
\newblock {Magnetism in the spiral galaxy NGC 6946: Magnetic arms,
  depolarization rings, dynamo modes, and helical fields}.
\newblock {\em Astron. Astrophys.} {\bf 2007}, {\em 470},~539--556.

\bibitem[{En{\ss}lin} and {Vogt}(2006)]{2006A&A...453..447E}
{En{\ss}lin}, T.A.; {Vogt}, C.
\newblock {Magnetic turbulence in cool cores of galaxy clusters}.
\newblock {\em Astron. Astrophys.} {\bf 2006}, {\em 453}, 447--458.

\bibitem[{Vazza} \em{et~al.}(2014)]{2014MNRAS.445.3706V}
{Vazza}, F.; {Br{\"u}ggen}, M.; {Gheller}, C.; {Wang}, P.
\newblock {On the amplification of magnetic fields in cosmic filaments and galaxy clusters}.
\newblock {\em Mon. Not. R. Astron. Soc.} {\bf 2014}, {\em 445},~3706--3722.

\bibitem[{Browning}(2008)]{2008ApJ...676.1262B}
{Browning}, M.K.
\newblock {Simulations of dynamo action in fully convective stars}.
\newblock {\em Astrophys. J.} {\bf 2008}, {\em 676},~1262--1280.

\bibitem[{Kuang} and {Bloxham}(1997)]{1997Natur.389..371K}
{Kuang}, W.; {Bloxham}, J.
\newblock {An Earth-like numerical dynamo model}.
\newblock {\em Nature} {\bf 1997}, {\em 389},~371--374.

\bibitem[Li \em{et~al.}(2010)]{ComplexDyn}
Li, K.; Livermore, P.W.; Jackson, A.
\newblock An optimal Galerkin scheme to solve the kinematic dynamo eigenvalue problem in a full sphere.
\newblock {\em J. Comput. Phys.} {\bf 2010}, {\em 229},~8666--8683.

\bibitem[Manning(2006)]{BookEntropy}
Manning, A.
\newblock Topological Entropy and the First Homology Group.
\newblock In {\em Dynamical Systems}; Springer: Berlin/Heidelberg, Germany, 2006;  pp.~185--190.

\bibitem[Lecomte \em{et~al.}(2005)]{TopEntropy1}
Lecomte, V.; Appert-Rolland, C.; van Wijland, F.
\newblock Chaotic properties of systems with Markov dynamics.
\newblock {\em Phys.~Rev.~Lett.} {\bf 2005}, \emph{95},~010601. 

\bibitem[Gaspard(2004)]{TopEntropy2}
Gaspard, P.
\newblock Time-reversed dynamical entropy and irreversibility in Markovian random processes.
\newblock {\em J.~Stat.~Phys.} {\bf 2004}, {\em 117},~599--615.

\bibitem[Muratore-Ginanneschi (2003)]{PathStochastics}
Muratore-Ginanneschi, P.
\newblock Path integration over closed loops and Gutzwiller's trace formula.
\newblock {\em Phys. Rep.  {\bf 2003}, 383},~299--397.

\bibitem[Sekimoto (2010)]{Sekimoto}
Sekimoto, K.
\newblock {\em Stochastic Energetics}; Springer: Berlin/Heidelberg, Germany, 2010.

\bibitem[Gallavotti (2007)]{Therm1}
Gallavotti, G. 
\newblock Fluctuation relation, fluctuation theorem, thermostats and entropy creation in nonequilibrium statistical physics.
\newblock {\em Comptes Rendus Physique} {\bf 2007}, {\em 8},~486--494. 

\bibitem[Maes (1999)]{Therm2}
Maes, C. 
\newblock The fluctuation theorem as a Gibbs property.
\newblock {\em J. Stat. Phys.} {\bf 1999}, {\em 95}, ~367--392.

\bibitem[Polettini and Esposito (2014)]{Therm3}
Polettini, M.; Esposito, M. 
\newblock Transient fluctuation theorem for the currents and initial equilibrium ensembles.
\newblock {\em J. Stat. Mech.} {\bf 2014}, \emph{2014}, P10033.

\bibitem[Altaner (2014)]{Therm4}
Altaner, B.
\newblock Foundations of stochastic thermodynamics.
\newblock {\bf 2014}, arXiv:1410.3983.

\end{thebibliography}
\bibliographystyle{mdpi}
\renewcommand\bibname{References}


\end{document}